\documentclass[aps,prd,twocolumn]{revtex4}
\pdfoutput=1
\usepackage{graphicx}
\usepackage{bm}

 \newcommand{\nc}{\newcommand}
 \nc{\mb}[1]{\makebox[#1]{}}
 \nc{\V}{{\rm v}}
 \nc{\scV}{{\scriptscriptstyle V}}
 \nc{\W}{{\scriptscriptstyle W}}
 \nc{\X}{{\scriptscriptstyle X}}
 \nc{\Z}{{\scriptscriptstyle Z}}
 \nc{\D}{{\scriptscriptstyle D}}
 \nc{\F}{{\scriptscriptstyle F}}
 \nc{\G}{{\scriptscriptstyle G}}
 \nc{\A}{{\scriptscriptstyle A}}
 \nc{\N}{{\scriptscriptstyle N}}
 \nc{\Ss}{{\scriptscriptstyle S}}
 \nc{\CSV}{{\scriptscriptstyle CSV}}
 \nc{\CS}{{\scriptscriptstyle CS}}
 \nc{\SV}{{\scriptscriptstyle SV}}
 \nc{\PV}{{\scriptscriptstyle PV}}
 \nc{\be}{\begin{equation}}
 \nc{\ee}{\end{equation}}
 \nc{\bea}{\begin{eqnarray}}
 \nc{\eea}{\end{eqnarray}}
 \nc{\ra}{{\rightarrow}}
 \nc{\ppg}{\pi^+\pi^-\gamma}
 \nc{\Rcsv}{{R_{\CSV}}}
 \def\Apv{{A_{\PV}}}
 \def\ApveD{{A^{eD}_{\PV}}}
 \def\Gf{{G_{\F}}}

 \def\geA{{g^e_{\A}}}
 \def\geV{{g^e_{\scV}}}
 \def\gqA{{g^q_{\A}}}
 \def\gqV{{g^q_{\scV}}}
 \def\aOd{{a_1^d}}
 \def\aTd{{a_3^d}}
 \def\aOdz{{a_1^{d(0)}}}
 \def\aTdz{{a_3^{d(0)}}}
 \nc{\ovnu}{{\overline{\nu}}}
 \nc{\nuN}{{\nu N_0}}
 \nc{\nubN}{{\ovnu N_0}}
 \nc{\delmn}{{\delta_{m_{\N}}}}
 \nc{\ovdlmn}{{\overline{\delta}_{m_{\N}}}}
 \nc{\snuNC}{{\langle \sigma^{\nuN}_{\NC}\rangle }}
 \nc{\snubNC}{{\langle \sigma^{\nubN}_{\NC}\rangle }}
 \nc{\snuCC}{{\langle \sigma^{\nuN}_{\CC}\rangle }}
 \nc{\snubCC}{{\langle \sigma^{\nubN}_{\CC}\rangle }}
 \nc{\Rnu}{{R^{\nu}}}
 \nc{\Rnub}{{R^{\ovnu}}}
 \nc{\Ftp}{{F_2^{\mu p}}}
 \nc{\Ftn}{{F_2^{\mu n}}}
 \nc{\sintW}{{\sin^2 \theta_{\W} }}
 \nc{\costW}{{\cos \theta_{\W} }}
 \nc{\costtW}{{\cos^2 \theta_{\W} }}
 \nc{\WTdelt}{{\widetilde{\delta}}}
 \nc{\mWsq}{{M_{\W}^2}}
 \nc{\mZsq}{{M_{\Z}^2}}
 \nc{\xF}{{x_{\F}}}
 \nc{\sigS}{{\sigma_{\Ss}}}
 \nc{\Afb}{{A_{fb}}}
 \nc{\vp}{{\bf p}}
 \nc{\rz}{{1\over \rho_0^2}}
 \nc{\uv}{{u_{\V}}}
 \nc{\dv}{{d_{\V}}}
 \nc{\sv}{{s_{\V}}}
 \nc{\cv}{{c_{\V}}}
 \nc{\Uv}{{U_{\V}}}
 \nc{\Dv}{{D_{\V}}}
 \nc{\Sv}{{S_{\V}}}
 \nc{\yW}{{y_{\W}}}
 \nc{\delu}{{\delta u}}
 \nc{\deld}{{\delta d}}
 \nc{\dels}{{\delta s}}
 \nc{\dwtilm}{{\delta \widetilde{m}}}
 \nc{\delCSV}{{\delta^{(\CSV)}}}
 \nc{\ubar}{{\overline{u}}}
 \nc{\dbar}{{\overline{d}}}
 \nc{\sbar}{{\overline{s}}}
 \nc{\dubar}{{\delta\ubar}}
 \nc{\ddbar}{{\delta\dbar}}
 \nc{\dsbar}{{\delta\sbar}}
 \nc{\guv}{{g_{\scV}^u}}
 \nc{\gdv}{{g_{\scV}^d}}
 \nc{\gua}{{g_{\A}^u}}
 \nc{\gda}{{g_{\A}^d}}
 \nc{\pis}{{\pi_{\Ss}}}
 \nc{\piv}{{\pi_{\V}}}
 \nc{\piz}{{\pi^0}}
 \nc{\xpi}{{x_\pi}}
 \nc{\deluv}{{\delta \uv}}
 \nc{\deldv}{{\delta \dv}}
 \nc{\delsv}{{\delta \sv}}
 \def\CC{{\scriptscriptstyle CC}}
 \def\NC{{\scriptscriptstyle NC}}
 \def\CS{{\scriptscriptstyle CS}}

 \def\GLS{{\scriptscriptstyle GLS}}

 \def\SA{{S_{\A}}}
 \def\ovSA{{\overline{S}_{\A}}}
 \def\IE{{\it i.e.}}
 \def\EG{{\it e.g.,}}
 \def\EA{{\it et al.}}

\begin{document}
 \thispagestyle{empty}
 \begin{center}
 \textbf{Charge Symmetry at the Partonic Level} \\
  \vspace{0.8cm} 
  J.T. Londergan \\
 \vspace{0.4cm} \textit{Department of Physics and Nuclear Theory
 Center, Indiana University, Bloomington, IN 47404 USA  } \\
 \vspace{0.8cm} J.-C. Peng \\
 \vspace{0.5cm} \textit{Department of Physics, University of Illinois at 
 Urbana-Champaign, Urbana, IL, 61801 USA} \\
 \vspace{0.8cm} A.W. Thomas \\
 \vspace{0.5cm} \textit{Thomas Jefferson National Laboratory, 12000 
 Jefferson Avenue, Newport News, VA, 23606 USA} \\
  \vspace{0.6cm} \today 
\vspace{0.6cm}
 \end{center}

 \begin{flushleft}
This review article discusses the experimental and theoretical status of 
partonic charge symmetry. It is shown how the partonic content of 
various structure functions gets redefined when the assumption of charge 
symmetry is relaxed. We review various theoretical and phenomenological 
models for charge symmetry violation in parton distribution functions. 
We summarize the current experimental upper limits on charge symmetry 
violation in parton distributions. A series of experiments are presented, 
which might reveal partonic charge symmetry violation, or alternatively 
might lower the current upper limits on parton charge symmetry violation. 
 \end{flushleft}

 \newpage

 \newpage

 \section{Charge Symmetry and Parton Distributions}
 \label{Sec:six}
 \mb{.5cm}

The notion of isotopic spin was introduced to account for the strong 
similarity between the proton and neutron. In this picture the proton and 
neutron are defined as two components of a single object, the 
\textit{nucleon}. If the strong interaction $H_s$ does not distinguish between 
the proton and neutron; then $H_s$ will 
commute with the isospin vector \textbf{T}, such that 
\be 
[H_s ,{\bf T}] = 0 \ .
\label{eq:Hstrong}
\ee
A strong interaction satisfying Eq.~(\ref{eq:Hstrong}) is said to satisfy  
charge independence. \textit{Charge symmetry} is a specific operation 
involving the isospin vector. It is defined as a rotation of $180^\circ$ 
about the ``2'' axis in isospin space. Thus the charge symmetry operator 
$P_{\CS}$ is defined as 
\be 
P_{\CS} = e^{i\pi T_2} \ . 
\label{eq:Pcs}
\ee
Charge symmetry involves interchanging a proton and neutron. When operating 
on light quarks, the charge symmetry operator interchanges 
up and down quarks, namely 
\be 
P_{\CS} |u\rangle = -|d\rangle \ ; \hspace{1.0cm} P_{\CS} |d\rangle 
  = |u\rangle \ . 
\label{Pcsud}
\ee
In nucleon isospin space, the operation of charge symmetry thus interchanges 
up and down quarks (also up and down antiquarks), while interchanging proton 
and neutron labels.  
 
 As is well understood, inclusive processes 
 at high energies can be described in terms of a small number of structure 
 functions, and these structure functions can be characterized in 
 terms of \textit{parton distribution functions}, or PDFs, 
 that describe the probability of finding a given flavor quark or antiquark
 with a fraction $x$ of the nucleon's momentum. Over the past 30 years, 
 increasingly precise measurements have been made of the PDFs and their 
 dependence on $x$ and $Q^2$.  If we assume that charge symmetry is obeyed 
 at the level of parton distributions, this implies the relations
 \bea
 u^p(x,Q^2) &=& d^n(x,Q^2)\,; \nonumber \\  
 d^p(x,Q^2) &=&  u^n(x,Q^2)\,; \nonumber \\
 s^p(x,Q^2) &=& s^n(x,Q^2) \equiv s(x,Q^2) \,;  \nonumber \\
 c^p(x,Q^2) &=& c^n(x,Q^2) \equiv c(x,Q^2) \, . 
 \label{eq:parton1}
 \eea
 In Eq.~(\ref{eq:parton1}), the superscript describes the target nucleon and 
the quantities $u,d,s$ and $c$ represent the flavor of the struck quark. 
Relations analogous to Eq.~(\ref{eq:parton1}) are obtained by replacing all 
quark distributions by antiquarks.

 Until recently, all quark/parton phenomenological models assumed the validity 
 of charge symmetry at the outset. This was a sensible assumption for several  
 reasons. First, charge symmetry is obeyed to a very high precision at low 
 energies; whereas in many nuclear reactions isospin symmetry is obeyed only 
 to the level of a few percent, in most cases 
 charge symmetry is valid to better than one percent \cite{Mil90,Hen79}.   
 Recent precise measurements of charge symmetry violation (CSV) in single-pion 
 production in few-body systems \cite{Opp03,Ste03} have led to better 
 understanding of charge symmetry at low energies 
 \cite{Mil05,Nis99,VK00,Hen90,Fra89}. 
 The high precision of charge symmetry at low energies makes it natural to 
 assume that charge symmetry is valid at high energies; 
 indeed, it is difficult to imagine a scenario with large charge symmetry 
 violation at the partonic level, which would lead to  
 very small CSV at low energies. Second, the assumption of charge 
 symmetry reduces by a factor of two the number of independent quark 
 PDFs that must be determined.  Third, early measurements of high energy 
 structure functions showed that the requirements of charge symmetry were 
 at least qualitatively obeyed \cite{Whi90,Meyers,Macfar,Ben90,BCDMS,Whi90b}.

 In 1998 the current situation regarding parton charge symmetry was 
 reviewed~\cite{Lon98a}. Since then there have been 
 several developments that warrant an updated review. At that time, 
 comparison of the $F_2$ structure functions from charged lepton deep 
 inelastic scattering (DIS) and 
 neutrino charge-changing DIS \cite{Sel97,Sel97a} suggested substantial CSV 
 contributions in the nucleon sea \cite{Bor97,Bor99}. However, re-analysis of 
 the neutrino reactions \cite{Yan01,Bor99a} removed the discrepancies that 
 appeared at that time to indicate the possibility of surprisingly large CSV 
 effects~\cite{Bor97}.  

 In the past few years CSV terms have for the first time been included in 
 global fits to high energy data~\cite{MRST03}. Although these global fits 
 contain some model dependence, nonetheless such fits allow one to 
 set phenomenological limits on CSV contributions to PDFs. In addition, 
 another mechanism for isospin violation in PDFs (quantum 
 electromagnetic or ``QED splitting'' effects) 
 has now been included in calculations of PDFs \cite{MRST05,Glu05} through 
 modification of what is termed DGLAP evolution~\cite{Gri72,Alt77,Dok77}. 
 Inclusion of these QED splitting terms also leads to  
 CSV effects in parton distribution functions.   

 The phenomenological limits for PDFs obtained from global fits to high 
 energy data provide effective upper limits for the magnitude of CSV effects. 
 As we shall see, these limits are somewhat larger than those obtained from 
 theoretical estimates of partonic CSV contributions. Using these 
 phenomenological estimates provides limits to the size of CSV effects that 
 might reasonably be observed in certain experiments. We will use the 
 phenomenological limits obtained from global fits to estimate the 
 maximum value of CSV effects that might be seen in dedicated experiments. 
 This will provide at least qualitative estimates of the size of effects 
 that could be observed in various experiments. It will also provide 
 guidance as to the most promising experiments that could tighten the 
 existing upper limits on parton CSV. 
  
 Since the publication of our previous review on parton CSV, the NuTeV 
 group has measured total cross sections for $\nu$ and $\ovnu$ charged-current 
 and neutral-current reactions on an iron target \cite{Zel02a,Zel02b}. These 
 measurements allow them to extract an independent measurement of the 
 Weinberg angle. Their measurement differs by $3\sigma$ from the extremely 
 precise values for the Weinberg angle measured at the $Z^0$ mass 
 \cite{Abb01}. The 
 publication of the NuTeV measurement has led to a great deal of investigation 
 of effects that might explain this result. We 
 will review various `QCD corrections' to the NuTeV result (\IE, corrections 
 within the Standard Model) and in particular we will summarize the 
 potential corrections to the NuTeV measurement from parton CSV.   

 Our review is organized as follows. In Sect. \ref{Sec:genform} we review 
 the general form of high-energy cross sections in terms of structure 
 functions. In Sect. \ref{Sec:sixtwo} we define parton distributions 
 when charge symmetry violation is included. In Sect. \ref{Sec:PDFs} we review 
 the definitions of structure functions in terms of quark/parton 
 distributions. We list the general form of structure functions when one 
 relaxes the assumption of charge symmetry. In Sect. \ref{Sec:sixtwob} we 
 write down relations between leading-order structure functions, and show how 
 the possible presence of parton CSV affects those relations.   

 Sect.~\ref{Sec:seven} reviews both the experimental and theoretical 
 situation regarding CSV in valence quark PDFs. 
 In Sect.~\ref{Sec:MRSTcsv} we will summarize recent global fits of PDFs that 
 allow for charge symmetry violation. Sect.~\ref{Sec:sevenone} reviews 
 various theoretical estimates of CSV in valence quark distributions. 
 We will argue that one can make reasonably model-independent estimates 
 of the magnitude and sign of valence parton CSV. 
 In Sect.~\ref{Sec:QEDsplt} we summarize the phenomenon of `QED splitting,' 
 a new source of partonic CSV that results from inclusion of terms where 
 a quark radiates a photon; this is the electromagnetic analog of the 
 familiar terms where quarks radiate gluons. 

 In Sect.~\ref{Sec:seventwo} we summarize the experimental limits on 
 valence parton CSV. The most rigorous upper limits on partonic CSV come 
 from the ``charge ratio,'' which compares the $F_2$ structure functions 
 measured in charge-changing reactions induced by neutrinos and antineutrinos, 
 with the $F_2$ structure function from charged lepton DIS, in principle 
 both measured on isoscalar targets. This is reviewed in 
 Sect.~\ref{Sec:seventwoone}. In Sect.~\ref{Sec:sixthreetwo} we review 
 at length the NuTeV measurements of $\nu$ and $\bar{\nu}$ reactions on 
 iron targets, and the resulting extraction of the Weinberg angle. We 
 particularly examine potential contributions to this result from partonic 
 CSV. 

 A new nuclear effect has been proposed that will mimic the effects of 
 partonic charge symmetry violation. We call this ``pseudo CSV,'' This 
 is defined in Sec.~\ref{Sec:pseudoCSV}. We show that pseudo CSV effects 
 could make significant contributions to analyses of the NuTeV experiment. We 
 also discuss how this effect could be observed by measuring the nuclear 
 dependence of the EMC effect. 

 Existing and proposed new experimental facilities offer several 
 opportunities for dedicated precision experiments that could significantly 
 improve our chances of observing partonic CSV effects, or alternatively of 
 lowering the current upper limits on such effects. In 
 Sect.~\ref{Sec:seventhree} we summarize four such experiments. We show the 
 size of the effects that are compatible with the current limits on 
 partonic CSV, and we discuss those experimental 
 facilities that would be best suited to such measurements.   

 In Sect.~\ref{Sec:eight} we review the situation regarding sea quark CSV. 
 In contrast to valence quark CSV, where there are reliable and rather 
 model-independent estimates of the magnitude and sign of such effects,  
 it is substantially more difficult either to 
 make theoretical predictions of sea quark CSV, or to conceive of 
 experiments to measure such effects. In Sect.~\ref{Sec:phenCSV} we review 
 theoretical and phenomenological estimates of sea quark CSV. Perhaps 
 surprisingly, the phenomenological fit by Martin, Roberts, Stirling and 
 Thorne (MRST) \cite{MRST03} 
 found evidence for a rather large sea quark CSV effect. One potentially 
 promising way to test sea quark CSV effects involves partonic sum rules. 
 The most popular QCD sum rules involve the first moment of some combination 
 of structure functions. For the purposes of this review, 
 we define the $n^{th}$ moment of a parton distribution $q(x)$ as 
 \be
 \int_0^1 \,x^{n-1}q(x) \,dx \ . 
 \label{eq:nthmom}
 \ee
 The first moment of valence quark CSV effects is necessarily zero, in order 
 to preserve valence quark normalization. Thus the only 
 CSV effects to survive in this integration are from sea quark CSV. In 
 Sect.~\ref{Sec:eightfour} we review the contributions of partonic CSV 
 to the Gottfried, Adler and Gross-Llewellyn Smith sum rules. In 
 Sect.~\ref{Sec:eightfourfour} we review a new sum rule proposed by 
 Ma \cite{Ma92}. Such a sum rule would be uniquely sensitive to sea quark 
 CSV effects.   
 
 In Sect.~\ref{Sec:Summary} we provide a summary and outlook. 

\section{Relations Between High Energy Cross Sections and 
Parton Distributions\label{Sec:Xsects}}
\mb{.5cm}

\subsection{General form of high energy cross sections\label{Sec:genform}}

We can write the cross sections for deep inelastic scattering
in terms of a set of structure functions, which depend on
the relativistic kinematics of the reaction.  Through the
quark/parton model, these structure
functions can in turn be written in terms of quark/parton
distributions \cite{Lea96}. For simplicity, we write the cross sections 
in leading order (LO) in QCD. By now, all phenomenological analyses 
of high energy reactions and structure functions work in next to 
leading order (NLO) or higher \cite{MSTW07,CTEQ6}. At sufficiently high 
energies, quark mass effects are small and can be accounted for with quite 
good precision. Current issues in partonic analyses involve data, particularly 
in neutrino experiments, in the region where the charm quark mass cannot be 
neglected. One way to deal with these issues is to work in a variable
flavor number scheme (VFNS), where one increases the number of active quark 
flavors at various matching points \cite{ACOT}. Recently the CTEQ group 
has examined the effects of quark masses in global fit analyses, 
particularly in extracting the strong coupling constant $\alpha_s$ from 
such global fits \cite{Tun07}. The MRST group has produced a new set 
of parton distributions at next to next to leading order (NNLO) 
\cite{MSTW07}. In their VFNS scheme they introduce discontinuities into 
their coefficient functions that counter the discontinuities that arise 
in their parton distributions at the matching points.     

\begin{figure}
\center
\hspace{-1.2cm} 
\includegraphics[width=1.9in]{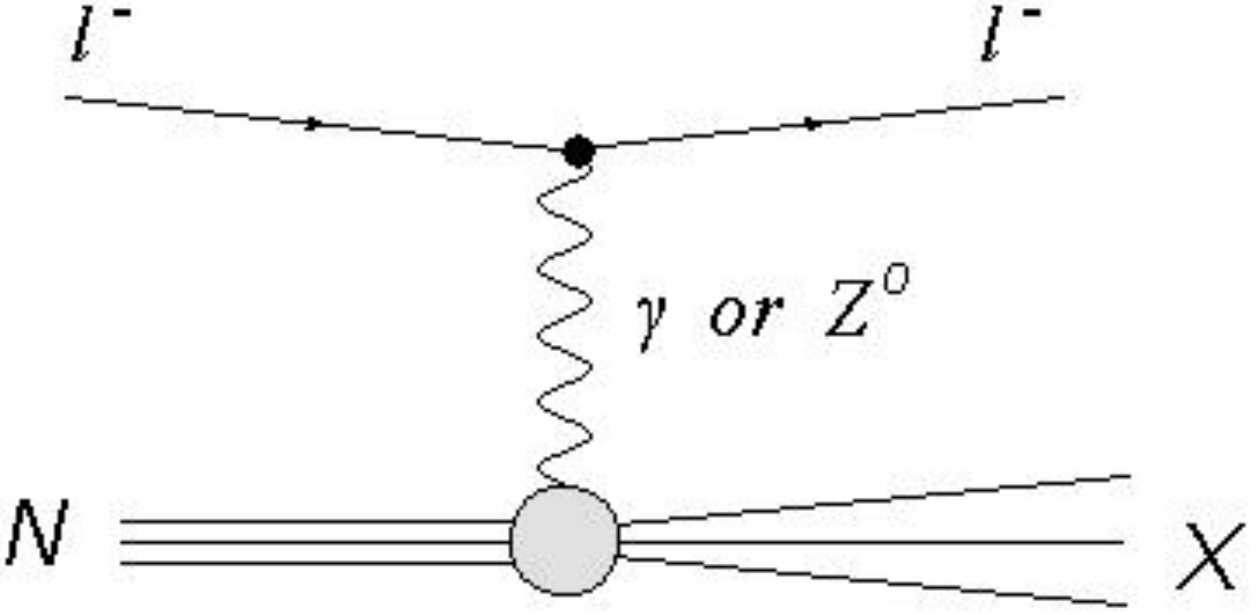}
\caption{Schematic picture of deep inelastic scattering of charged 
leptons from a nucleon. Neutral-current electroweak interactions 
involve exchange of a photon or $Z^0$.}
\vspace{0.1truein}
\label{fig:NClept}
\end{figure}

The cross section for scattering of a left (L) or
right (R) handed charged lepton in neutral current (NC) deep inelastic 
scattering reactions has the form 
\bea
 &\,& {d^2\sigma^{L,R}_{\NC} \over dx\,dy} = {4\pi\alpha^2 s \over
   Q^4}\,\biggl( \bigl[ xy^2 F_1^\gamma(x, Q^2) \nonumber \\ &+& 
  f_1(x,y)F_2^\gamma(x, Q^2) \bigr] - {Q^2\over (Q^2 + \mZsq)}
  {v_\ell \pm a_\ell \over 2\sin\theta_{\W} \costW} 
  \nonumber \\ &\times& \bigl[ xy^2 F_1^{\gamma Z}(x, Q^2) + 
  f_1(x,y)F_2^{\gamma Z}(x, Q^2) \nonumber \\ 
   &\pm& f_2(y)xF_3^{\gamma Z}(x, Q^2) \bigr] 
  + \left({Q^2\over Q^2 + \mZsq}\right)^2 \nonumber \\ 
  &\times& {v_\ell \pm a_\ell \over 2\sin \theta_\W \costW} 
  \bigl[ xy^2 F_1^Z(x, Q^2) 
  \nonumber \\ &+& f_1(x,y)F_2^Z(x, Q^2)  \pm 
  f_2(y)xF_3^Z(x, Q^2) \bigr] \biggr) . \nonumber \\
\label{sigLRNC}
\eea  
This process is shown schematically in Fig.~\ref{fig:NClept}.  In 
Eq.~(\ref{sigLRNC}) $\alpha$ is the electromagnetic coupling, $M_{\Z}$ is the 
mass of the $Z^0$ boson, and $\theta_{\W}$ is the Weinberg angle. 
We define the quantities 
\bea
f_1(x,y) &\equiv& 1 - y - \frac{xyM^2}{s} \ , \nonumber \\ 
f_2(y) &\equiv& y - \frac{y^2}{2} \ . 
\label{eq:f12}
\eea
In Eq.~(\ref{eq:f12}), $M$ is the nucleon mass. These equations are usually 
evaluated at very high energies where 
$xyM^2 << s$, so we generally neglect this term; in this case for the 
remainder of this paper we will approximate $f_1(x,y) \approx f_1(y) = 
1-y$. 
  
Either a photon or $Z$ boson can be exchanged in this process.  
The relativistic invariants in Eq.~(\ref{sigLRNC}) 
are $Q^2 =-q^2$, the square of the four 
momentum transfer for the reaction, $x$ and $y$.  For four
momentum $k$ ($p$) for the initial state lepton (nucleon), 
we have the relations 
\bea 
  x &=& {Q^2 \over 2p\cdot q}; \qquad y = {p\cdot q \over p\cdot k};   
  \nonumber \\ s &=& (k+p)^2 \ . 
\label{kinem}  
\eea
Explicit expressions for the structure functions $F_i$ in terms of 
parton distribution functions are given in Sec.~\ref{Sec:PDFs} below. 

\begin{figure}
\center
\hspace{-1.2cm} 
\includegraphics[width=1.9in]{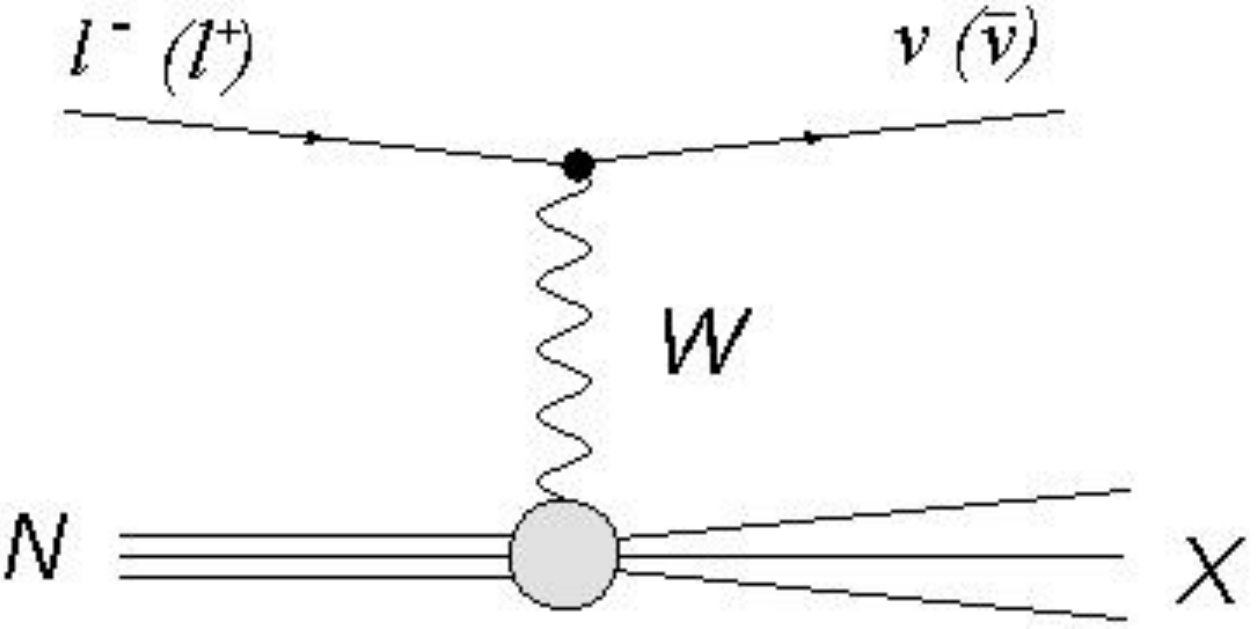}
\caption{Schematic picture of deep inelastic scattering involving the 
charged-current weak interaction initiated by charged leptons.  An 
intermediate $W$ is exchanged between the leptons and the nucleon.}
\vspace{0.1truein}
\label{fig21}
\end{figure}

\begin{figure}
\center
\hspace{-1.2cm} 
\includegraphics[width=1.9in]{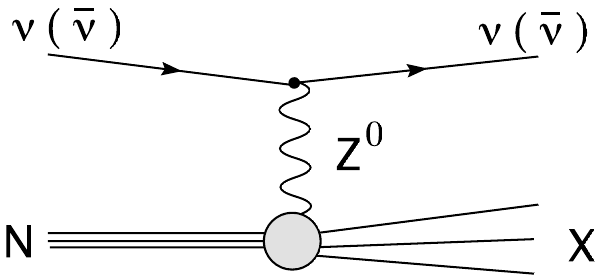}
\caption{Schematic picture of deep inelastic scattering of neutrinos 
through neutral-current interactions mediated by $Z^0$ exchange.}
\vspace{0.1truein}
\label{fig:NCnu}
\end{figure}

In Eq.~(\ref{sigLRNC}), we have 
\bea
  v_{\ell} &=& {-1 + 4 \sintW \over 4 \sin\theta_\W \costW}; 
  \nonumber \\ a_{\ell} &=&  \frac{-1}{4 \sin\theta_\W \costW}\ . 
\eea

The most general form of the cross section for charged current (CC) 
interactions initiated by charged leptons on nucleons can be written 
\bea
 &\,& {d^2\sigma^{l^+ (l^-)}_{\CC} \over dx\,dy} = \frac{\pi s}{2}
   \left( \frac{\alpha}{2 \sintW (\mWsq + Q^2)} \right)^2
  \nonumber \\ &\times& \biggl[ xy^2 F_1^{W^\pm}(x, Q^2) + 
  f_1(y)F_2^{W^\pm}(x, Q^2) \nonumber \\ &\mp& f_2(y)xF_3^{W^\pm}(x, Q^2) 
  \biggr]. 
\label{siglCC}
\eea   
This process is shown schematically in Fig.~\ref{fig21}.  It involves a 
charged virtual $W^\pm$ of momentum $q$ being interchanged between the 
lepton/neutrino vertex, and the hadronic vertex.  In Eq.~(\ref{siglCC}), $M_W$ 
is the mass of the charged weak vector boson.  

Similarly, the cross section for charged current interactions initiated by 
neutrinos or antineutrinos on nucleons has the form 
\bea
  &\,& {d^2\sigma^{\nu (\bar\nu)}_{\CC} \over dx\,dy} = \pi s 
   \left( \frac{\alpha}{2 \sintW (\mWsq + Q^2)} \right)^2
  \nonumber \\ &\times& \biggl[ xy^2 F_1^{W^\pm}(x, Q^2) + 
  f_1(y)F_2^{W^\pm}(x, Q^2) \nonumber \\ &\pm& f_2(y) xF_3^{W^\pm}(x, Q^2) 
  \biggr]\ . 
\label{signuCC}
\eea  
This process is obtained by interchanging the initial and final
state leptons in Fig.~\ref{fig21}.  

Finally, NC reactions initiated by neutrinos or
antineutrinos have the form 
\bea
  &\,& {d^2\sigma^{\nu (\bar\nu)}_{\NC} \over dx\,dy} =  \pi s 
   \left( \frac{\alpha}{2 \sintW \costtW (\mZsq + Q^2)} \right)^2
  \nonumber \\ &\times& \biggl[ xy^2 F_1^Z(x, Q^2) + 
  f_1(y)F_2^Z(x, Q^2) \nonumber \\ &\pm& f_2(y) xF_3^Z(x, Q^2) \biggr]\ .  
\label{signuNC}
\eea   
This process is shown schematically in Fig.~\ref{fig:NCnu}. 

 \subsection{Charge Symmetry and Parton Distribution Functions}
 \label{Sec:sixtwo}

 To obtain the charge symmetry violating parton distributions, we
 introduce the CSV parton distributions for up and down quarks via
 \be
 d^n(x) \equiv u^p(x) - \delta u(x) \,; \hspace{0.5cm} 
 u^n(x) \equiv d^p(x) - \delta d(x) ~~,
 \label{csvdef}
 \ee
 and analogous relations for the antiquark distributions. If the quantities 
 $\delta u(x)$ and $\delta d(x)$ vanish, then charge symmetry is exact.  We 
 assume that the strange quark distributions are the same in both the proton
 and neutron, as are the antistrange distributions.  There is no 
 theoretical or experimental reason to expect strange or charm 
 distributions to vary significantly from proton to neutron. Except at low 
 $x$, the strange and charm distributions are also rather small. 

 It is useful to divide light quark parton distributions into
 valence quark and sea quark parts.  For a given flavor $q$, the
 valence quark distributions in a nucleon are defined by
 \bea
 u_{\rm v}(x) &\equiv& u(x) - \bar{u}(x) ~~, \nonumber \\ 
 d_{\rm v}(x) &\equiv& d(x) - \bar{d}(x) ~~\ .
\label{eq:qvaldef}
 \eea  
We will also use a capital letter to denote 
the second moment of a parton distribution, \IE 
\be 
U_{\V} \equiv \int_0^1 \,dx\, x\uv (x) \ .
\label{eq:momtot}
\ee 
The quantity $U_{\V}$ in Eq.~(\ref{eq:momtot}) gives the total momentum 
carried by up valence quarks in a nucleon.  

For heavy quarks and sea quarks, these often appear as linear combinations 
of the sum and difference of quark and antiquark distributions. For these 
we use the notation 
 \be
 q^{\pm}(x) \equiv q(x) \pm \bar{q}(x) \ .
\label{eq:qpm}
 \ee 
From Eq.~(\ref{eq:qpm}) it is obvious that $\uv(x) = u^-(x)$. However as we 
will examine in detail features of the up and down valence quark PDFs, in 
this review we will use the notation of Eq.~(\ref{eq:qvaldef}) for the 
light valence quarks. 

 The first moments of the valence quark distributions obey the quark 
 normalization conditions; 
 \bea
&\,&  \frac{1}{2}\int_0^1 \,u^p_{\rm v}(x) \,dx = \frac{1}{2}\int_0^1 
 \,d^n_{\rm v}(x) \,dx \nonumber \\ &=& \int_0^1 \,d^p_{\rm v}(x) \,dx = 
  \int_0^1 \,u^n_{\rm v}(x) \,dx = 1 \ \ ; \nonumber \\
&\,&  \int_0^1 \,s^-(x) \,dx = \int_0^1 \,c^-(x) \,dx = 
  0 ~~.
 \label{qnorm}
 \eea
 The CSV quantities defined in Eq.~(\ref{csvdef}) can also be 
 decomposed into valence and sea pieces.  From the definitions of valence 
 quark CSV, and the valence quark 
 normalization from Eq.~(\ref{qnorm}), it is straightforward to
 show that the first moment of the valence quark CSV distributions
 must vanish,  \IE 
 \be
  \int_0^1 \,\delta u_{\rm v}(x) \,dx = 
  \int_0^1 \,\delta d_{\rm v}(x) \,dx = 0 ~~.
 \label{csvnorm}
 \ee
If Eq.~(\ref{csvnorm}) was not true, this would mean that the 
valence quark normalization conditions of Eq.~(\ref{qnorm}) could not be 
satisfied. A consequence of Eq.~(\ref{csvnorm}) is that 
 \bea
  \int_0^1 \,\delta u(x) \,dx &=& \int_0^1 \,\delta \bar{u}(x) \,dx 
  \nonumber \\ {\rm and} \ \ 
  \int_0^1 \,\delta d(x) \,dx &=& \int_0^1 \,\delta \bar{d}(x) \,dx.
 \label{eq:flavint}
 \eea

 The most precise limits on sea quark CSV are derived from QCD sum rules, 
 which involve various moments of the structure functions integrated over 
 all $x$. Some observables will involve the first moment of sea 
 quark parton CSV distributions. Thus we can distinguish two different types 
 of partonic charge symmetry.  The first, or  ``strong form'' of charge 
symmetry, is the statement that charge symmetry violating parton distributions 
 vanish at all $x$.  The ``weak form'' of charge symmetry corresponds to the 
 assumption that the first moment of the CSV sea quark parton distributions 
 is zero, \IE 
 \be
 \int_0^1 \,\delta\bar{u}(x) \, dx = \int_0^1 \,\delta\bar{d}(x) \, dx 
 = 0 ~~, 
\label{eq:weakform}
 \ee 
 even if the parton CSV distributions themselves do not necessarily vanish. 

 Note that valence quark normalization requires that the first moments of 
 heavy quark and antiquark distributions must be identical. From 
 Eq.~(\ref{qnorm}) we see that 
 \be
 \int_0^1 \,s(x) \, dx = \int_0^1 \,\bar{s}(x) \, dx 
\label{eq:sint}
 \ee 
with an analogous relation for charm quarks. This simply reflects the 
statement that the nucleon contains no net strangeness or charm.  
From Eq.~(\ref{eq:sint}), it is tempting to conclude that the strange quark 
and antiquark distributions should be equal for all values
of $x$, \EG 
\be 
s^- (x) \equiv s(x) - \bar{s}(x)= 0 \ , 
\label{eq:sminZ}
\ee
with an identical relation to Eq.~(\ref{eq:sminZ}) for the charm and  
anticharm distributions. If all strange quarks arise from gluon radiation, 
Eq.~(\ref{eq:sminZ}) would be satisfied since $s$ and $\bar{s}$ would always 
be produced in pairs.  However, there is no compelling theoretical reason why 
strange quarks cannot arise from other sources. For example, in 
 ``meson-cloud'' models \cite{Sig87,Ji95,Hol96,Bro96,Mel97,Glu96,Spe97}, which 
 include virtual transitions of a nucleon into a baryon and meson, the quark 
 resides in the nucleon and the antiquark in the meson. Such models naturally 
 lead to differences between quark and antiquark PDFs. We will return to this 
 issue in Sect.~\ref{Sec:sixthreetwo}, where it will 
 be relevant in interpreting the ``NuTeV anomaly'' in the determination of the 
 Weinberg angle.  

Eventually, we would expect partonic charge symmetry violation to be 
calculated directly from lattice gauge theory. This would require 
two additional inputs into current lattice calculations. The first would 
be to input different up and down current quark masses. The second will 
be to include electromagnetic interactions into lattice calculations. 
The first of these should be straightforward. As far as electromagnetic 
interactions are concerned, there are presently lattice calculations that 
include electromagnetic effects. For example, Blum \EA~estimate light quark 
masses by including electromagnetic interactions and calculating pion and kaon 
mass splittings\cite{Blu07}, using two flavors of domain-wall quarks. However, 
for the purpose of testing partonic charge symmetry, one must include 
electromagnetic interactions to sufficient accuracy to account for all of 
the major effects in the nucleon. A good test would be the degree to which 
the inclusion of electromagnetic effects on the lattice can reproduce the 
experimental neutron-proton mass difference of 1.3 MeV.

 \subsection{Structure Functions in Terms of Parton Distribution Functions}
 \label{Sec:PDFs}

 Introducing the CSV parton distributions from Eq.~(\ref{csvdef}),
 we can write the leading order expressions for structure functions without 
 assuming 
 charge symmetry. Most tests of charge symmetry involve deep inelastic
 scattering on isoscalar targets, which we label as $N_0$.  Such
 reactions involve contributions from equal numbers of protons and
 neutrons.  So we write the electromagnetic and weak structure functions per 
 nucleon on an isoscalar target. These expressions are true under the 
 following conditions. First, we have neglected contributions from small 
 components of the CKM quark mixing matrix \cite{Cab63,Kob73}. Second, we 
 assume that we are 
 working at sufficiently high energies and $Q^2$ such that the quark mass can 
 be neglected (this assumption may be inappropriate for the charm quark 
 mass, particularly in the case of charged-current interactions initiated by 
 neutrinos).  These expressions neglect higher-twist contributions 
 to the structure functions. The PDFs depend on the starting scale $\mu^2$ 
 at which they are evaluated; in the following equations we do not explicitly 
 include the dependence upon the starting scale.  

 First we provide expressions for the 
 structure functions relevant to NC reactions induced by charged leptons, 
 given in Eq.~(\ref{sigLRNC}), 
\bea
  &\,& 36F_1^{\gamma N_0}(x,Q^2) = 5\,[ u^+(x) + d^+(x)] 
  \nonumber \\ &+& 2s^+(x)+ 8c^+(x)- 4\deld^+(x) - \delu^+(x)  .
  \nonumber \\  
\label{eq:F1Nzero}
\eea
 In the lowest order quark/parton model, the structure function
 $F_2^{\gamma p}$ is related to the structure function $F_1^{\gamma p}$
 by
 \be
 F_2^{\gamma p}(x,Q^2) = {1+R(x,Q^2)\over 1+4M^2x^2/Q^2}\,2x\,
  F_1^{\gamma p}(x,Q^2)~~.
 \label{Rdenn}
 \ee
 In Eq.~(\ref{Rdenn}), $R = \sigma_L/\sigma_T$ is the ratio of
 the cross section for longitudinally to transversely polarized
 photons.  An empirical relation fit to the
 world's available data on $R$ has been made by Whitlow \EA
 \cite{Whi90}.  This fit covers the kinematic region $x > 0.1$ and
 $Q^2 < 125$ GeV$^2$.

For momentum transfers which are sufficiently small (relative
to $M_Z^2$) and for parity-conserving interactions, we can neglect the 
contribution from $Z$ bosons, in which case the scattering is a function only 
of the two electromagnetic structure functions, $F_1^\gamma$ and 
$F_2^\gamma$, respectively. The cross terms involving $Z$ bosons are 
important either at very large values of $Q^2$, or alternatively for 
parity-violating lepton scattering where the leading terms cancel. The 
structure functions involving photon-$Z$ interference have the form 
\bea 
  &\,&6F_1^{\gamma Z; N_0}(x,Q^2) =  (2\guv - \gdv)[u^+(x) + d^+(x)] 
  \nonumber \\ &+& 2\guv [2c^+(x) - \deld^+(x)] - 
  \gdv [2s^+(x) - \delu^+(x)]\ ; \nonumber \\ 
  &\,&2F_3^{\gamma Z; N_0}(x,Q^2) =  (\guv-\gdv)[\uv(x) + \dv(x)] 
   \nonumber \\  &-& \guv\deldv(x) + \gdv\deluv(x)  \ . 
\label{eq:FgammZ}
\eea  
 The structure functions corresponding to $Z^0$ exchange can be written 
\bea 
  &\,&4F_1^{Z N_0}(x,Q^2) =  (G_u^2 + G_d^2) [u^+(x) + d^+(x)] 
  \nonumber \\ &+& G_d^2 [2s^+(x) - \delu^+(x)] +   
  G_u^2 [ 2c^+(x) - \deld^+(x)]; \nonumber \\ 
  &\,&2F_3^{Z N_0}(x,Q^2) =  (\guv - \gdv)[ \uv(x) + \dv(x)] 
  \nonumber \\ &+& \guv[2c^-(x) - \deldv(x)] -\gdv[2s^-(x) - \deluv(x)] 
  \ . \nonumber \\ 
\label{eq:FZ}
\eea  
In Eqs.~(\ref{eq:FgammZ}) and (\ref{eq:FZ}) we have introduced the relations 
\bea
  \guv &=& \frac{1}{2} - \frac{4}{3}\sintW \ ; \hspace{0.5cm} G_u^2 = 
  (\guv)^2 + \frac{1}{4}; \nonumber \\ 
  \gdv &=& \frac{2}{3}\sintW - \frac{1}{2} \ ; \hspace{0.5cm} G_d^2 = 
  (\gdv)^2 + \frac{1}{4} . \nonumber \\ 
\label{eq:guvdef}
 \eea

For charged-current interactions initiated by either charged leptons or by 
neutrinos, for sufficiently low values of $Q^2$ one must take account of 
the masses of heavy quarks. We do not include bottom and top quark 
effects in this review; however one must account in some way for the 
nonzero charm quark mass. In this case the charged-current structure 
functions will depend on the CKM matrix elements \cite{Cab63,Kob73}. 
Expressions for the charged-current structure functions that take into 
account effects of heavy quark masses and CKM matrix elements can be 
found in the literature \cite{Lea96}. 
However, for sufficiently large values of $Q^2$, the structure functions in 
Eqs.~(\ref{siglCC}) and (\ref{signuCC}) will to a good approximation 
simplify to the form 
\bea
 &\,&2F_1^{W^+ N_0}(x, Q^2) \rightarrow u^+(x) + d^+(x) + 2s(x)
 \nonumber \\ &+& 2\bar{c}(x)- \delta u(x) - \delta\bar{d}(x) \ ;  \nonumber \\
&\,& 2F_1^{W^- N_0}(x, Q^2) \rightarrow  u^+(x) + d^+(x) + 2\bar{s}(x) 
 \nonumber \\ &+& 2c(x) - \delta d(x) - \delta\bar{u}(x) \ ; \nonumber \\
&\,& F_3^{W^+ N_0}(x, Q^2) \rightarrow \uv(x) + \dv(x) + 2s(x) \nonumber \\ 
  &-& 2\bar{c}(x) - \delta u(x) + \delta\bar{d}(x) \ ; \nonumber \\
&\,& F_3^{W^- N_0}(x, Q^2) \rightarrow \uv(x) + \dv(x) - 2\bar{s}(x) 
  \nonumber \\ &+& 2c(x) - \delta d(x) + \delta\bar{u}(x) ~. 
\label{eq:Fccdef}
 \eea
 In Eq.~(\ref{eq:F1Nzero}) and subsequent equations, we have suppressed 
 the nucleon index on the parton distributions.  Since we have 
 explicitly introduced the parton CSV amplitudes, the remaining PDFs  
 are now understood to be those for the proton. The relation between the 
 $F_1$ and $F_2$ structure functions for neutrinos is given by 
 an equation analogous to Eq.~(\ref{Rdenn}), where $R$ is now the 
 longitudinal to transverse ratio that holds for CC and NC neutrino 
 reactions. For the CC reactions initiated by neutrinos, the experimental 
 values for $R^{\nu}$ are summarized in Conrad, Shaevitz and Bolton 
 \cite{Con98}. For NC reactions, the value of $R$ is essentially unknown. 
 This provides some uncertainty in extracting parton distribution functions 
 from neutrino NC reactions \cite{MelPC}. 

 We have written the nuclear structure functions in terms of the parton 
 distributions for free nucleons. However, as is well known, parton 
 distributions are modified in nuclei.  At small $x$ there are shadowing 
 corrections, at intermediate $x$ there are `EMC effects,' 
 \cite{emc,emcb,Geesaman:1995yd}, and at large $x$ 
 Fermi-motion effects dominate.  Nuclear modifications of PDFs have been 
 reviewed recently by Kumano and collaborators \cite{Kum02,Hir03,Hir04}, 
 and also by Kulagin and Petti \cite{Kul06,Kul07,Kul07b}.  Consequently, in 
 any precision experiments these effects must be accounted for if we compare 
 to parton distributions taken from free protons (there should even be small 
 modifications arising in the deuteron \cite{Mel93,Mel94,Mel96}).  We will 
 discuss these effects later as they arise. 

 \subsection{Relations Between Structure Functions}
 \label{Sec:sixtwob}

 Using the relation between leading-order high-energy structure functions 
 given in Eqs.~(\ref{eq:F1Nzero}) and (\ref{eq:Fccdef}), we obtain the 
 following relation between the structure functions, including charge symmetry 
 violating effects.  
 \bea
 &\,&{5\over 18} \overline{F}_2^{W N_0}(x) - F_2^{\gamma N_0}(x) 
  = {1\over 12}\Bigl[ xF_3^{W^+ N_0}(x) \nonumber \\ &-& xF_3^{W^- N_0}(x) 
  \Bigr] \approx  {x\over 12}\,\biggl[ 2( s^+(x) - c^+(x)) 
  \nonumber \\ &+& \delta d^+(x) - \delta u^+(x) \biggr].   
  \label{five18} 
 \eea
 Eq.~(\ref{five18}), sometimes called the ``$5/18^{th}$ rule,'' relates 
 the $F_2$ structure function from charged-current neutrino reactions
 to the $F_2$ structure function from interactions of charged leptons,  
 with both quantities measured on isoscalar targets. In Eq. (\ref{five18}) 
 we have for simplicity neglected the longitudinal to transverse correction 
 factors $R$ given in Eq. (\ref{Rdenn}). However these correction factors 
 were included when the $F_2$ structure functions were extracted 
 from experimental cross sections.   

 In Eq.~(\ref{five18}), 
 $\overline{F}_2^{W N_0}$ is the average of the CC cross sections induced 
 by $\nu$ and $\ovnu$, 
\be
 \overline{F}_2^{W N_0}(x) = \frac{1}{2}\left[ F_2^{W^+ N_0}(x) + 
  F_2^{W^- N_0}(x) \right] \ \ .
\label{eq:F2bar}
\ee
 The right-hand side of Eq.~(\ref{five18}) includes 
 contributions from strange and charmed quarks, and is correct to lowest 
 order in CSV terms. Although the light quark contributions cancel 
 in this expression, there is a residual contribution from strange and 
 charm quarks. At small $x$ one has contributions from both the CSV 
 and heavy quark PDFs. However, at larger values of $x$ the strange and 
 charm contributions should be extremely small, and in this region 
 the only significant contribution to the right-hand side of 
 Eq.~(\ref{five18}) should come from valence quark CSV terms. 
 Furthermore, if the heavy quark contributions are known, Eq.~(\ref{five18}) 
 may be used to investigate the charge-symmetry 
 violating quark distributions for the light quarks. 

 The strange quark PDFs have been determined from the production of 
 opposite-charge muon pairs in neutrino-induced reactions 
\cite{Baz95,Gon01,Kre04,Mas07,Ste95,Fou90,Rab93}.   
 Thus comparison 
 of these two $F_2$ structure functions, combined with our knowledge of 
 strange and charm PDFs, has the potential to measure (or to place 
 strong upper limits on) parton CSV probabilities. The current experimental 
 and theoretical situation will be reviewed in 
 Sec.~\ref{Sec:seventwoone}. From Eq.~(\ref{eq:Fccdef}), we note that in 
 principle we could obtain the same information as in Eq.~(\ref{five18}),  
 by measuring the difference between the $xF_3$ 
 structure functions from charge-changing $\nu$ and $\ovnu$ interactions 
 on isoscalar targets.

 We can obtain another relation between structure functions by measuring 
 the $F_2$ structure function from $\nu$ and $\ovnu$ CC 
 reactions on isoscalar targets. Using Eq.~(\ref{eq:Fccdef}) we obtain
 \bea
&\,&F_2^{W^+ N_0}(x) - F_2^{W^- N_0}(x) = \nonumber \\ &\,& x\,[ 2( s^-(x)  
   - c^-(x)) + \deldv(x)  - \deluv(x) ] . \nonumber \\ 
 \label{Wplmn} 
 \eea
 The right-hand side of Eq.~(\ref{Wplmn}) contains ``valence'' contributions 
 (the difference between quark and antiquark probabilities) for strange 
 and charm quarks, as well as contributions from valence quark CSV terms. In 
 Sec.~\ref{Sec:seventhreethree}, we discuss 
 the experimental possibilities for measuring this quantity, and we 
 show theoretical predictions and phenomenological limits on this quantity. 

 Another relation for structure functions can be obtained by 
 comparing the $F_2$ structure functions obtained from charge-changing 
 interactions of antineutrinos and neutrinos on proton targets. 
 If one takes the difference between these structure functions 
 and divides by $2x$ one obtains 
 \bea 
&\,&{F_2^{W^- p}(x) - F_2^{W^+ p}(x)\over 2x} =  \nonumber \\ 
&\,& \uv(x) - \dv(x) - s^-(x) + c^-(x) . 
 \label{eq:Adler} 
 \eea
 Since the first moment of the strange and charmed ``valence'' 
 contributions must vanish, the difference between these two $F_2$ 
 structure functions, divided by $2x$ and integrated over all $x$, 
 should be equal to the difference between the up and down valence 
 quark occupation numbers in the proton, or one. This is the Adler sum rule 
 \cite{Adler}, which will be reviewed in Sec.~\ref{Sec:eightfourtwo}. 
    
 If the $F_3$ structure functions for neutrino and antineutrino 
 charge-changing reactions are measured on isoscalar targets, then 
 from Eq.~(\ref{eq:Fccdef}) the sum of these structure functions gives  
 \bea
&\,&{xF_3^{W^+ N_0}(x) + xF_3^{W^- N_0}(x)\over 2x} = \uv(x) 
  + \dv(x) \nonumber \\ &+& s^-(x) + c^-(x) - \frac{\deldv(x)+ 
  \deluv(x)}{2} ~~. 
  \label{F3plus}  
 \eea
 The sum of these structure functions includes only valence quark 
 probabilities plus valence CSV contributions.  Consequently integrating 
 Eq.~(\ref{F3plus}) over 
 all $x$, and applying valence quark normalization from Eqs.~(\ref{qnorm}) 
 and (\ref{csvnorm}) gives (modulo QCD corrections) just the sum of valence up 
 and down probabilities in the nucleon, or three.  This is the Gross-Llewellyn 
 Smith sum rule \cite{GLS}, reviewed in Sec.~\ref{Sec:eightfourthree}.

 One final relation can be obtained by comparing the $F_2$ structure 
 function from charged lepton DIS on protons with that for neutrons. 
 One obtains 
 \bea
  &\,&\frac{F_2^{\gamma p}(x) - F_2^{\gamma n}(x)}{x}  = 
  \frac{u^+(x) - d^+(x)}{3} \nonumber \\ 
  &+& \frac{4\delta d^+(x) + \delta u^+(x)}{9} . 
 \label{Gttfrd}
 \eea
 If the quantity in Eq.~(\ref{Gttfrd}) is integrated over all $x$ then one 
 obtains the Gottfried sum rule \cite{Got67}.  
 The experimental and theoretical implications of this sum rule are discussed 
 in Sec.~\ref{Sec:eightfourone}.

 \section{Charge Symmetry in Valence Quark Distributions}
 \label{Sec:seven}
 \mb{.5cm}

 In this section, we will review both theory and experiment 
 regarding charge symmetry in valence quark distributions.  First, we will
 review phenomenological estimates of CSV for valence quarks. Next, we will
 review theoretical estimates of valence CSV, and then we will review the 
 experimental upper limits. Finally we will suggest new experiments that  
 could provide strong constraints on valence quark CSV. 

 \subsection{Phenomenological Estimates of Valence Quark CSV
 \label{Sec:MRSTcsv}}

 Recently, Martin, Roberts, Stirling and Thorne (MRST) \cite{MRST03} have 
 evaluated uncertainties in parton distributions arising 
 from a number of factors, including isospin violation. They  
 chose a specific model for valence quark charge 
 symmetry violating PDFs, adopting a function of the form
\bea 
 \delta u_{\V}(x) &=& - \delta d_{\V}(x) = \kappa f(x) \ ; \nonumber \\ 
  f(x) &=& (1-x)^4 x^{-0.5}\, (x - .0909)  \, .
\label{eq:CSVmrst}
\eea
 The quantity $f(x)$ in Eq.~(\ref{eq:CSVmrst}) was chosen so that its $x$ 
dependence had roughly the same form as the MRST valence quark parton 
distribution functions (at the starting scale for QCD evolution) in both 
the limits $x \rightarrow 0$ and 
$x \rightarrow 1$. The first moment of $f(x)$ was fixed to be zero, in 
 agreement with the valence quark normalization constraint of 
 Eqs.~(\ref{qnorm}) and (\ref{csvnorm}).  The valence quark normalization 
 condition requires that the CSV function $f(x)$ have at least one node.    

 Inclusion of valence quark CSV can in principle change the momentum carried 
by valence quarks in the neutron from those in the proton, since the 
total momentum carried by valence quarks is given by the second 
moment of the distribution. The total momentum carried by valence (up plus 
down) quarks in the neutron is determined experimentally to within about 2\%, 
so MRST chose a functional form that insured that equal momentum was 
carried by valence quarks in the proton and neutron.   
For this reason, they insisted that the valence CSV terms $\delta d_{\V}$ and 
$\delta u_{\V}$ be equal and opposite. With this constraint it is 
straightforward to show that 
\be
 u_{\V}^p(x) + d_{\V}^p(x) = u_{\V}^n(x) + d_{\V}^n(x) \ .
\label{eq:pnmom}
\ee
Multiplying both sides of Eq.~(\ref{eq:pnmom}) by $x$ shows that the 
momentum carried by valence quarks is identical for proton and neutron.  
The overall coefficient $\kappa$ in Eq.~(\ref{eq:CSVmrst}) was varied in a 
global fit to a wide range of high energy data. For simplicity, MRST neglected 
the $Q^2$ dependence of the CSV term in their global fit. Later, when we use 
these charge symmetry-violating PDFs to estimate potential effects of 
partonic CSV, this will introduce some uncertainty. To lowest order in 
QCD, typical CSV effects will have a form proportional to something like  
\be 
\frac{\delu(x) - \deld(x)}{u(x) + d(x)} \ .
\label{eq:CSVratio}
\ee
If we use the MRST CSV PDFs obtained using Eq.~(\ref{eq:CSVmrst}), for a given 
$Q^2$ we will be using parton distributions where the 
denominator has been evolved in $Q^2$ but the numerator is not evolved.

\begin{figure}[ht]
\vspace{0.5cm}
\center
\includegraphics[width=2.8in]{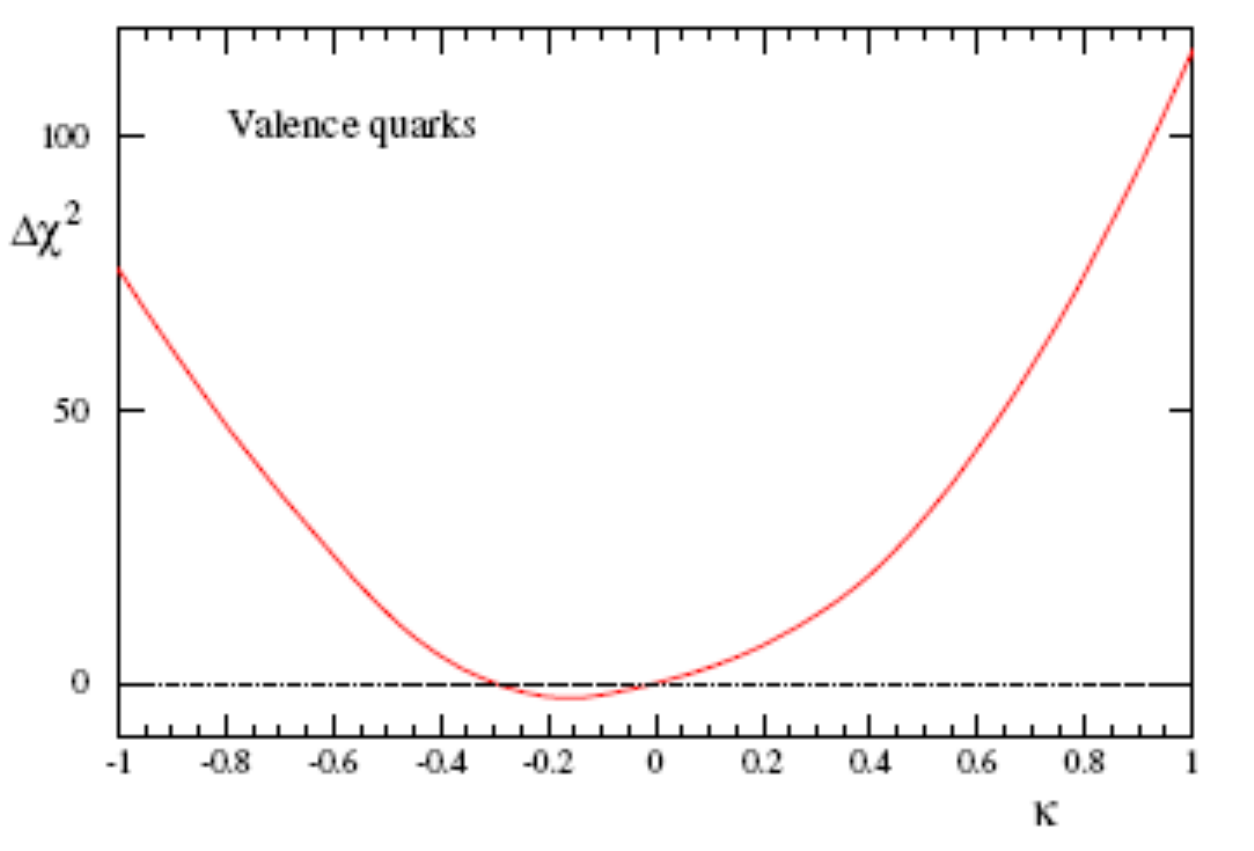}
\caption{[color online] The $\chi^2$ obtained by MRST, 
Ref.\ \protect{\cite{MRST03}} for a 
global fit to high energy data of parton distribution functions including 
valence quark CSV with the functional form defined in 
Eq.~(\protect\ref{eq:CSVmrst}). $\chi^2$ is plotted vs.~the free parameter 
$\kappa$.  
\label{Fig:MRSTval}}
\end{figure}

Including valence quark CSV in their global fit to high energy data, MRST 
obtained a very shallow minimum in $\chi^2$ 
with a best-fit value $\kappa = -0.2$, and a 90\% confidence level 
for the range $-0.8 \le \kappa \le +0.65$.  The $\chi^2$ for their fit 
vs.~the parameter $\kappa$ is shown in Fig.~\ref{Fig:MRSTval}. Since MRST 
chose a very 
specific functional form for valence quark CSV, their results could 
have a substantial model dependence. The MRST global fit guarantees 
that CSV distributions with this shape, and with values of $\kappa$ 
within the 90\% confidence range, will give reasonable agreement with all of 
the high energy data used to extract quark and gluon PDFs.

Since the MRST functional form for valence CSV PDFs requires that $\deldv$ 
be equal in magnitude to $\deluv$, this implies that at large $x$ the 
fractional charge symmetry violation is substantially larger for the 
``minority valence quark'' distribution $\dv$ than for $\uv$, since $\dv << 
\uv$ in this region. Similar results have been obtained for 
valence CSV distributions within a number of 
theoretical models, as will be 
discussed in the following section. 

\begin{figure}[ht]
\center
\includegraphics[width=2.6in]{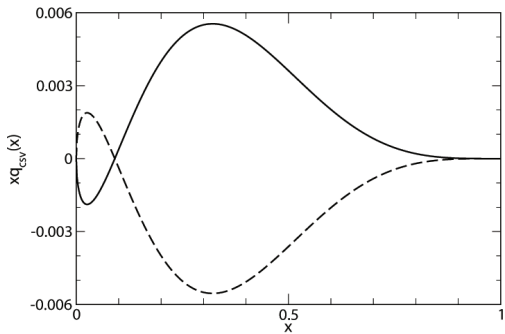}
\caption{The phenomenological valence quark CSV function from Ref.\ 
\protect{\cite{MRST03}}, corresponding to best fit value $\kappa = -0.2$ 
defined in Eq.~(\protect\ref{eq:CSVmrst}). Solid curve: 
$x\deldv$; dashed curve: $x\deluv$.   
\label{Fig:MRSTfx}}
\end{figure}

In Fig.~\ref{Fig:MRSTfx} we show the phenomenological valence quark CSV 
distributions obtained by MRST using the function 
$f(x)$ from Eq.~(\ref{eq:CSVmrst}) with the parameter $\kappa = -0.2$, 
which represents the best fit to the high energy data. The solid curve 
corresponds to $x\delta \dv(x)$ and the dashed curve corresponds to 
 $x\delta \uv(x)$. These valence CSV PDFs reach a maximum value of 
approximately $0.006$ at a value $x \sim 0.3$, and (by inspection of 
Eq.~(\ref{eq:CSVmrst})) they have a zero crossing at $x = 0.0909$. 

 \subsection{Theoretical Estimates of Valence Quark CSV
 \label{Sec:sevenone}}

The phenomenological MRST results of the preceding section can be compared 
with theoretical estimates of valence quark CSV. In a valence quark 
approximation, the nucleon can be considered 
as consisting of three valence quarks, with proton and neutron 
described as 
\bea 
|p \rangle &\sim& [uud] \ , \nonumber \\ 
|n \rangle &\sim& [udd] \ .
\label{eq:valnuc}
\eea   
In quark models evaluated on the light cone, the valence quark distribution 
can be expressed as \cite{Sig89a,Sch91} 
\bea
q_{\V}(x, \mu^2) &=& M \sum_X \, |\langle X |\frac{1+ \alpha_3}{2} \psi(0) 
  | N\rangle |^2  \nonumber \\ &\times& \delta (M(1-x) - p_X^+ ) \ . 
\label{eq:qvx}
\eea
In Eq.~(\ref{eq:qvx}) $\alpha_3 = \gamma^0\gamma^3$; this relation denotes 
the process where a valence quark is removed 
from a nucleon $|N\rangle$, and the result is summed over all final states 
$|X\rangle$. The quantity $p_X^+$ is the energy of the state following removal 
of a valence quark with momentum $k$. The quantity $\mu^2$ represents the 
starting value for the $Q^2$ evolution of the 
parton distribution. Eq.~(\ref{eq:qvx}) treats only the quark longitudinal 
momentum and neglects transverse quark momentum.  

 There are several potential sources of charge symmetry violation in 
 Eq.~(\ref{eq:qvx}). First, there is possible charge symmetry violation in 
 the quark wavefunctions. Second, there are mass differences in the spectator 
 multiquark system. And finally there are additional electromagnetic effects 
which break charge symmetry. Now, model quark wavefunctions are found to be 
almost invariant under the small mass changes typical of CSV \cite{Rod94}, 
so we do not discuss these effects further. Electromagnetic effects are of 
order $\alpha$ where $\alpha$ is the electromagnetic 
 coupling constant; hence such effects are expected to be at the 1\% level.  
 At the large values of $Q^2$ characteristic of high-energy reactions, 
 typically one to 
 many GeV$^2$, such effects should be small.    

 For simplicity, one can consider charge symmetry violation to arise 
 from two effects. The first is the $n-p$ mass difference $\delta M \equiv 
 M_n - M_p = 1.3$ MeV. This mass difference is the result of a number of 
 electromagnetic effects, in addition to the $u-d$ quark mass difference. A 
 second effect giving rise to charge symmetry 
 violation is differences in diquark masses arising from the current 
 quark mass difference between up and down quarks. We define the quantity 
\be 
\dwtilm = m_{dd} - m_{uu}
\label{eq:mtilde}
\ee
One has a robust estimate for this mass difference, $\dwtilm \sim 4$ MeV 
\cite{Bic82}. With these approximations one can write charge symmetry 
violating valence parton distributions in terms of 
\be 
  \delta q_{\V} \approx {\partial q_{\V}\over \partial \dwtilm}\dwtilm 
  + {\partial q_{\V}\over \partial M}\delta M \ .
\label{eq:iso}
\ee
From Eq.~(\ref{eq:iso}) the valence charge symmetry violating parton 
distributions are obtained by taking variations with respect to diquark 
and nucleon masses on valence parton distributions from quark models.
  
If we take the simple valence quark picture of the nucleon as given by 
Eq.~(\ref{eq:valnuc}), then we can consider diquark mass differences 
following the removal of one quark from the nucleon. If we remove a 
``majority'' valence quark (a $u$ quark in the proton or $d$ quark in the 
neutron), then for both proton and neutron  
one is left with a $ud$ diquark. Thus for the majority quark 
distribution, there is no quark mass asymmetry for the residual diquark. 
For removal of a ``minority'' quark, (a $d$ quark in the proton or $u$ quark 
in the neutron), the remainder is a $uu$ diquark in the proton and a $dd$ 
diquark in the neutron. Thus the diquark mass asymmetry is just given 
by the quantity $\dwtilm$ in Eq.~(\ref{eq:mtilde}). 

 This technique was used by Sather \cite{Sat92}, who investigated these 
effects in a static quark model. Sather obtained an analytic approximation 
relating valence quark CSV to derivatives of the valence PDFs   
\bea 
 \delta d_{\V}(x) &=& -\frac{\delta M}{M} \frac{d}{dx} \left[ 
  x d_{\V}(x)\right] - \frac{\delta \widetilde{m}}{M} \frac{d}{dx} d_{\V}(x) 
 \, ,  \nonumber \\  \delta u_{\V}(x) &=& \frac{\delta M}{M} 
 \left( - \frac{d}{dx}\left[ x u_{\V}(x)\right] + \frac{d}{dx} u_{\V}(x) 
 \right) \, . \nonumber \\  
\label{eq:Satanl}
\eea   
Note that Sather's equations agree with our earlier arguments. The 
``majority'' quark CSV distributions $\delta u_{\V}(x) = u_{\V}^p(x) - 
 d_{\V}^n(x)$ are functions only of $\delta M$ and do not depend on 
$\dwtilm$, while the ``minority'' valence quark CSV distributions 
$\delta \dv(x)$ depend on both $\delta M$ and $\dwtilm$. 

\begin{figure}[ht]
\center
\includegraphics[width=2.5in]{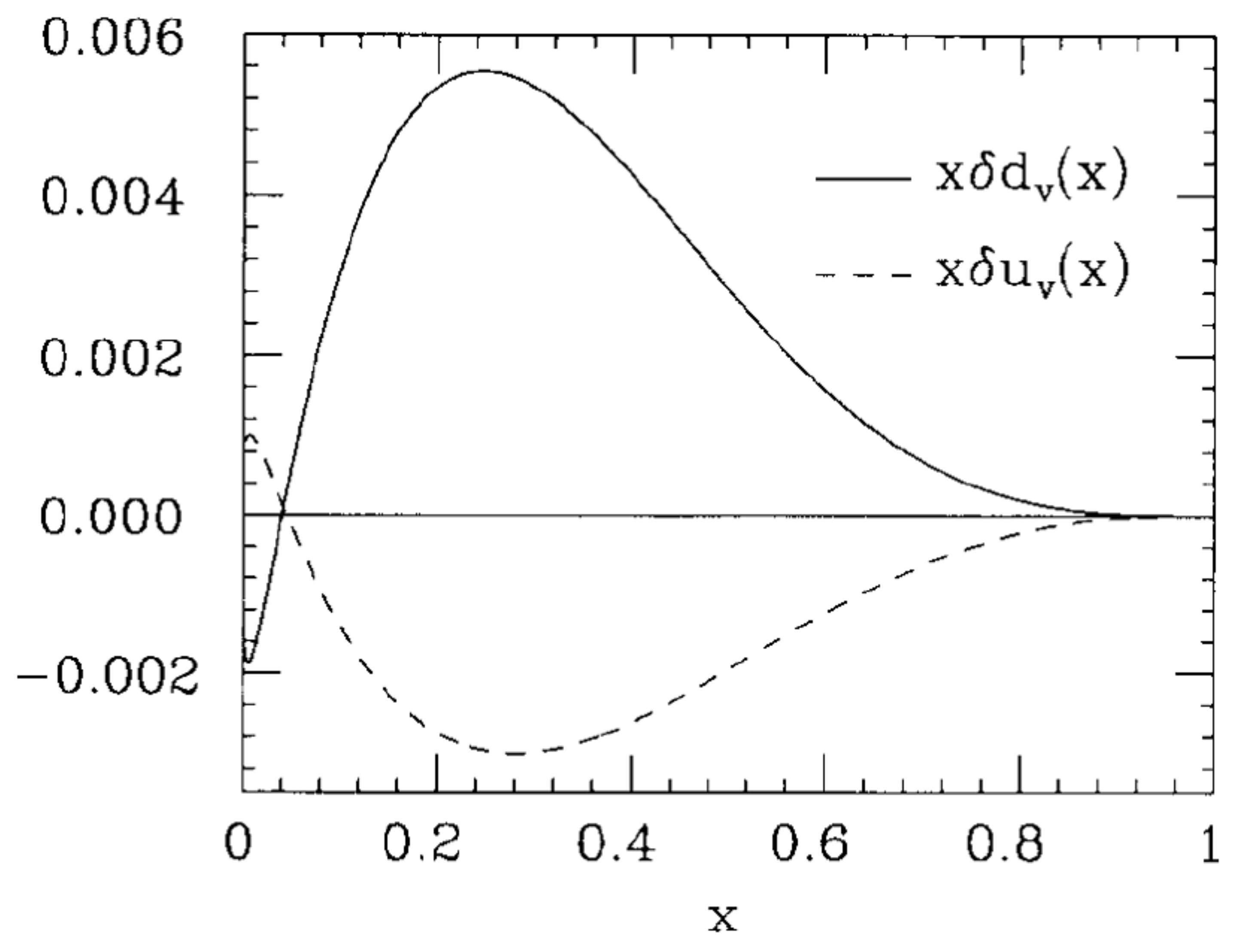}
\caption{Theoretical valence CSV PDFs from Sather, 
Ref. \protect\cite{Sat92} and Eq. (\protect\ref{eq:Satanl}). Solid curve: 
$x\deldv$; dashed curve: $x\deluv$. The valence PDFs were preliminary fits to 
CCFR Tevatron structure functions evolved to $Q^2 = 12.6$ GeV$^2$. 
\label{Fig:Sather}}
\end{figure}

In Fig.~\ref{Fig:Sather} we plot the CSV valence parton distributions 
of Sather \cite{Sat92}. The valence parton distributions used by Sather were 
from preliminary fits to Tevatron structure functions obtained by the CCFR 
group. The solid curve is $x\deldv (x)$ vs.~$x$, while the dashed curve plots 
$x\deluv (x)$. The PDFs have been evolved to $Q^2 = 12.6$ GeV$^2$. The 
qualitative features of these charge symmetry violating valence PDFs 
are similar for all models that we will review. Since the first moment 
of the valence CSV parton distributions has to vanish (in order to 
maintain valence quark normalization), the valence CSV PDFs must change 
sign at least once. For Sather's model this occurs at $x \sim 0.05$.
In general, if one calculates the CSV valence PDFs by inserting 
phenomenonogical parton distribution functions into Eq.~(\ref{eq:Satanl}) 
or other analytic formulae, the resulting CSV parton distributions will 
not obey the valence quark normalization condition. However, 
Sather obtained his PDFs from moments of the quark distributions; for 
his valence CSV parton distributions Sather set the first moment to zero, thus 
guaranteeing valence quark normalization. 
 
For larger values of $x$, $\deldv (x)$ is positive while $\deluv (x)$ is 
negative. The distributions peak at $x \sim 0.3$. In Sather's model 
$\deldv (x)$ is roughly 50\% larger in magnitude than $\deluv (x)$. 
By observing the quark model wavefunctions we can understand these 
qualitative features of the valence quark CSV distributions. 
The ``minority valence quark'' CSV term is defined by 
$\deldv = d_{\V}^p - u_{\V}^n$.  Removing a 
minority valence quark from a nucleon with 3 valence quarks 
leaves a diquark system that is $uu$ for the proton and $dd$ for 
the neutron. Simple theoretical arguments \cite{Rod94} suggest that the 
down quark distribution in the proton will be 
shifted to higher $x$ and the up quark distribution in the neutron 
will be shifted to lower $x$. This predicts that, at large $x$, 
$\deldv$ should be positive.  

Conversely for the ``majority valence quark'' CSV term 
$\deluv = u_{\V}^p - d_{\V}^n$, removing a majority quark leaves 
intermediate states with the same quark configuration ($ud$) for both neutron 
and proton.  From Eq.~(\ref{eq:Satanl}) the majority valence CSV term should 
depend only on the $n-p$ mass difference, and one expects that $\deluv$ should 
be negative at large $x$. These qualitative predictions agree with the quark 
model CSV valence distributions shown in Fig. \ref{Fig:Sather}, and 
they also are in agreement with the phenomenological best-fit CSV PDFs 
shown in Fig.~\ref{Fig:MRSTfx}. Although the down valence distribution in 
the proton is less than half the up valence distribution, these qualitative 
arguments suggest that 
\be
\delta d_{\V}(x) > |\delta u_{\V}(x)|
\label{eq:lgx}
\ee
at large $x$. This is observed in the theoretical model calculations by 
Sather, however Eq.~(\ref{eq:lgx}) is not satisfied by 
the phenomenological MRST parameterization of Eq.~(\ref{eq:CSVmrst}), which 
requires by definition that $\delta d_{\V}(x)$ and 
$\delta u_{\V}(x)$ should be equal in magnitude and opposite in sign.     

There is another way to understand the qualitative features of these 
valence CSV distributions. It was 
pointed out by Londergan and Thomas \cite{Lon03b} that from Sather's 
expression Eq.~(\ref{eq:Satanl}) one can obtain 
an analytic expression for the second moment of the CSV parton distributions, 
\bea
\delta U_{\V} &=& \frac{\delta M}{M}\left( U_{\V} -2 \right) \ ,\nonumber \\ 
  \delta D_{\V} &=& \frac{\delta M}{M}D_{\V} + \frac{\delta \widetilde{m}}{M} 
 \ , \nonumber \\   
 \delta D_{\V} &\approx& \frac{\delta M}{M}\left( D_{\V} + 3 \right) \ .
\label{eq:LTint}
\eea
In Eq.~(\ref{eq:LTint}), $U_{\V}$ and $D_{\V}$ are respectively the 
total momentum carried by up and down valence quarks.  The final line of 
Eq.~(\ref{eq:LTint}) follows from the fact that 
$\delta \widetilde{m} \sim 3\delta M$. Since the valence 
CSV distributions are required by valence quark normalization to have zero 
first moment (see Eq.~(\ref{csvnorm})), the valence CSV distributions must 
change sign at least once. Eq.~(\ref{eq:LTint}) predicts that at large 
$x$ $\delta \uv$ will be negative and $\delta \dv$ will be positive. The  
sign of the valence CSV distributions is the same for all parton distributions 
derived from quark models. Furthermore, from the relative magnitude of 
$\delta U_{\V}$ and $\delta D_{\V}$, we expect the maximum of $\delta d_{\V}$ 
to be larger than that for $\delta u_{\V}$. 

Benesh and Londergan \cite{Ben98} also considered parton charge symmetry 
violation from quark models using Eq.~(\ref{eq:qvx}). They related the 
change in PDFs due to charge symmetry violation in minority valence quark 
distributions (the term proportional to the diquark mass difference 
$\dwtilm$), to the color 
hyperfine splitting between $N$ and $\Delta$ states in quark models of baryons 
that initially assume $SU(4)$ spin-flavor symmetry (these arise from diquark 
spin splittings). Using the work of Close and Thomas \cite{Clo88}, 
they obtained 
\bea 
 \deldv (x) &=& \frac{\dwtilm}{\delta_{hf}} \left[ \frac{ \uv (x) - 
  2\dv (x)}{6}\right] - \frac{\delta M}{M} \frac{d}{dx}\dv(x)   
 \, ,  \nonumber \\  \delta u_{\V}(x) &=& 
   -\frac{\delta M}{M}\frac{d}{dx} u_{\V}(x) \ . \nonumber \\  
\label{eq:BenLon}
\eea   
In Eq.~(\ref{eq:BenLon}), $\delta_{hf} = 50$ MeV is the $S=1$ color hyperfine 
splitting in the $SU(4)$ limit. These equations also differ from Sather's 
result of Eq.~(\ref{eq:Satanl}) in that Benesh and 
Londergan considered variations of the nucleon mass $M$ while keeping 
the quantity $Mx$ constant, following arguments by Benesh and Goldman 
\cite{Ben97a}. 

\begin{figure}[ht]
\center 
\includegraphics[width=2.65in]{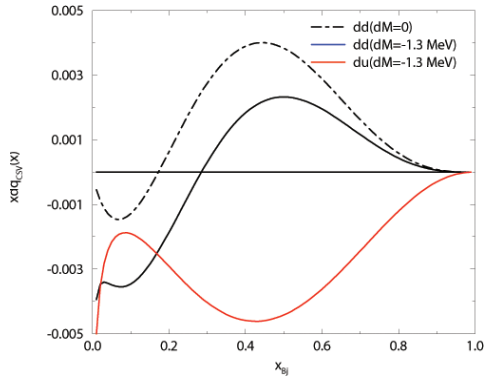}
\caption{Theoretical valence CSV PDFs from Benesh and Londergan, 
Ref. \protect\cite{Ben98} and Eq. (\protect\ref{eq:BenLon}). Thick solid line: 
$x\deldv$; thin solid line: $x\deluv$. Dash-dot line: $x\deldv$ when the n-p 
mass difference is set to zero. The PDFs were the CTEQ LQ (low Q) parton 
distributions from Ref. \protect\cite{Lai96}, evaluated at the low momentum 
starting scale $Q^2 = 0.49$ GeV$^2$. 
\label{Fig:BenLon}}
\end{figure}

In Fig.~\ref{Fig:BenLon} we plot the theoretical valence CSV parton 
distributions calculated by Benesh and Londergan \cite{Ben98}, using 
Eq.~(\ref{eq:BenLon}). The PDFs used were the 
phenomenological CTEQ LQ (low Q) parton distributions from the CTEQ group 
\cite{Lai96}, evaluated at the low momentum starting scale $Q^2 = 0.49$ 
GeV$^2$. 

As we have mentioned (see Eq.~(\ref{csvnorm}), valence parton CSV 
distributions should respect valence quark normalization, and hence 
$\langle \delta d_{\V}\rangle = \langle \delta u_{\V} \rangle = 0$.  
By inspection of Fig.~\ref{Fig:BenLon}, the valence CSV PDFs of Benesh
and Londergan do not satisfy the quark normalization condition. Although 
the sign of the CSV PDFs obtained by Benesh and Londergan agrees with that 
of Sather, and also with the predictions of Eq.~(\ref{eq:LTint}), the 
magnitudes are somewhat different. Here the magnitude of 
$\deluv$ is larger than $\deldv$, a result opposite from Sather and in 
disagreement with Eq.~(\ref{eq:LTint}). The 
difference between these theoretical PDFs is likely related to the fact 
that Benesh and Londergan used an additional approximation to relate quark 
CSV terms to mass splittings in SU(4) symmetric quark models. Note that 
the Benesh-Londergan PDFs are evaluated at a considerably lower value of 
$Q^2$ than for Sather. One could evolve these CSV PDFs to  
higher $Q^2$ by inserting these parton distributions into the QCD 
evolution equations. 

Rodionov, Thomas and Londergan \cite{Rod94} also calculated charge symmetry 
violating parton 
distributions using Eq.~(\ref{eq:qvx}). They also included the quark 
relativistic energy (if one ignores this the resulting parton distributions 
do not have the correct support; the PDFs are then defined on the range 
$ 0 \le x < \infty$, rather than from $0 \le x \le 1$). Rodionov \EA~also 
evaluated Eq.~(\ref{eq:qvx}) including the effects of quark transverse 
momentum. In this case one can no longer obtain analytic expressions for the 
CSV valence parton distributions. 

\begin{figure}[ht]
\includegraphics[width=2.7in]{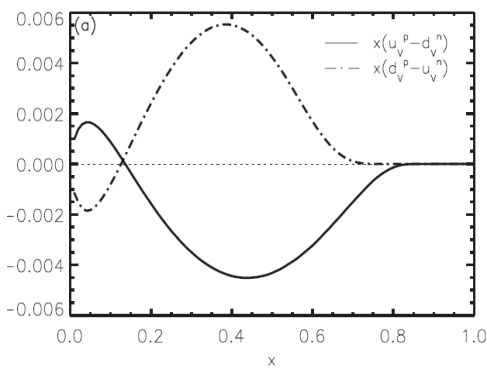}
\caption{Theoretical CSV PDFs by Rodionov \EA, Ref. \protect\cite{Rod94}. 
Solid line: $x\deluv$; dash-dot line: $x\deldv$. The PDFs have been 
evolved to $Q^2 = 10$ GeV$^2$. 
\label{Fig:Rodion}}
\end{figure}

Fig.~\ref{Fig:Rodion} plots the theoretical quark model calculations of 
valence CSV by Rodionov \EA \cite{Rod94}. The dash-dot curve in 
Fig.~\ref{Fig:Rodion} represents the quantity $x\deldv (x)$ while 
the solid curve is $x\deluv (x)$. The curves were initially calculated at the 
low $Q^2$ appropriate for quark models, and evaluated at $Q^2 = 10$ GeV$^2$ 
through DGLAP evolution \cite{Dok77,Gri72,Alt77}. The valence CSV parton 
distributions obtained by Rodionov \EA~\cite{Rod94} are quite similar to 
those of Sather \cite{Sat92}, as seen by comparison of Figs. \ref{Fig:Sather} 
and \ref{Fig:Rodion}. The sign and magnitude of both $\delta d_{\V}(x)$ and 
$\delta u_{\V}(x)$ are very similar, and the second moments of both 
distributions agree to within about 20\%. The zero crossing in Sather's model 
appears at a smaller value of $x$ than that for Rodionov.  

We can also compare the theoretical valence CSV 
distributions with the phenomenological valence CSV distributions obtained 
by MRST from their global fit to high energy data. These are plotted in 
Fig.~\ref{Fig:MRSTfx} for the best fit value $\kappa = -0.2$ in 
Eq.~(\ref{eq:CSVmrst}) . The solid (dashed) curve in Fig.~\ref{Fig:MRSTfx} 
represents $x\delta \dv(x)$ ($x\delta \uv(x)$). The sign and relative 
magnitude of both $\delta \dv$ and $\delta \uv$, and the point 
where they cross zero, are remarkably similar in both the MRST 
phenomenological CSV PDFs, and the results obtained by Rodionov \EA~and shown 
in Fig. \ref{Fig:Rodion}. The second moments 
of both of the Rodionov quark model CSV PDFs are equal to the moments of the 
corresponding MRST values to better than 10\%.  

The valence CSV parton distributions $\deluv$ obtained by Benesh 
and Londergan \cite{Ben98} and shown in Fig.~\ref{Fig:BenLon} are quite 
similar to those of Sather and Rodionov, while the CSV valence distribution 
$\deldv$ is roughly a factor of two smaller 
than the others. Benesh and Goldman \cite{Ben97a}  
calculated parton CSV distributions from a quark potential model, and 
their CSV PDFs have the same sign and a similar shape to those derived by 
Sather and Rodionov, but the Benesh-Goldman CSV PDFs are roughly 
a factor two smaller in magnitude.    

The qualitative agreement between the phenomenological valence quark PDFs 
obtained 
by MRST, using the best value $\kappa = -0.2$ from their global fit and 
shown in Fig.~\ref{Fig:MRSTfx}, and the theoretical CSV PDFs obtained 
by Rodionov \EA~and shown in Fig.~\ref{Fig:Rodion} is rather remarkable, 
especially considering that the theoretical results were obtained some 
ten years earlier and used relatively simple bag model quark wavefunctions. 
The excellent agreement with the phenomenological results provides some 
theoretical support for the functional form chosen by MRST.  However, 
within the 90\% confidence region for the global fit, the valence quark CSV 
PDFs could be either four times as large as those predicted by Sather 
and Rodionov, or they could be three times as large with the opposite sign.  

One feature of the theoretical CSV distributions is the prediction that 
for moderately large values of $x$ (\IE~for $x$ above the zero crossings)   
\be 
| \deluv(x) + \deldv(x)| << | \deluv(x) - \deldv(x)| \ .
\label{eq:csvdiff}
\ee
Consequently, valence quark CSV observables that depend upon 
the difference between the minority and majority CSV terms should be 
substantially larger than those that depend on the sum of these terms. 
Eq.~(\ref{eq:csvdiff}) is satisfied trivially for the MRST phenomenological 
 valence CSV distributions of Eq.~(\ref{eq:CSVmrst}), since by definition 
their sum is zero.

Cao and Signal \cite{Cao00} calculated partonic charge symmetry violation 
assuming that partonic CSV arises through mesonic fluctuations of the 
nucleon. They used a meson-cloud model to estimate partonic CSV 
\cite{Spe97,Kum98}. In the meson-cloud model, mass splittings in the 
baryon and meson multiplets lead to charge symmetry violating parton 
distributions. The resulting CSV distributions obtained by Cao and Signal 
are substantially smaller than those obtained by Sather \cite{Sat92} or 
Rodionov \cite{Rod94}, and peak at substantially smaller values 
$x \sim 0.1$. This could be expected from the splitting functions for 
baryons in meson-cloud models. 

Cao and Signal break 
up the parton distributions into three parts: a bare part, a perturbative 
part and a non-perturbative part. The first two of these are assumed to 
be charge symmetric. The perturbative part is assumed to arise from gluon 
splitting. In the next section we will show that there is an additional 
perturbative part arising from photon splitting. This additional part will 
contribute to parton charge symmetry violation since the photons couple 
differently to up and down quarks, by virtue of their different charges. 

 \subsubsection{``QED Splitting:'' Another Source of Parton CSV
 \label{Sec:QEDsplt}}

 Recently, another source of parton charge symmetry violation has 
been included 
 in calculations of PDFs by both MRST \cite{MRST05} and Gl\"uck, 
 Jimenez-Delgado and Reya \cite{Glu05}. The most important terms in 
 the usual QCD evolution involve gluon radiation, where a quark 
 radiates a gluon leaving a quark with a lower $x$ value.  These authors 
 suggested that one assume charge symmetry at some initial extremely 
 low-mass scale, 
 and include in the QCD evolution equations the effect of photon radiation. 
 This involves the explicit coupling of quarks to photons, the analog of 
 quark coupling to gluons. 

\begin{figure}[ht]
\includegraphics[width=2.8in]{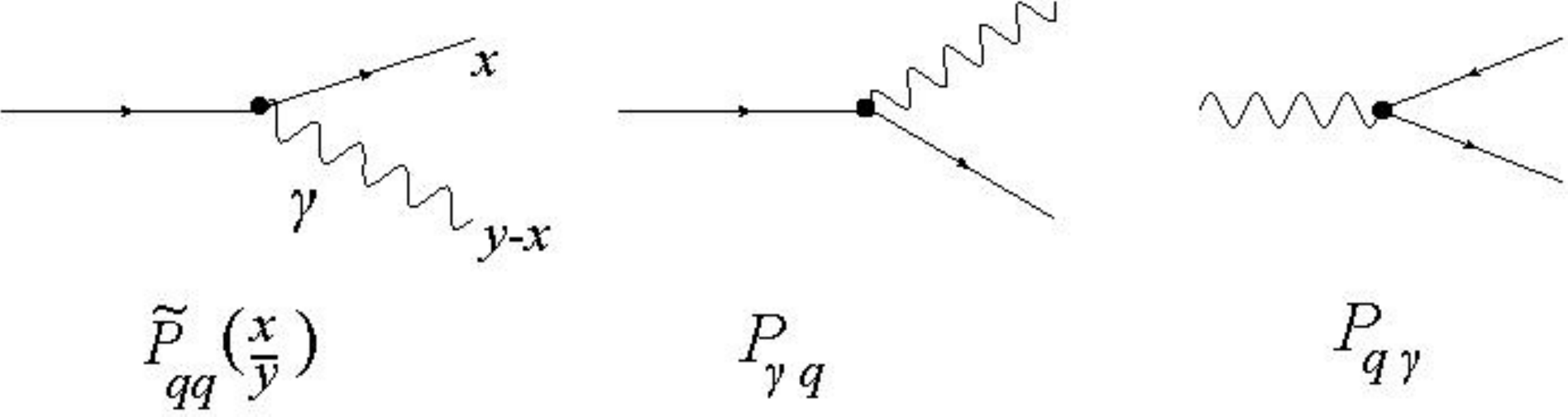}
\caption{Schematic picture of quarks coupling to photons. Replacing a  
gluon line by a photon everywhere (except for the gluon self-coupling) 
produces the electromagnetic coupling of photons to partons. This gives the 
origin of QED splitting that produces additional CSV effects in parton 
distribution functions.
 \label{Fig:QEDsplt}}
\end{figure}

 Fig.~\ref{Fig:QEDsplt} presents a schematic 
 picture of the coupling of quarks to photons. The electromagnetic couplings 
 are obtained by replacing gluon lines with photons, except for the 
 gluon self-coupling which is not present for photons, and replacing the 
 gluon splitting functions with the appropriate coupling for photons. This 
 QED coupling changes the parton distribution functions in two distinct ways.  
 First, it introduces an additional source of charge symmetry violation, 
 since the photon couples differently to up and down partons because of their 
 different electromagnetic charges. Second, 
 radiation of the photon produces a ``photon parton distribution.'' This 
 photon PDF must be accounted for in the evolution equations. The photon 
 PDF also makes a contribution to the total momentum carried by the nucleon. 

When one includes QED contributions in this way, to lowest order in both the 
strong coupling $\alpha_{\Ss}$ and the electromagnetic coupling $\alpha$, the 
so-called DGLAP evolution equations due to Dokshitzer \cite{Dok77}, Gribov and 
Lipatov \cite{Gri72} and Altarelli and Parisi \cite{Alt77} are modified. To 
lowest order in the QED coupling $\alpha$ the evolution equations obtained by 
MRST have the form   
\bea
\frac{\partial q_i (x,\mu^2)}{\partial {\rm log} \,\mu^2} &=& 
 \frac{\alpha_{\Ss}}{2\pi} \left[ P_{qq}\otimes q_i + P_{qg}\otimes g 
 \right] \nonumber \\ &+& \frac{\alpha}{2\pi}
  \widetilde{P}_{qq}\otimes e_i^2 q_i \ , \nonumber \\ 
  \frac{\partial g (x,\mu^2)}{\partial {\rm log} \,\mu^2} &=& 
 \frac{\alpha_{\Ss}}{2\pi} \left[ P_{gq}\otimes \sum_j q_j + P_{gg}\otimes g 
 \right] \ , \nonumber \\ 
 \frac{\partial \gamma (x,\mu^2)}{\partial {\rm log} \,\mu^2} &=& 
 {\alpha\over 2\pi} P_{\gamma q}\otimes \sum_j e_j^2 q_j \ . \nonumber \\ 
\label{eq:QEDevol}
\eea
The convolution integral in Eq.~\ref{eq:QEDevol} is defined by 
\be
 P \otimes f \equiv \int_x^1 \,\frac{dy}{y} P(y) f(\frac{x}{y}) \ .
\label{eq:convolut}
\ee
In Eq.~(\ref{eq:QEDevol}), the right hand side of the schematic evolution 
equations represents a convolution of the splitting functions with the quark 
and gluon distributions (which have an explicit dependence on the 
factorization scale parameter $\mu^2$).  Inclusion of the electromagnetic 
contribution to the evolution equations introduces a ``photon parton 
distribution'' $\gamma(x,\mu^2)$ which is coupled to the quark and 
gluon distributions. The new splitting functions that occur in 
Eq.~(\ref{eq:QEDevol}) are related to the standard QCD splitting functions by 
\bea
\widetilde{P}_{qq}(y) &=& P_{qq}(y)/C_F; \nonumber \\ P_{\gamma q}(y) &=& 
  P_{gq}(y)/C_F;  \nonumber \\ 
  C_F &=& \frac{N_c^2 -1}{2N_c} \ . 
\label{eq:split}
\eea
In Eq.~(\ref{eq:split}), $N_c$ is the number of colors. Conservation of 
momentum is assured by the relation 
\be
\int_0^1 dx x \left[ \sum_i q_i(x,\mu^2) + g(x,\mu^2) + 
   \gamma(x, \mu^2) \right] = 1 \ . 
\label{eq:momcons}
\ee

It is necessary to simplify Eq.~(\ref{eq:QEDevol}).  First, since the 
electromagnetic interaction is not asymptotically free, it is difficult to 
determine a model-independent method for  
setting the starting values for the various PDFs that are coupled by 
these QED effects. In particular, it is not clear where the QED effects 
should be assumed to vanish. Second, inclusion of the QED couplings could in 
principle more than double the number of parton distribution functions (one 
must now differentiate between proton and neutron PDFs, in addition to the 
new photon PDFs).  

 Two groups have adopted somewhat 
 different strategies, with similar overall results. 
 Gl\"uck \EA~\cite{Glu05} adopt the standard convention for DIS 
 reactions of setting the scale $\mu^2 = Q^2$.  The most important 
 contribution from the photon coupling occurs in the valence quark PDFs. To 
 lowest order in the electromagnetic coupling $\alpha$, the convolution 
 equations for the CSV  
 valence quark distributions arising from QED coupling have the form 
\bea
  \frac{d\,\delta u_{\V}(x,Q^2)}{d\, {\rm ln}Q^2} &=& \frac{\alpha}{2\pi} 
  \,\int_x^1 \frac{dy}{y}\, P(y) u_{\V}\left(\frac{x}{y},Q^2 \right)   
   \nonumber \\ &=& \frac{\alpha}{2\pi} P \otimes u_{\V} ; \nonumber \\ 
  \frac{d\,\delta d_{\V}(x,Q^2)}{d\, {\rm ln}Q^2} &=& 
  -\frac{\alpha}{2\pi} \,\int_x^1 \frac{dy}{y}\, P(y) 
  d_{\V}\left(\frac{x}{y},Q^2 \right) \nonumber \\ &=& 
  -\frac{\alpha}{2\pi} P \otimes d_{\V} ; \nonumber \\ 
  P(z) &=& (e_u^2 - e_d^2)\, \widetilde{P}_{qq}(z) \nonumber \\ 
  &=& (e_u^2 - e_d^2) \left(\frac{1+z^2}{1-z} \right)_+ \ .
\label{eq:Gluevol}
\eea
Similar relations hold for the antiquark distributions. Gl\"uck \EA~assume 
that the average current quark mass $\overline{m}_q = 10$ MeV is 
the kinematical lower bound for a quark to emit a photon. This is 
analogous to taking the electron mass as the lower limit for radiation 
of photons in the earliest calculations of the Lamb shift (before the 
advent of renormalization group arguments) \cite{Bet57}.      
Eq.~(\ref{eq:Gluevol}) is then integrated from $\overline{m}_q^2$ to $Q^2$.  
QED evolution effects are evaluated while keeping 
the QCD effects fixed.  The quark distributions appearing on the right 
hand side of Eq.~(\ref{eq:Gluevol}) are the GRV leading-order parton 
distributions \cite{GRV98}. In the resulting integrals, in the region 
$\overline{m}_q^2 \leq q^2 < \mu_{LO}^2= 0.26$ GeV$^2$ corresponding to 
momentum transfers below the input scale for GRV, the PDFs are ``frozen,'' 
\IE~in this region they are assumed to be equal to their value at the input 
scale $\mu_{LO}^2$.  

\begin{figure}[ht]
\center
\includegraphics[width=3.3in]{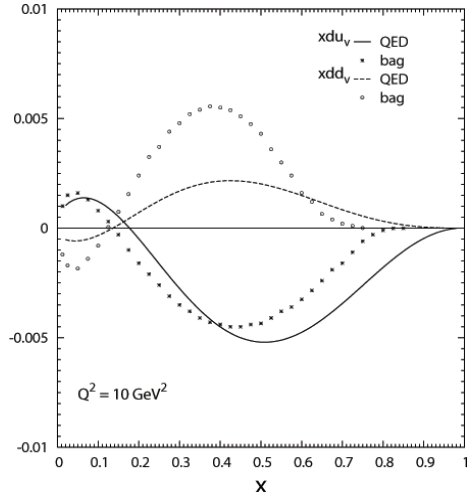}
\caption{The isospin-violating majority $x\delta u_{\V}$ (solid curve) 
and minority $x\delta d_{\V}$ (dashed curve) valence parton distributions 
obtained by Gl\"uck \EA \ \protect{\cite{Glu05}} at $Q^2 = 10$ GeV$^2$, 
assuming QED evolution from a scale set by the current quark mass. 
These are compared with majority (solid points) 
and minority (open circles) CSV distributions obtained from theoretical 
quark model calculations \protect{\cite{Rod94}}. 
\label{Fig:Glueck}}
\end{figure}

The resulting valence isospin asymmetries $x\delta u_{\V}$ and 
$x\delta d_{\V}$ are plotted in Fig.~\ref{Fig:Glueck} at $Q^2 = 10$ GeV$^2$.  
For comparison, they are plotted along with the valence quark CSV  
asymmetries obtained from quark model calculations by Rodionov 
\EA~\cite{Rod94,Lon03}.  In the Rodionov calculations CSV 
distributions arose from diquark mass differences $\delta \widetilde{m} = 
m_{dd} - m_{uu} \approx 4$ MeV~\cite{Sat92,Rod94}, and from the target nucleon 
mass difference $\delta M = M_n - M_p$. While 
the quantity $\delta u_{\V}$ is quite similar in both sign and magnitude for 
both the bag model and the QED calculations, the QED results for 
$\delta d_{\V}$ are roughly half as large as the bag model results. 
This can be understood from the evolution equations of 
Eq.~(\ref{eq:Gluevol}). The coefficients of the QED evolution are equal 
and opposite for up and down valence quarks, but since $\uv$ is roughly 
twice $\dv$, one expects the CSV effects for up quarks to be approximately 
twice the magnitude and the opposite sign as those for down quarks.  
As a result, the CSV effects obtained from ``QED splitting'' will not obey 
the relation of Eq.~(\ref{eq:pnmom}) assumed by MRST in their phenomenological 
fit. As noted 
previously, the bag model results for valence quark CSV are extremely close to 
those obtained by MRST using the phenomenological form of 
Eq.~(\ref{eq:CSVmrst}), for the best-fit value $\kappa = -0.2$.     

The MRST group \cite{MRST05} solved the evolution equations of 
Eq.~(\ref{eq:QEDevol}) with assumptions about the parton distributions 
at the starting scale $Q_0^2 = 1$ GeV$^2$. At the starting scale, the sea 
quark and gluon distributions are assumed to be charge symmetric. The 
photon PDFs at the starting scale were taken as those due to one-photon 
radiation from valence quarks in leading-logarithm approximation, evolved from 
current quark masses $m_u = 6$ MeV and $m_d = 10$ MeV  to $Q_0$. This 
produces different photon PDFs for the neutron and proton at the starting 
scale. Enforcing overall valence parton momentum conservation from 
Eq.~(\ref{eq:momcons}) requires valence quark charge asymmetry at the 
starting scale. MRST assumed a simple phenomenological form chosen to obey the 
valence quark normalization condition, which produced charge symmetry 
violating 
distributions that resemble the valence PDFs at large and small $x$, and 
with an overall magnitude chosen to enforce quark momentum conservation. 
MRST then determine the proton's quark and gluon distributions at the starting 
scale $Q_0^2$ by a global fit to an array of high energy data.  The only 
difference from other MRST global fits is the use of the modified DGLAP 
evolution equations of Eq.~(\ref{eq:QEDevol}). The MRST results for valence 
quark CSV are quantitatively quite similar to the Gl\"uck analysis. 

The contribution to CSV arising from QED splitting would occur even if 
the up and down quark masses, and the neutron-proton masses, were 
initially identical. This is different from the CSV terms which were 
calculated from quark models and described in the preceding section. From 
Eq.~(\ref{eq:iso}), it is clear that those 
CSV terms were proportional to the up-down quark mass difference and the 
$n-p$ mass difference. 

Because the two types of parton charge symmetry 
violation tend to arise from different sources, and both CSV effects are 
quite small, we have evaluated them independently and we add them together.  
Note, however, that the CSV contributions from QED splitting cannot be 
treated as being completely independent of the 
quark-model CSV terms.  This is because the quark-model calculations used 
estimates of electromagnetic effects in calculating $m_{dd} - m_{uu}$ as 
described following Eq.~(\ref{eq:Satanl}). 

The quark PDFs calculated using the QED splitting terms in 
Eq.~(\ref{eq:QEDevol}) have explicitly included photon radiation by the 
quarks. These PDFs are relevant for the quark distribution prior to a 
hard interaction. Thus, it would be double-counting if one included radiative 
corrections for a quark prior to a hard interaction, since these represent 
the same terms that were included in Eq.~(\ref{eq:QEDevol}). Such a procedure  
corresponds to the ``DIS factorization'' scheme, which assumes that the 
${\cal O}(\alpha)$ corrections arising from photon emission from 
incoming quarks are included in the definition of the quark PDFs. 
The consistent treatment of partonic radiative corrections is discussed in 
some detail by Diener, Dittmaier and Hollik \cite{Die04,Die05}.  

 \subsection{Experimental Limits on Valence Quark Charge Symmetry}
 \label{Sec:seventwo}
 \mb{.5cm}

There have been no direct observations of any violation of partonic charge 
symmetry. As a result we have at present only upper limits on the magnitude of 
parton CSV. We also have the indirect evidence for partonic CSV from the 
global fits carried out by the MRST group and discussed in 
Sect.~\ref{Sec:MRSTcsv}. From Eq.~(\ref{five18}) in Sec.~\ref{Sec:sixtwob}, 
we can obtain a relation between the $F_2$ structure function in 
charged-lepton DIS, and the average of the $F_2$ structure functions 
for neutrinos and antineutrinos, both on isoscalar targets. The difference 
between these two (appropriately normalized) structure functions is given 
by two components. The first is a contribution from strange and charmed 
quarks, and the second is a contribution from partonic CSV. 

At small Bjorken $x$, comparison of these two structure functions will 
provide a linear combination of heavy quark PDFs and charge symmetry 
violating parton distributions. Extracting limits on parton CSV then requires 
very accurate knowledge of heavy quark parton distributions. This is further 
complicated by the fact that sea quark distributions increase quite 
rapidly at very small $x$; so the fractional contribution of partonic 
CSV is likely to be small in this region. 

One has two 
possibilities for placing stronger experimental limits on partonic CSV. The 
first is to go to large Bjorken $x$. Since the heavy quark PDFs are quite 
small at large $x$, the relative contribution of charge symmetry violating 
parton distributions will be significantly larger. The second possibility 
is to take the first moment by integrating the parton distributions over 
all $x$. One then uses the valence quark normalization condition. This 
condition, given in Eqs.~(\ref{qnorm}) and (\ref{csvnorm}), requires that the 
first moment 
of the valence quark CSV terms vanish. Consequently, if one integrates 
parton distributions over all $x$ the valence quark CSV terms must give 
zero. Following this integration the only remaining contributions will 
be from the first moment of the sea quark CSV. 

 \subsubsection{The ``Charge Ratio:'' Comparison of $F_2$ Structure 
 Functions in Reactions of Muons and Neutrinos
 \label{Sec:seventwoone}}

 Currently, the strongest upper limit on parton CSV distributions is obtained 
by comparing the $F_2$ structure functions measured in CC reactions 
induced by $\nu$ and $\ovnu$, and the $F_2$ structure function for charged 
lepton DIS, both measured on isoscalar targets.  Using the relation 
derived in Eq.~(\ref{five18}), at sufficiently high energies we can 
construct the ratio 
 \bea
 &\,&R_c(x) \equiv \frac{F_2^{\gamma N_0}(x) + 
  x[s^+(x) + c^+(x)]/6}{5\overline{F}_2^{W N_0}(x)/18};  
   \nonumber \\ &\,&R_c(x) \approx 1+ \frac{3 \left( \delta u^+(x) - 
  \delta d^+(x)\right)}{10 \sum_j \,q_j^+(x)} \ .
\label{Rc}
\eea
In Eq.~(\ref{Rc}) the function $\overline{F_2}^{W N_0}(x)$ is the average of
 the CC $F_2$ structure functions induced by $\nu$ and $\ovnu$ and defined 
in Eq.~(\ref{eq:F2bar}). In the denominator of the second line of 
Eq.~(\ref{Rc}) the sum is taken over all quark flavors. 
Eq.~(\ref{Rc}) shows that in the limit of 
 exact charge symmetry, the ratio of the muon and neutrino $F_2$ structure 
 functions,when corrected for heavy quark contributions and the factor
 $5/18$, should be one independent of $x$ and $Q^2$, in the naive parton
 model.  The factor $5/18$ in Eq.~(\ref{five18}) and in 
 Eq.~(\ref{Rc}) reflects the fact that the virtual photon 
 couples to the squared charge of the quarks while the weak interactions 
 couple to the weak isospin. The quantity $R_c$ is sometimes called the 
 ``charge ratio,'' and the relation between the $F_2$ structure functions 
 is often termed the ``5/18$^{th}$ rule.''  In Eq.~(\ref{Rc}) the final line 
 is expanded to lowest order in the (presumably small) CSV terms.  

 The quantity $R_c(x)$ requires knowledge of the heavy quark PDFs.  For 
 example, the observables most sensitive to strange quark distributions are 
 cross sections for opposite sign dimuon events produced from neutrino DIS 
 on nuclei \cite{Baz95,Gon01}.  Once the strange quark distributions have 
 been extracted from the dimuon production process, they can be inserted in 
 Eq.~(\ref{Rc}). The intrinsic charm PDFs are generally quite small; however 
 a significant amount of data is collected near charm quark threshold, where 
 it is important to take proper account of the charm mass. 
 Comparing the $F_2$ structure functions for lepton-induced processes
 with the $F_2$ structure functions from weak processes mediated by
 $W$-exchange, one can in principle measure both the magnitude and $x$ 
 dependence of parton CSV. Clearly, since extraction of parton CSV 
 distributions depends on precise knowledge of strange and charm PDFs, 
 our knowledge of these quantities will be strongly correlated.  
 Certainly this is the case at low $x$, where the sea quark distributions 
 (including strange quarks) are large.  

 The charge ratio provides the strongest direct limits to
 date on parton CSV.  There should be no additional QCD corrections to this
 relation so it should be independent of $Q^2$, {\em provided} that
 the structure functions are calculated in the so-called ``DIS scheme,''
 where the $F_2$ structure functions are defined to 
 have the form $F_2(x) = x\sum_i e_i^2 q_i^+(x)$ to
 all orders, where
 $e_i$ is the quark charge appropriate for either the electromagnetic
 or weak interactions.  For example, the CTEQ4D parton distributions
 \cite{Lai96} were determined in the DIS scheme. 

 Ever since one has been able to extract the $F_2$ structure functions and 
 hence the parton distribution functions from both muon and neutrino DIS, one 
 has had the possibility of constructing the charge ratio using 
 Eq.~(\ref{Rc}). Within error bars, the results have always been consistent 
with the assumption of parton charge symmetry.  
 However, until a few years ago the charge ratio gave only qualitative upper 
 limits on CSV, because of the great difficulty in obtaining precise 
 absolute neutrino cross sections, and because of the number of corrections 
 that must be taken into
 account. These corrections include: relative normalization 
 between lepton and neutrino cross sections; contributions from strange and 
 charm quarks; higher twist effects on PDFs; and heavy quark threshold 
effects.  In addition, one must be able to separate $F_2$ 
 and $F_3$ structure functions in $\nu$ CC reactions. 
 Another potentially important effect is heavy target corrections in 
 neutrino reactions.  The most precise lepton structure functions
 are obtained from deuteron targets, while the most accurate neutrino 
 cross sections are extracted from experiments 
 on heavy targets like iron, so it is necessary to correct the neutrino $F_2$
 structure functions for heavy target effects, and also for effects arising 
 from the fact that iron is not an isoscalar target. These effects include 
 shadowing and antishadowing at small $x \le 0.1$, EMC effects for 
 $0.2 \le x \le 0.6$ \cite{emc,emcb} and Fermi motion at large $x$.

 Earlier analyses compared the muon $F_2$ structure functions of Meyers 
 \EA~on iron \cite{Meyers} to $F_2$ obtained from CCFRR neutrino 
 measurements \cite{Macfar}. The extracted ratio $R_c$ of Eq.~(\ref{Rc}) was 
 consistent with unity, except possibly at the largest value $x= 0.65$.  
 The experimental data was consistent with zero charge symmetry violation and 
 ruled out very large violation of parton charge symmetry. However, the 
 extracted charge ratio had errors of several percent. Because of the 
 factor of $3/10$ in the last line of Eq.~(\ref{Rc}), an error of 5\% in the 
 charge ratio would lead to upper limits on parton CSV at roughly the 15\% 
 level.  From our discussion of the phenomenological and theoretical estimates 
 of parton CSV summarized earlier, we expect that the CSV contribution to the 
 charge ratio will not exceed a few percent at any value of $x$.
 Consequently, a measurable deviation of the charge ratio from unity, at any 
 value of $x$, would be very interesting but experimentally quite 
 challenging.

\begin{figure}[ht]
\center 
\includegraphics[width=3.1in]{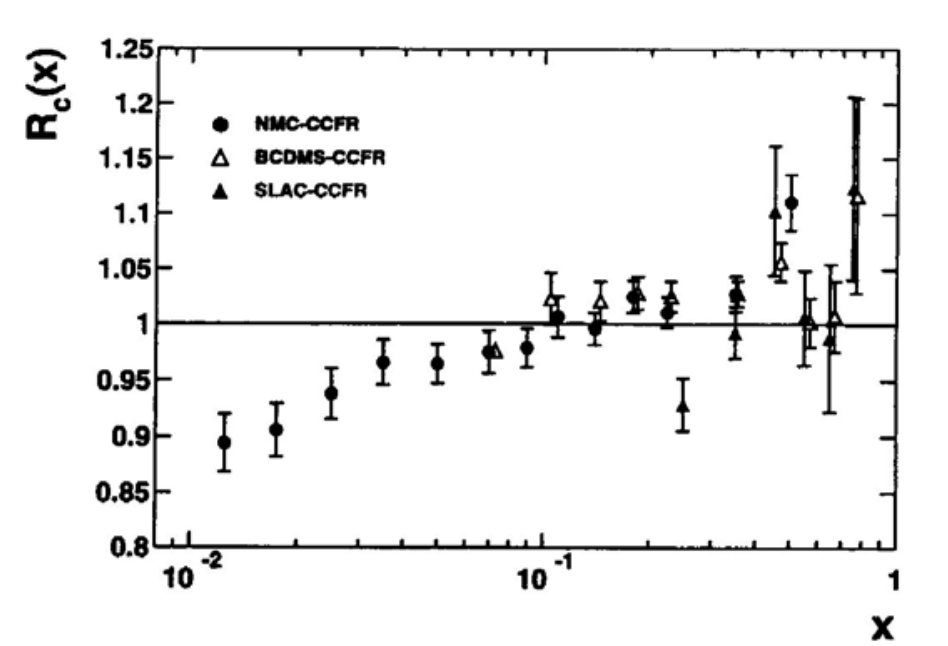}
 \caption{Charge ratio $R_c^\nu(x)$ of Eq.~\protect\ref{Rc} vs.\ $x$.
 Solid circles: CCFR $\nu-Fe$ data, Ref.\ \protect\cite{Sel97} and $\mu +D$ 
 measurements from NMC, Ref.\ \protect\cite{Ama91,Ama92,NMC97}. Open 
 triangles: CCFR $\nu$ data and $\mu  + D$ measurements from BCDMS, Ref.\ 
 \protect\cite{Ben90}. Solid triangles: CCFR and SLAC electron scattering 
 data, Ref.\ \protect\cite{Whi90,Whi90b}.}
 \vspace{0.1truein}
 \label{fig45ab}
 \end{figure}

 In recent years we have obtained significantly more precise DIS data for 
 both muons and neutrinos. This should allow us to make more stringent tests 
 of parton charge symmetry.  The NMC group \cite{Ama91,Ama92,NMC97} 
 measured the $F_2$ structure 
 function for muon interactions on deuterium at energy $E_\mu = 90$ and 280 
 GeV. The NMC measurements are more precise than the earlier BCDMS muon 
 scattering results on deuterium \cite{Ben90} and carbon \cite{BCDMS}, or the 
 SLAC electron scattering results \cite{Whi90,Whi90b}. The CCFR group 
 \cite{Sel97} extracted the $F_2$ structure function for $\nu$ and $\ovnu$ 
 interactions on iron using the Quadrupole Triplet Beam at Fermilab. They 
 also performed a comprehensive comparison of their neutrino data with the 
 NMC muon results \cite{Sel97,Sel97a}. In Fig.\ \ref{fig45ab} we plot the 
 charge ratio $R_c$ of Eq.~(\ref{Rc}) vs.\ $x$.  The solid circles give 
 the charge ratio comparing the NMC and CCFR measurements. The open 
 triangles give the charge ratio comparing CCFR with BCDMS data, and 
 the solid triangles compare CCFR neutrino data with the SLAC 
 electron scattering measurements. 
 
 Analysis of the charge ratio as a function of $x$ should in principle 
 provide a test of parton charge symmetry in both the valence and sea regime. 
 In the region $0.1 \le x \le 0.4$, where valence quarks should 
 dominate, the charge ratio is consistent with unity, with errors on  
 the charge ratio at about the 3\% level.  From Eq.~(\ref{Rc}), this would 
 provide upper limits on valence quark CSV of about 10\%. For larger values of
 $x$ the upper limit on errors in the charge ratio is in the 5-10\% level, due 
 primarily to the poorer statistics and the large Fermi motion corrections 
 that become important at very large $x$. Particularly after including heavy 
 target corrections for the $\nu$-iron measurements, the charge ratio appeared 
 to deviate significantly from one at the smallest
 values $x < 0.1$ \cite{Bor98}.  The deviation appeared to grow with 
 decreasing $x$ and reached values as large as 15-20\%. This is apparent 
 from Fig.\ref{fig45ab} which shows that $R_c^\nu$ is definitely less than 
 one for small $x < 0.1$. Boros, Londergan and Thomas  
 \cite{Bor97,Bor99} studied the origin of this discrepancy.  They 
 suggested that, assuming that the identification of the neutrino $F_2$ CC 
 structure functions was reliable, the most likely explanation for this 
 anomaly was a substantial violation of charge symmetry in the nucleon sea. 

\begin{figure}[ht]
 \includegraphics[width=3.30in]{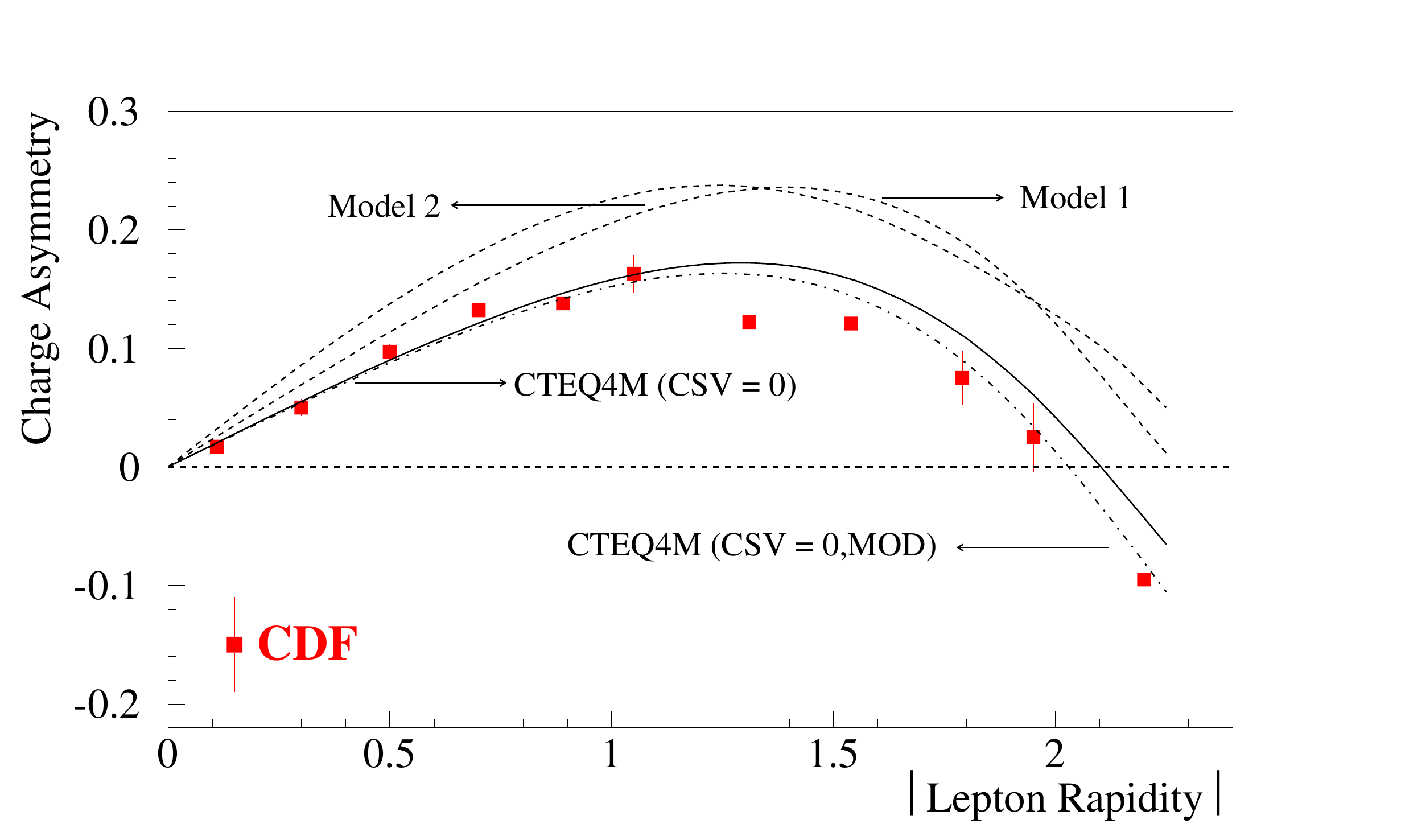}
\caption{[color online] The $W$ charge asymmetry from $p \bar{p}$ 
 reactions at the Tevatron. The experimental points are those of the CDF 
 group, Ref.~\protect{\cite{Abe98}}. The solid and dash-dot curves represent 
 two fits using the CTEQ4M PDFs \protect{\cite{Lai96}} with no CSV terms. The 
 dashed and dotted curves represent two 
 different assumptions by Bodek \EA, Ref.~\protect{\cite{Bod99}} using 
 sea quark CSV distributions calculated by Boros \EA, 
 Ref.~\protect{\cite{Bor97,Bor99}}.
     }
 \vspace{0.1truein}
 \label{Bodfig}
 \end{figure}

 Boros \EA~showed that one needed sea quark CSV of at least 25\% in order to 
 explain this discrepancy in the charge ratio. This apparent violation of 
 charge symmetry was extremely surprising, as it was at least an order of 
 magnitude larger than theoretical estimates. Bodek \EA~\cite{Bod99} 
 questioned whether such a large CSV effect was consistent with other 
 experiments. They analyzed the $W$ boson charge asymmetry obtained in 
 $p \bar{p}$ experiments from the CDF group at the Fermilab Tevatron 
 \cite{Abe98}. Since this experiment involves proton-antiproton scattering, 
 CSV effects do not enter directly. However, Bodek and collaborators 
 argued that the most precise PDFs arise from charged lepton DIS on 
 isoscalar targets. Because the $F_2$ structure 
 functions are weighted by 
 the squared charge of the quarks, they are most sensitive to the up 
 quarks in the proton and neutron. Thus to a significant degree our 
 identification of $d^p$ is obtained from $u^n$ plus the assumption of 
 parton charge symmetry. Bodek examined two different methods for 
 extracting CSV distributions from the data, and calculated the effect 
 on the $W$ charge asymmetry which would arise from CSV effects of the 
 magnitude assumed by Boros \EA The results are shown in Fig.~\ref{Bodfig}. 

 In Fig.~\ref{Bodfig}, the solid and dash-dot curves are fits to the 
 CDF data \cite{Abe98} using the CTEQ4M parton distributions \cite{Lai96} 
 with no parton CSV terms. The dashed and dotted curves resulted from 
 two different assumptions by Bodek \EA~for the large sea quark CSV 
 terms of Boros \EA~\cite{Bor97,Bor99}. The CDF measurements are very 
 sensitive to the sea quark distributions, and Bodek argued that the 
 large sea quark CSV was incompatible with those experimental results. 
 Although Bodek and collaborators examined only two potential ways of 
 defining parton CSV distributions, it is hard to imagine that sea quark 
 CSV of this magnitude could be made consistent with the $W$ charge 
 asymmetry data. 

 This issue was eventually resolved when the CCFR 
 collaboration re-analyzed 
 its neutrino data \cite{Yan01} and the low-$x$ discrepancy disappeared. 
 There were two primary reasons for this change. 
 The first was an improved treatment of charm mass corrections. This 
 was particularly important for the low-$x$ data, which were taken in 
 a region close to charm threshold. In analyzing this data it is necessary 
 to take into account accurately the charm quark mass. The  
 initial analysis used a ``slow rescaling'' hypothesis due to 
 Georgi and Politzer \cite{Geo76,Bar76} to account for charm mass corrections. 
 The re-analysis involved NLO 
 calculations, which account for massive charm production using 
 variable-flavor techniques \cite{Aiv94,Tho98,Bor99a}. 

 The second significant effect involved the separation of structure functions 
 in charged-current $\nu$ DIS. The sum of $\nu$ and $\ovnu$ charged-current 
 DIS cross sections gives a linear combination of $F_2$ and $F_3$ 
 structure functions, 
\bea 
 {d^2 \sigma^{\nu}_{\CC} \over dx dy} + 
 {d^2 \sigma^{\bar{\nu}}_{\CC} \over dx dy} &\sim& 
 2(1- y - y^2/2) \overline{F}_2^{W\,N_0}(x,Q^2) \nonumber \\ 
  &+& (y- y^2/2)\Delta xF_3(x,Q^2) \ . 
\label{eq:sigmaCC} 
\eea
In Eq.~(\ref{eq:sigmaCC}), the quantity $\Delta x F_3(x)$ is the 
difference in the $F_3$ charged-current structure functions for neutrino 
and antineutrino beams, 
\bea
\Delta xF_3(x) &=& xF_3^{W^+}(x) - xF_3^{W^-}(x) \ ; \nonumber \\ 
\Delta xF_3^{N_0}(x) &\rightarrow& x\Bigl[ 2(s^+(x) - c^+(x)) \nonumber \\ 
  &+& \deld^+(x) - \delu^+(x) \Bigr].  
\label{eq:DelF}
\eea
The second line in Eq.~(\ref{eq:DelF}) is valid to leading order in QCD 
for an isoscalar target and for sufficiently high $Q^2$. In these limits, 
assuming the validity of charge symmetry, the quantity $\Delta x F_3$ 
is sensitive only to heavy quark distributions. 

For simplicity in Eq.~(\ref{eq:sigmaCC}) we have dropped terms of order 
$M^2/s$ and have set the longitudinal to transverse ratio $R$ to zero 
(these approximations were not made in re-analyzing the data). 
In the initial 
analysis \cite{Sel97}, data for a given $x$ bin was averaged over all $y$, and 
 the $\Delta xF_3$ structure function was estimated 
using phenomenological PDFs.  In the re-analysis the data was binned in 
$x$ and $y$ so that both $\overline{F}_2$ and $\Delta xF_3$ could be 
extracted \cite{Yan01}.  The experimental values for 
$\Delta xF_3$ differed substantially from the phenomenological predictions.  
From Eq.~(\ref{eq:sigmaCC}), a change in $\Delta xF_3$ will affect the 
values extracted for the charged-current $F_2$ neutrino structure functions. 
The combined effect of the NLO treatment of charm production, and 
the model-independent extraction of $\Delta xF_3$ removed the small-$x$ 
discrepancy.  The charge ratio $R_c$ of Eq.~\ref{Rc} 
is now unity to within experimental error, even at small $x$.     

 \begin{figure}[ht]
 \hspace{-1.0cm}
\includegraphics[width=3.4in]{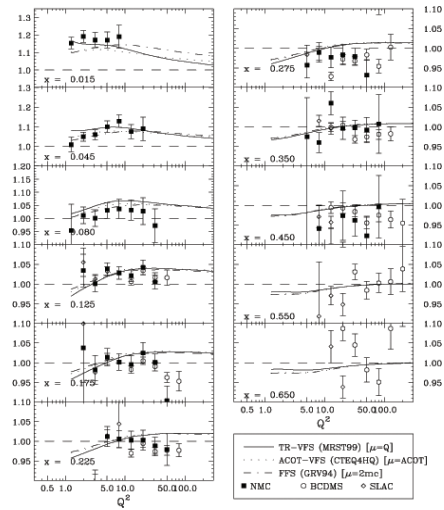}
 \caption{The ratio $5F_2^{\nu}/(18F_2^{\mu})$, calculated by the 
 CCFR Collaboration, Ref.~\protect\cite{Yan01}. Curves are for various 
 NLO parton calculations. Solid curve: Ref.~\protect\cite{Tho98}; 
 dotted curve: Ref.~\protect\cite{Aiv94}; dash-dotted curve: 
 Ref.~\protect\cite{Grv95}. Neutrino structure functions from the CCFR 
 group, Ref.~\protect\cite{Yan01}. Solid points, NMC muon data, 
 Ref.~\protect\cite{NMC97}; open circles: BCDMS data, 
 Ref.~\protect\cite{BCDMS}; diamonds: SLAC data,
 Ref.~\protect\cite{Whi90b}.}
 \vspace{0.1truein}
 \label{Fig:yang}
 \end{figure}

 The results are shown in Fig.~\ref{Fig:yang}.  These graphs plot the 
 ratio $5F_2^{\nu}/(18 F_2^{\mu})$ from the CCFR re-analysis, 
 vs.\ $Q^2$ for various values of $x$. The different data points involve 
 muon DIS experiments from NMC, BCDMS, and SLAC 
 \cite{NMC97,Ben90,BCDMS,Whi90b}. The 
 curves are NLO analyses using various methods for including charm 
 mass effects \cite{Grv95,Aiv94,Tho98,Bor99a}. The previous low-$x$ 
 discrepancy between theory and experiment has largely disappeared.  
 This allows one to place qualitative limits of $< 10$\% on  
 the magnitude of CSV effects in the sea, for values $x \ge 0.015$.  
 To obtain more quantitative 
 limits on CSV, it will be necessary to obtain reliable estimates 
 for the few remaining uncertainties in this comparison.  Perhaps the 
 largest undetermined correction remains the shadowing of parton 
 distributions for $\nu-Fe$ interactions.  Experimental analyses 
 have assumed that the nuclear shadowing corrections are the same for 
 neutrinos (virtual $W$'s) as for charged leptons (virtual photons).  
 Boros \EA \cite{Bor98} showed that one could expect 
 substantially different shadowing for $W$'s than for photons, primarily
 because the $W$s couple to axial currents as well as to vector 
 currents. This was further expanded by Brodsky \EA \cite{Bro04,Kov02} who 
 calculated both shadowing and anti-shadowing effects for neutrino 
 DIS.   

 \subsubsection{Charge Symmetry and Determination of the Weinberg Angle}
 \label{Sec:sixthreetwo}

 The NuTeV group \cite{Zel02a,Zel02b} have measured total  
 charged-current and neutral-current cross sections for $\nu$ and $\ovnu$
 on an iron target. From these measurements they made an independent 
 determination of the Weinberg angle, motivated by a procedure 
 initially suggested by Paschos and Wolfenstein \cite{Pas73}. 
 Paschos and Wolfenstein showed that a ratio of total cross sections for 
 neutral current (NC) and charged current (CC) interactions for $\nu$ and 
 $\ovnu$ on an isoscalar target $N_0$ gave the remarkably simple 
 Paschos-Wolfenstein (PW) relation 
 \be 
 R^- \equiv {\langle \sigma^{\nu \,N_0}_{\NC}\rangle 
 - \langle \sigma^{\bar{\nu} \,N_0}_{\NC} \rangle  
  \over \rho_0^2 \left[ \langle \sigma^{\nu \,N_0}_{\CC} \rangle -  
 \langle \sigma^{\bar\nu \,N_0}_{\CC}\rangle \right]} 
 = {1\over 2} - \sin^2 \theta_W ~~.
 \label{PW}
 \ee
 In Eq.~(\ref{PW}), the quantities are the total NC and CC cross sections 
 for $\nu$ and $\ovnu$ on an isoscalar target, and the quantity 
 $ \rho_0 \equiv M_W/(M_Z \cos \theta_W)$ is one in the Standard Model.  
 The brackets denote integration of the cross sections over all Bjorken $x$. 
 Although the individual cross sections depend upon details of parton 
 distributions, the ratio of these combinations contains no dependence 
 upon PDFs, and in addition a number of experimental effects cancel.     
 
 The NuTeV collaboration used the Sign Selected Quadrupole Train beamline 
 at Fermilab to separate $\nu$ and $\ovnu$ 
 arising from pion and kaon decays following the 
 interaction of 800 GeV protons.  The resulting 
 interaction events were observed in the NuTeV detector, where they were 
 required to deposit between 20 GeV and 180 GeV 
 in the calorimeter.  CC and NC events were distinguished by the 
 event length in the counters, as CC events contained a final muon 
 that penetrated substantially farther than the hadron shower. The 
 NuTeV collaboration measured the individual ratios 
 $R^\nu$ and $R^{\bar\nu}$ defined by 
 \bea 
 R^\nu &=& {\langle \sigma^{\nu \,N_0}_{\NC} \rangle \over 
  \rho_0^2 \langle \sigma^{\nu \,N_0}_{\CC} \rangle } ~; \hspace{0.6cm} 
 R^{\ovnu} = {\langle \sigma^{\ovnu \,N_0}_{\NC} \rangle \over 
  \rho_0^2 \langle \sigma^{\ovnu \,N_0}_{\CC} \rangle } ~; \nonumber \\  
 r &=& {\langle \sigma^{\ovnu \,N_0}_{\CC} \rangle \over 
  \langle \sigma^{\nu \,N_0}_{\CC} \rangle }\ .
 \label{Rnu}
 \eea
 In terms of the ratios defined in Eq.~(\ref{Rnu}), the PW ratio has the 
 form 
 \be
 R^- = {R^\nu - r\,R^\ovnu \over 1 - r} \ .
 \label{eq:RnuNT}
 \ee 

 The NuTeV group measured $R^\nu = 0.3916 \pm 0.0007$ and 
 $R^{\overline\nu} = 0.4050 \pm 0.0016$. The quantity $r = 0.499 \pm 0.005$ 
 was taken from the world average of $\nu-Fe$ charged-current DIS reactions 
 \cite{blair,berge} and from measurements by the CCFR collaboration 
 \cite{Sel97}. Since acceptances and cuts 
 differ for $\nu$ and $\ovnu$ reactions, they did not directly construct 
 the Paschos-Wolfenstein ratio via Eq.~(\ref{PW}); instead the measured 
 NC/CC ratios were compared 
 with a Monte Carlo simulation of the experiment, from which they 
 extracted the on-shell value for the Weinberg angle $\sin^2 \theta_W = 
 0.2277 \pm 0.0013 ~(stat) \pm 0.0009 ~(syst)$.  This value is 
 three standard deviations above the measured fit to other 
 electroweak processes, $\sin^2 \theta_W = 0.2227 \pm 0.00037$ \cite{Abb01}. 

\begin{figure}[ht]
\resizebox{0.455\textwidth}{!}{%
\includegraphics{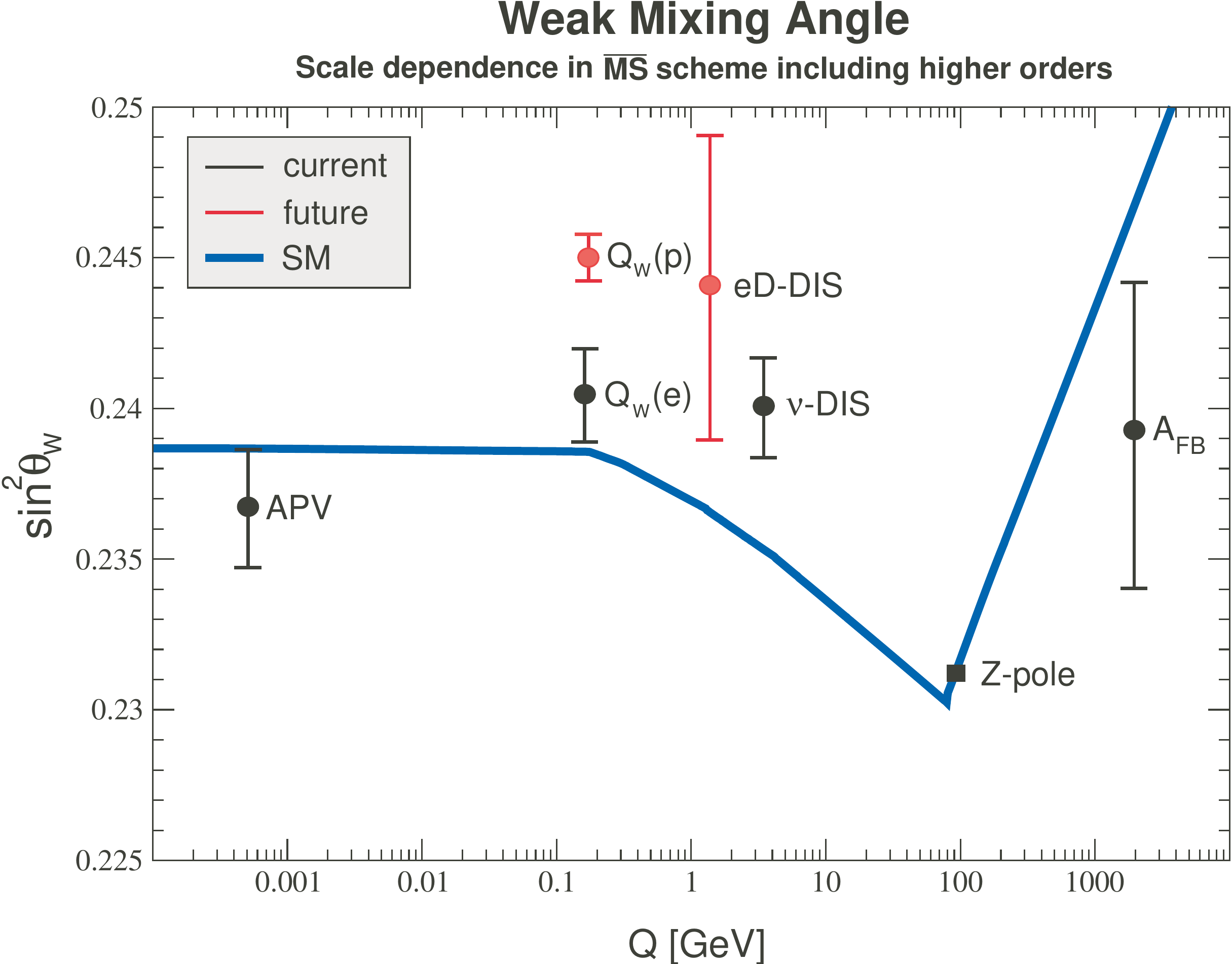}
}
\caption{[color online] Effective value for $\sintW$ vs. $Q$. Curve is from 
Ref.~\protect{\cite{Erl05}}, calculated in $\overline{MS}$ scheme. 
Experimental points represent atomic parity violation (APV) 
\protect{\cite{Ben99}}, M\o ller scattering ($Q_w(e)$) \protect{\cite{Ant05}}, 
measurements at the $Z$ pole \protect{\cite{Abb01}}, NuTeV ($\nu$-DIS) 
\protect{\cite{Zel02a}}, and forward-backward asymmetry from CDF ($A_{FB}$) 
\protect{\cite{Aco05}}. Points in red represent error estimates for two 
proposed experiments, $e-D$ parity violating scattering ($eD$-DIS) 
\protect{\cite{Arr06}}, and the Q-weak experiment ($Q_w(p)$) 
\protect{\cite{QWeak}}.}  
\label{Fig:Weinrun}
\end{figure}

 In a given renormalization scheme, the effective Weinberg angle $\sintW$ 
 will acquire a $Q^2$-dependence from radiative and loop corrections 
 \cite{Cza96,Cza00,Erl05}. In Fig.~\ref{Fig:Weinrun} we plot the effective 
 value for $\sintW$ vs.~$Q$ with the results of several experiments. The 
 curve is that of Erler and Ramsey-Musolf \cite{Erl05}, calculated in 
 $\overline{MS}$ scheme. The experimental points represent atomic parity 
 violation in Cesium (APV) \cite{Ben99}, a M\o ller scattering measurement 
 from experiment E158 at SLAC ($Q_w(e)$) \cite{Ant05}, a series of 
 measurements at the $Z$ pole at LEP and SLD ($Z$-pole) \cite{Abb01}, the 
 forward-backward asymmetry of $e^+e^-$ pairs produced from $\overline{p}p$ 
 collisions at the Tevatron by the CDF group ($A_{FB}$) \cite{Aco05}, and the 
 NuTeV result ($\nu$-DIS) \cite{Zel02a}. Additional points in red give error 
 estimates for two proposed experiments, $e-D$ parity-violation in DIS 
 ($eD$-DIS) \cite{Arr06}, and an approved $Q$-weak experiment involving DIS 
 for polarized electrons on protons at Jefferson Laboratory ($Q_w(p)$) 
 \cite{QWeak}. Although the error bars on some of the experiments are rather 
 large, the E158 experiment establishes the $Q^2$ dependence of 
 the effective Weinberg angle at the $6\sigma$ level, and the NuTeV 
 experiment is claimed to differ from the expected value by $3\sigma$.   

 The NuTeV result, which implies an effective left-handed coupling of 
 light quarks to the neutral current that is about 1.2\% smaller than  
 obtained from other electroweak data, is rather surprising. The status of 
 what has been termed the ``NuTeV anomaly'' has recently been summarized 
 \cite{Lon05}.  Davidson \EA~\cite{Dav02} considered a number of corrections 
 from physics outside the Standard Model.  It is quite difficult for new 
 physics to explain the NuTeV finding since such effects have to agree with
 both the NuTeV result and also with the very precise measurements of EW 
 effects at LEP, which constrain some parameters to a few parts per thousand. 
 As a result, most recent efforts have focused on effects within the 
 Standard Model. At present the three most likely ``QCD effects'' are:  
 effects due to radiative corrections or nuclear effects in the neutrino 
 reactions; contributions from strange quark 
 momentum asymmetry; or charge symmetry violation in parton 
 distributions.  
 
To lowest order in the strong coupling $\alpha_s$, one can calculate 
various analytic corrections to the Paschos-Wolfenstein relation. These can be 
written in the form 
\bea 
\delta R^- &=& \frac{(-1+ \frac{7}{3}\sintW)}{\Uv + \Dv}\biggl[ 
 \frac{(N-Z)}{A}(\Uv -\Dv) \nonumber \\ &+& S^- + \frac{\delta \Dv - 
\delta \Uv}{2} \biggr]
\label{eq:delR}
\eea
Eq.~(\ref{eq:delR}) gives estimates of the corrections to the NuTeV result. 
Terms of the form $Q_{\V}$ denote the second moment (integral over all $x$) 
of a given flavor valence distribution, \EG~$\Uv = \langle x[u(x)-
\bar{u}(x)]\rangle$ represents the total fraction of the proton momentum 
carried by up valence quarks. The first term is an isoscalar correction due 
to excess neutrons in the iron target. The NuTeV collaboration have taken 
this correction into account. The correction is large (of order -.008), 
but should be known to within a couple percent. The second and third terms 
represent respectively contributions from a possible strange quark momentum 
asymmetry and from parton charge symmetry violation. 

Because the NuTeV group did not directly construct the Paschos-Wolfenstein 
ratio, Eq.~(\ref{eq:delR}) gives only an estimate of the effects of these 
contributions to the NuTeV experiment. The NuTeV group has provided 
functionals that give the sensitivity of their experiment (in Bjorken $x$) 
to various quantities \cite{Zel02b}, \EG~charge symmetry violation or a 
strange quark momentum asymmetry. To obtain a quantitative result for a 
particular effect, one multiplies the effect in question by the appropriate 
functional and integrates over $x$. For example, corrections to the 
Paschos-Wolfenstein relation depend only on valence quark properties; sea 
quarks give no contribution to that relation. Sea quark corrections to the 
NuTeV experiment are much smaller than the 
corresponding valence quark contributions, but they are not zero. 

Radiative corrections, 
which involve coupling of soft photons to the final muon line, are 
important for CC events, and constitute a substantial 
correction. Recently, Diener, Dittmaier and Hollik have re-calculated the 
radiative corrections, including all corrections of order ${\cal O}(\alpha)$, 
and a number of additional higher-order corrections \cite{Die04,Die05}. They 
find some differences from the older radiative correction program of Bardin 
and Dokuchaeva \cite{Bar86}. Diener \EA~also include new terms that result 
from including electromagnetic coupling in the QCD evolution equations. Such 
terms have recently been included by the MRST group \cite{MRST05}, and by 
Gluck \EA~\cite{Glu05}; these contributions were reviewed in 
Sect.~\ref{Sec:QEDsplt}. Diener \EA~estimate that radiative correction 
effects would remove about one-fourth of the NuTeV anomaly \cite{Die05}. 
Note that these corrections are renormalization scheme dependent. The NuTeV 
group is currently re-analyzing their data, using a radiative 
corrections code provided by Diener \EA \cite{BerPC}.  

The NuTeV measurements require nuclear corrections for the structure 
functions. Kumano \cite{Kum02} calculated a ``modified PW 
relation'' for nuclei, and Hirai \EA \cite{Hir04} estimate that nuclear 
effects could remove up to one-third of the NuTeV anomaly. They assume 
that nuclear shadowing for neutrinos is identical to that for charged 
leptons \cite{Hir03}. Kulagin and Petti \cite{Kul06,Kul07} also considered 
nuclear effects, particularly in neutrino deep inelastic scattering 
reactions. Miller and Thomas \cite{Mil02} point out that $\nu$ 
shadowing effects 
could differ significantly from shadowing of muons \cite{Bor98}. They also 
emphasize that shadowing produces different effects for CC and NC events. 
Brodsky \EA~\cite{Bro04,Kov02} made a detailed calculation of shadowing and 
antishadowing arising from multigluon exchange. They conclude that nuclear 
shadowing effects could account for roughly 20\% of the NuTeV anomaly. 

 Another effect that might contribute to the NuTeV result arises from a 
 strange quark momentum asymmetry; this is the second term in 
 Eq.~(\ref{eq:delR}). As was discussed in Sect.~\ref{Sec:sixtwo}, 
 it is possible that $s(x) \ne \bar{s}(x)$.  Although the first moment 
 of $s-\bar{s}$ must be zero (there is zero net strangeness in the proton), 
 the second moment 
\be
 S^- \equiv \langle x[s(x) - \bar{s}(x)] \rangle = \langle xs^-(x) \rangle 
\label{eq:SVdef}
\ee
 (see Eqs.~(\ref{eq:qvaldef})-(\ref{eq:qpm})) need 
 not vanish. A nonzero value for $S^-$ would mean that the net momentum 
 carried by strange quarks and antiquarks was unequal. From 
 Eq.~(\ref{eq:delR}) we see that if the strange quark momentum asymmetry 
 $S^-$ is positive (in this case, strange quarks would carry more of the 
 nucleon's momentum than strange antiquarks), this would decrease the 
 extracted value of $\sin^2 \theta_W$, and decrease the discrepancy with the 
 expected value of the Weinberg angle; conversely, a negative value of $S^-$ 
 would increase the discrepancy.  

 \begin{figure}[ht]
\includegraphics[width=2.8in]{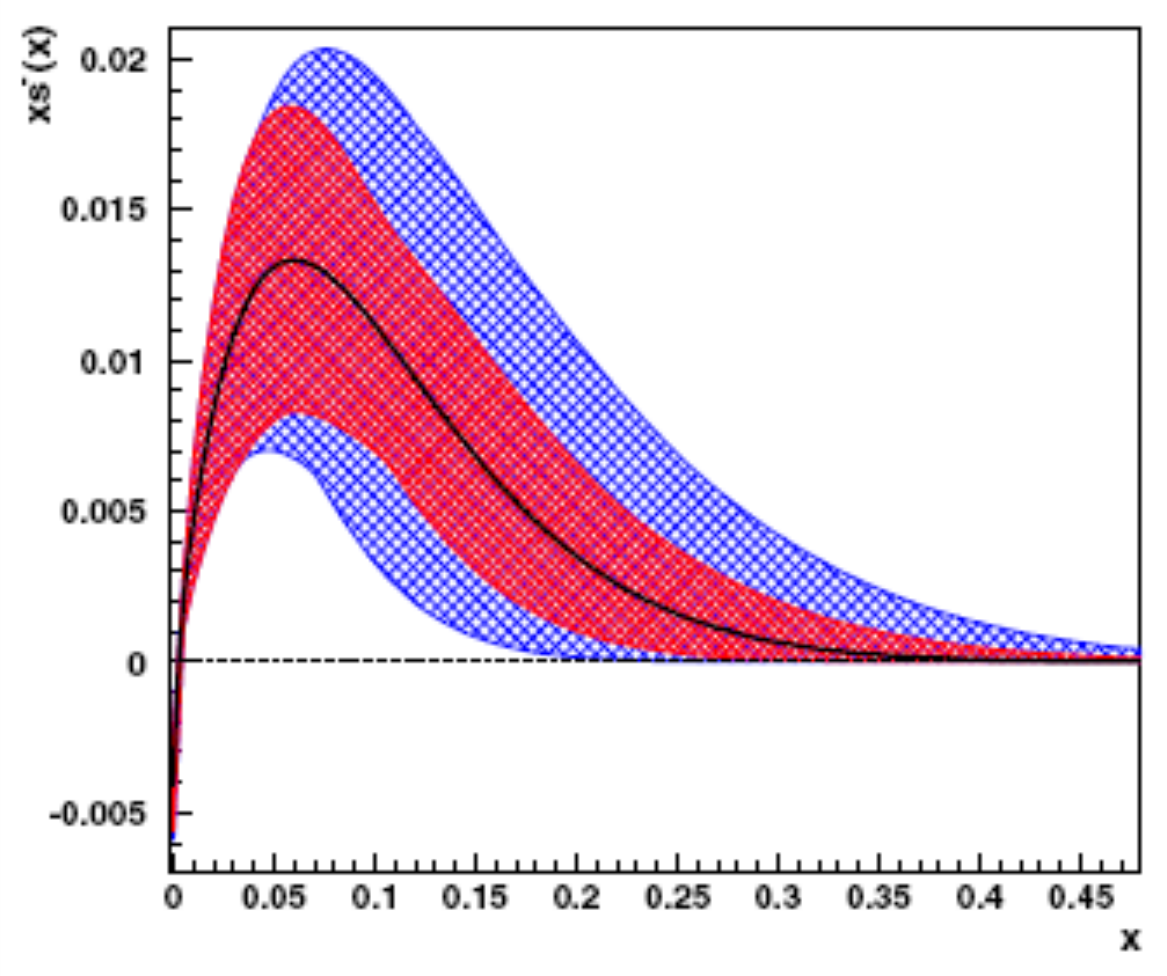}
 \vspace{0.15truein}
 \caption{[color online] The quantity $xs^-(x) = x[s(x) - \bar{s}(x)]$, 
 vs.~$x$, as extracted by the NuTeV Collaboration, Ref.~\protect\cite{Mas07}. 
 Values are obtained for $Q^2 = 16$ GeV$^2$. The outer error band is the 
 combined error, while the inner band is without the uncertainty in the 
 semileptonic branching ratio $B_c$.} 
 \label{Fig:Mason}
 \end{figure}

 The most direct knowledge of strange quark distributions comes
 from measurements of opposite sign dimuons produced in
 neutrino-induced nuclear reactions.  In such reactions, dimuon production 
 from $\nu$ ($\ovnu$) beams is sensitive to the $s$ ($\bar{s}$) distribution, 
 so that in principle comparison of these cross sections could enable one 
 to determine differences between $s$ and $\bar{s}$ PDFs.  These cross 
 sections have been extracted by the CCFR \cite{Baz95} and NuTeV \cite{Gon01} 
 collaborations. In the CCFR experiment the $\nu$ and $\ovnu$ beams are 
 not separated and the type of reaction is inferred from the charge of the 
 faster muon, while 
 the NuTeV experiment uses separated $\nu$ and $\ovnu$ beams. 

 For some time there was disagreement as to the interpretation of the
 dimuon experiments and extraction of the strange quark PDFs. The NuTeV group 
 analyzed the dimuon cross sections and extracted strange distributions 
 \cite{Zel02b}.  Their results were consistent with a small value for 
 $s^-(x)$, with a second moment that was zero or slightly negative 
 \cite{Zel02b}; the value that they extracted would increase 
 the discrepancy in the Weinberg angle to about $3.7 \sigma$. On the other 
 hand, the CTEQ group \cite{Kre04} estimated that $S^-$ was most likely 
 positive, and they suggested that this could remove roughly 1/3 of the NuTeV 
 anomaly. The CTEQ global analysis of $s^-(x)$ was dominated by the 
 CCFR/NuTeV data \cite{Baz95,Gon01} for opposite-sign dimuon production in 
 neutrino DIS, so it was unclear why the two groups obtained differing 
 results. Since then the CTEQ and NuTeV groups have collaborated on the data 
 analysis, recently obtaining consistent results.

 The latest NuTeV result obtained by Mason \EA \cite{Mas07} yields a  
 best value for $S^-$ that is positive. Fig.~\ref{Fig:Mason} plots the 
 quantity $xs^-(x)$ vs.~$x$ 
 from the latest NuTeV analysis. This results in a quantity
\bea 
  S^- &=& 0.00196 \pm 0.00046(stat) \pm 0.00045(syst) \nonumber \\ 
 &\,& {+0.00148 \atop -.00107}(external) . 
\label{eq:Sv}
\eea
In Eq.~(\ref{eq:Sv}), the quantity ``external'' refers to the contribution 
due to uncertainties on external measurements. The strange quark asymmetry 
of Eq.~(\ref{eq:Sv}) would remove roughly 
 one-third of the NuTeV anomaly. The NuTeV group provides a detailed error 
 analysis of the quantity $S^-$. It is quite sensitive to two quantities. 
 First is the semileptonic branching ratio $B_c$; the outer band in 
 Fig.~\ref{Fig:Mason} shows the result for $S^-$ with the $B_c$ uncertainty, 
 and the inner band is the result without the $B_c$ uncertainty. The second 
 is the point at which the quantity $xs^-(x)$ crosses zero (it must cross zero 
 at least once so the first moment of $s-\bar{s}$ is zero). The current best 
 fit crosses zero at a very small value $x \sim 0.004$. This means that the 
 quantity $s^-(x)$ would have a very large negative spike at very low $x$. 
 It is very difficult to imagine a physical mechanism that would cause 
 $s^-(x)$ to change sign at such a small value of $x$.   
 
\begin{figure}[ht]
\includegraphics[width=2.8in]{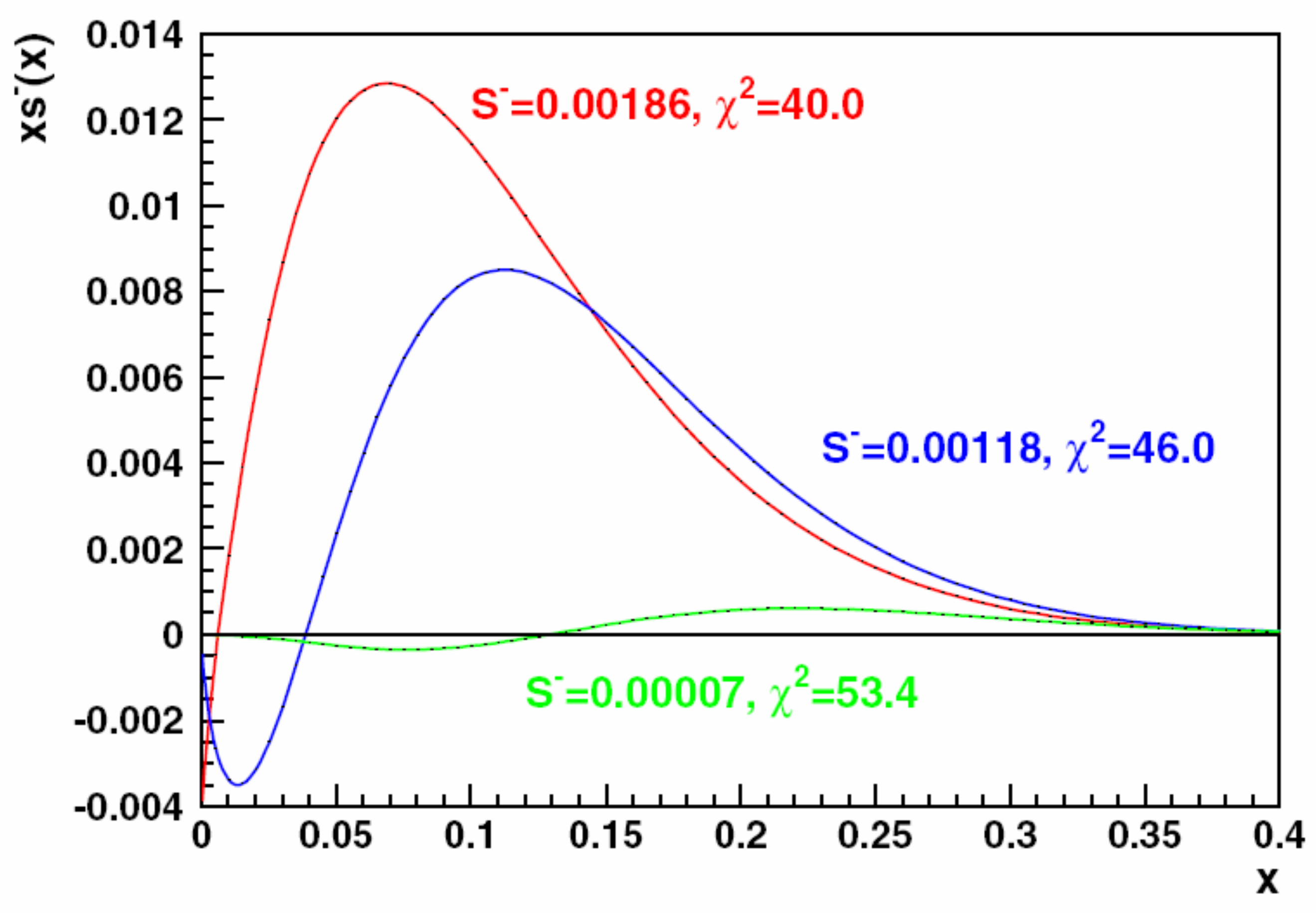}
 \vspace{0.15truein}
 \caption{[color online] The quantity $xs^-(x) = x[s(x) - \bar{s}(x)]$, 
 vs.~$x$, as extracted by the NuTeV Collaboration, Ref.~\protect\cite{Mas07}. 
 Three different results are shown, corresponding to different values of 
 the zero-crossing point. The $\chi^2$ value is listed for each curve.} 
 \label{Fig:Mason2}
 \end{figure}

 If one allows the zero-crossing point to increase, then the resulting value 
 of $S^-$ decreases, but the $\chi^2$ value also increases 
 somewhat. The best value obtained by Mason \EA~$S^- = 0.00196$ 
 occurs for a zero crossing of $x = 0.004$ and $\chi^2 = 38.2$ for 37.8 
 effective degrees of freedom. Fig.~\ref{Fig:Mason2} gives examples of 
 the relation between the zero-crossing point, the resulting curve of 
 $s^-(x)$ vs.~$x$, and the resulting $\chi^2$. For example, when the zero 
 crossing moves to $x = 0.15$, then one obtains $S^- = 0.00007$ but 
$\chi^2$ increases to 53.4. Fig.~\ref{Fig:Mason2} shows the strong 
correlation between strange quark momentum asymmetry $s^-(x)$ and the 
zero-crossing point.
     
 The contribution from charge symmetry violating parton distributions to 
 the NuTeV anomaly (the 
last term in Eq.~(\ref{eq:delR})) can be estimated by folding quark CSV 
distributions with the functionals provided by the NuTeV group.  This  
contribution is dominated by valence CSV distributions. Using the 
phenomenological CSV PDFs obtained by the MRST global fit \cite{MRST03}, 
valence CSV with $\kappa = -0.6$ would completely remove the NuTeV anomaly, 
whereas the value $\kappa = + 0.6$ would make it twice as large.  Both of 
these values are within the 90\% confidence level in the MRST global fit.
Thus the uncertainty in parton charge symmetry violation as calculated 
by MRST is capable of removing completely the NuTeV anomaly in the 
Weinberg angle.  

We can also investigate parton CSV contributions to the NuTeV result 
from theoretical calculations. From Eq.~(\ref{eq:delR}), the 
contribution from CSV to the PW ratio is proportional to 
\be 
\delta R^-_{\CSV} \propto \frac{\delta D_{\V} - \delta U_{\V}}
  {U_{\V}+ D_{\V}} \ .
\label{eq:delRCS}
\ee 
Thus the contribution from CSV to the PW ratio is related to the second 
moment of the CSV valence PDFs, divided by the total momentum carried by up 
and down valence quarks. In Sect.~\ref{Sec:sevenone} we showed that Sather's 
analytic approximation for valence charge symmetry violation gave an 
analytic expression for the second moment of these distributions (see 
Eq.~(\ref{eq:LTint})). Since these quantities depend only on total momentum 
carried by valence up and down quarks, quantities which are reasonably 
well determined, it was argued that the second moments of valence parton 
CSV were essentially model-indepedent quantities \cite{Lon03}. This 
partonic CSV correction would decrease the anomaly in the PW ratio by 
roughly 40\%. 

However, as we have stated the NuTeV group did not measure the 
Paschos-Wolfenstein ratio. If instead one uses the theoretical CSV 
distributions from Rodionov \EA~\cite{Rod94} with the functionals provided 
by NuTeV, then one finds that valence parton CSV removes  
 about $1/3$ of the anomaly in $\sin^2 \theta_W$ \cite{Lon03,Lon03b}.  
 Charge symmetry violation arising from the QED splitting mechanism described 
in Sect.~\ref{Sec:QEDsplt} would 
 remove another $1/3$ of the anomaly.  

Davidson and Burkardt \cite{Dav97} have estimated the effect on the 
Paschos-Wolfenstein relation arising from nuclear charge symmetry violation, 
\IE~the fact that protons are more weakly bound than neutrons due to Coulomb 
effects. Their results suggest that these nuclear Coulomb effects would 
increase the magnitude of the NuTeV anomaly by roughly 20\%. A new nuclear 
mechanism has recently been suggested by Clo{\"e}t \EA~\cite{Cloet:2009qs}. 
We call this ``pseudo CSV,'' since it produces effects that mimic those 
arising from charge symmetry violation. We discuss this effect and the 
implications for the NuTeV experiment in the following section.   

 We have shown that it is necessary to consider a number of ``QCD effects'' 
 within the Standard 
 Model, in order to obtain precise results for the NuTeV experiment. 
 Small but non-negligible contributions are likely from nuclear effects on 
 parton distributions and strange quark effects. At this time, charge symmetry 
 violation appears to be the only mechanism capable of single-handedly 
 removing the entire NuTeV anomaly. An additional nuclear effect which we term 
 ``pseudo CSV'' is discussed in the following section; this effect is 
 capable of making a substantial contribution to the NuTeV measurement. 
 Another possibility would be that a new 
 treatment of radiative corrections might produce significant corrections to 
 the extracted value for the Weinberg angle. However, a re-analysis of the 
 NuTeV data using the newer radiative corrections \cite{Die04,Die05} has not 
 been published at this time \cite{BerPC}.        

\subsection{Pseudo-CSV Nuclear Effects}
\label{Sec:pseudoCSV}

Many of the tests of partonic charge symmetry violation, and some applications 
that rely on CSV have been 
carried out with neutrino beams and often with an Fe target - simply to 
increase the event rate. The NuTeV anomaly, as an example, was carried out 
with an Fe target, even though the Paschos-Wolfenstein relation is only 
valid for an isoscalar target. Of course, the cross sections for $\nu$ and 
$\bar{\nu}$ scattering were corrected for the small number of excess neutrons. 
However, as pointed out in a recent paper by Clo\"et, Bentz and 
Thomas~\cite{Cloet:2009qs}, this will in general not be sufficient. It has 
been understood for some time~\cite{Geesaman:1995yd} that the famous 
``EMC effect'' \cite{emc,emcb}, the nuclear modification of the 
$F_2$ structure function in electromagnetic DIS reactions, cannot be 
understood simply in terms of the Fermi motion and 
binding of free nucleons, but the actual quark structure of the bound nucleon 
must also be modified in a significant way. A number of relatively successful 
models have been constructed~\cite{Cloet:2005rt,Cloet:2006bq,Saito:1992rm}, 
based upon the self-consistent modification of the bound nucleon structure in 
the relativistic mean scalar and vector potentials generated in a nuclear 
medium. The new realization in the case of Fe and, indeed any other nucleus 
with N $\neq$ Z, is that there will be an isovector piece of the EMC 
modification of the bound nucleon structure associated with the extra 
neutrons. Most important, {\it this effect will modify the structure of 
all of the neutrons and protons in the nucleus, not just the excess neutrons}. 

As the dominant piece of the isovector interaction in a relativistic mean 
field theory is usually associated with the $\rho$ meson, it will have a 
Lorentz vector character, with the $d$-quarks feeling more repulsion and the 
$u$-quarks more attraction. For this reason the sign of the effect is exactly 
the same as that found in the calculations of CSV which we have described 
earlier. If one ignores this medium modification it will {\it appear} 
as though the CSV is enhanced in a nucleus with $N > Z$. We stress that 
there is {\it no} violation of charge symmetry -- the isovector interaction is 
completely consistent with isospin invariance -- but to an observer unaware 
of the isovector EMC effect it will appear like CSV. An estimate of the impact 
of this additional EMC effect on the NuTeV analysis \cite{Cloet:2009qs} 
based on a nuclear matter calculation, reduces the NuTeV result for 
$\sin^2 \theta_W$ from 0.2277 to 0.2245, within 1 $\sigma$ of the Standard 
Model value. If the effect of true CSV is added \cite{Lon03,Lon03b} (as was 
discussed in the preceding section), the discrepancy is removed entirely.

 \begin{figure}[ht]
\includegraphics[width=3.0in]{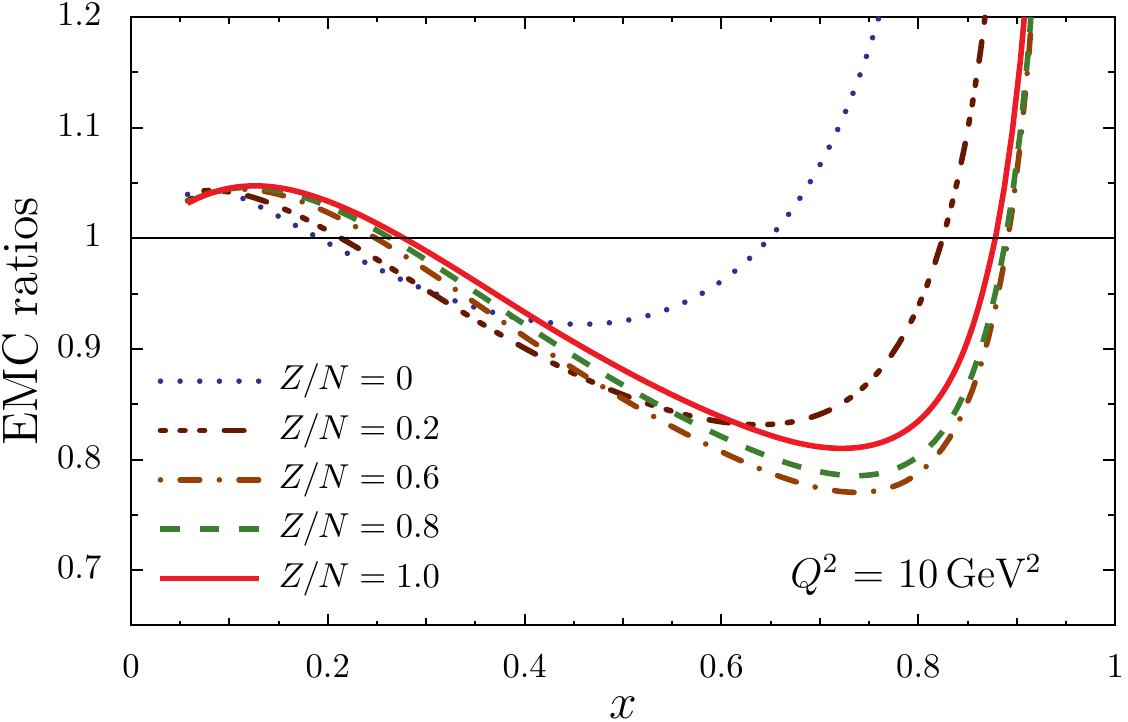}
 \caption{[color online] The EMC ratio $F_2^A(x)/F_2^D(x)$ vs. $x$, at a 
value $Q^2 = 10$ GeV$^2$, predicted 
by the model of Clo{\"e}t, Bentz and Thomas \protect\cite{Cloet:2009qs} as a 
function of the proton/neutron ratio $Z/N \leq 1$. Solid curve: $Z/N = 1$; 
dashed curve: $Z/N = 0.8$; dash-dot curve: $Z/N = 0.6$; triple dot-dashed 
curve: $Z/N = 0.2$; dotted curve: neutron matter ($Z/N = 0$).} 
 \label{Fig:Cloet}
 \end{figure}

It will clearly be very important to look for specific processes which could 
confirm this theoretical analysis of the isovector EMC effect. This model 
predicts a significant and characteristic $A$ dependence of the ratio of 
the nuclear $F_2$ electromagnetic structure function with that for the 
deuteron. Fig.~\ref{Fig:Cloet} shows the EMC ratio, $F_2^A(x)/F_2^D(x)$, 
vs.~$x$ at $Q^2 = 10$ GeV$^2$, for various values of $N \geq Z$. For 
a neutron excess, the medium modification of the $u$ quarks should be 
enhanced by coupling to the $\rho^0$ field, while the $d$ quark distribution 
should be less modified. For small neutron excess the EMC effect, which is 
initially dominated by the $u$ quarks, increases. However eventually the $d$ 
quark distribution dominates and the EMC ratio is predicted to decrease 
in the valence quark region. For example, in Au where $N \sim 1.5 Z$, a 
very large difference is predicted between the ratio of $u(x)$ 
in Au to that in the deuteron, compared with the same ratio for $d$-quarks. 
This could be investigated in experiments at Jefferson Laboratory following 
the 12 GeV upgrade.   

If the ``pseudo-CSV'' nuclear effect outlined here is confirmed 
experimentally, then it would seem that rather than presenting evidence for 
new physics beyond the Standard Model, the NuTeV result rather confirms in a 
fairly dramatic fashion the concept that the partonic structure of a 
bound nucleon is modified in a profound way.

 \subsection{Dedicated Experiments Sensitive to Valence Quark Charge Symmetry}
 \label{Sec:seventhree}

  In the preceding Section, we reviewed existing experiments
 and showed the limits they placed on charge symmetry and flavor symmetry 
 violation in parton distributions.  In this section we 
 discuss various dedicated experiments that might tighten the limits on 
 parton charge symmetry, and we review the conditions that would be necessary 
 in order that these experiments could detect parton CSV at levels that 
 are allowed from current phenomenological limits. 

 As will become clear, the experiments described are looking for quite 
 small effects, of the order of one or a few percent. In addition, one 
 generally has additional terms arising from other effects such as heavy 
 quark contributions. These must be under control before one can isolate 
 parton CSV effects. Finally, several of these require subtracting cross 
 sections from two separate measurements. These experiments are then 
 very sensitive to relative normalizations. Although these are not 
 easy experiments, it is also true that even tighter upper limits on 
 CSV contributions to parton distribution functions could improve 
 dramatically our understanding of these effects.

 \subsubsection{Drell-Yan Processes Initiated by Charged Pions}
 \label{Sec:seventhreefive}

\begin{figure}[ht]
\includegraphics[width=3.3in]{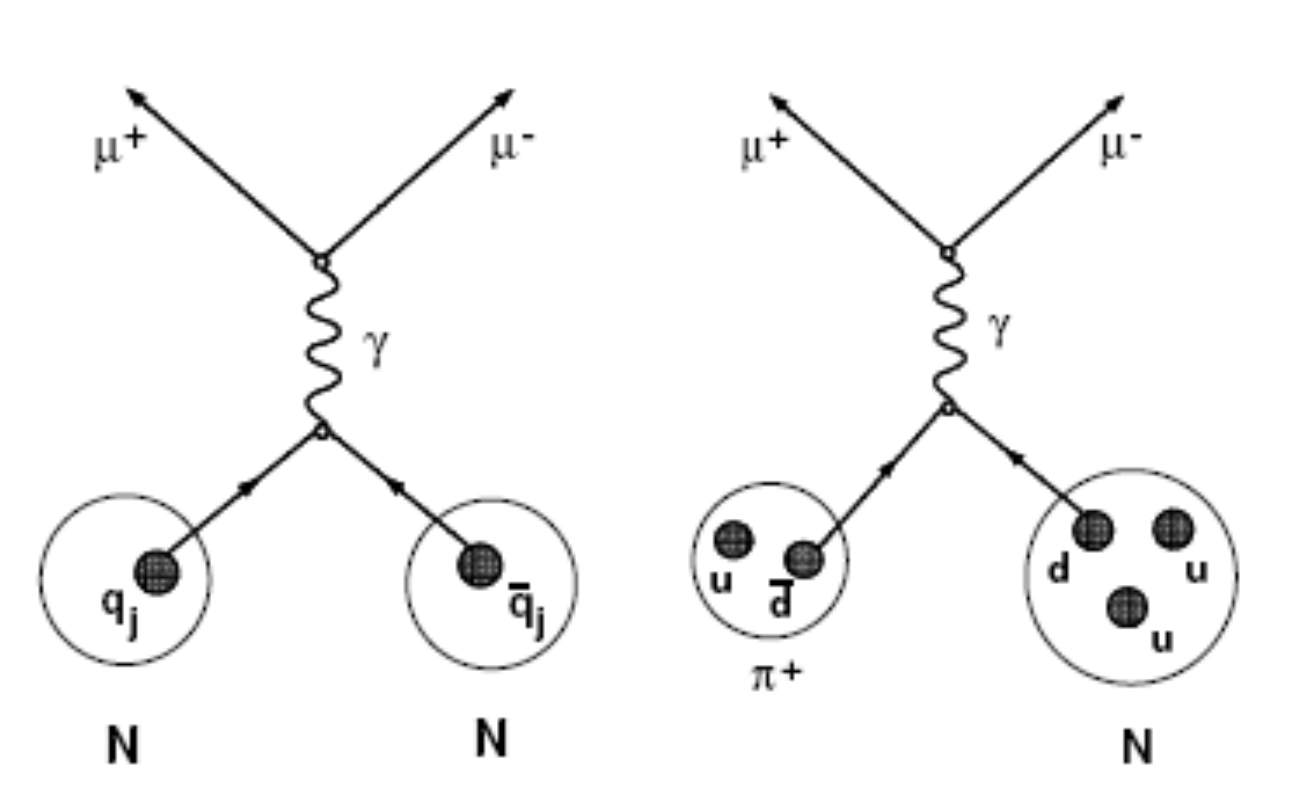}
\caption{Schematic picture of the Drell-Yan (DY) process, production of a 
 $\mu^+-\mu^-$ pair with high invariant mass through a virtual photon. 
 Left: $NN$ DY process; a quark in one nucleon annihilates with an 
 antiquark of the same flavor in the second nucleon. Right: $\pi^+p$ DY 
 process, in the valence-dominated region of $x$ for both $\pi^+$ and $p$. 
\label{Fig:DYan}}
\end{figure}

 A suitable probe for charge symmetry effects should differentiate 
 between up quarks in the proton and down quarks in the neutron.  This can
 be accomplished by comparing Drell-Yan (DY) processes induced by charged pions
 on isoscalar targets. Drell-Yan processes \cite{DY70,DY2} proceed via a quark 
 (antiquark) from the projectile annihilating an antiquark (quark) of the 
 same flavor from the 
 target, producing a virtual photon that eventually decays into 
 a pair of oppositely charged muons with large $Q^2$. The process is shown 
 schematically in Fig.~\ref{Fig:DYan}. The left figure shows the process 
 for $NN$ Drell-Yan processes. The right figure shows a schematic diagram  
 for $\pi^+-p$ Drell-Yan processes, in the kinematic regime where valence 
 quarks dominate for both the pion and nucleon.  

 We review here the calculation of Londergan \EA \cite{Lon94,Lon95}, who 
 suggested using Drell-Yan processes initiated by pions to study partonic 
 CSV. At large momentum fraction
 $x$, the nucleon distribution is dominated by its three valence
 quarks, while at large $x_\pi$ the pion is predominantly
 a valence $q-\bar{q}$ pair.  For DY processes induced by charged pion beams 
 on nucleon targets, in the kinematic region of reasonably large Bjorken $x$ 
 for both projectile and target quarks, the
 annihilating quarks will come predominantly from the nucleon and the
 antiquarks from the pion.  The
 $\pi^+$ contains a valence $\bar{d}$ (and will annihilate a $d$ quark
 in the nucleon) and $\pi^-$ a valence $\bar{u}$ (and will annihilate
 a nucleon $u$ quark). 

 Therefore, comparison of $\pi^+$ and $\pi^-$ induced DY processes on an 
 isoscalar target such as the deuteron should provide a sensitive method for
 comparing $d$ and $u$ valence distributions in the nucleon. As the
 $x$ and $x_\pi$ values of interest for the proposed measurements are
 large, a beam of 50 GeV pions will produce sufficiently
 massive dilepton pairs that the Drell-Yan mechanism is applicable.
 A flux of more than $10^9$ pions/second is desirable.  These experiments 
 might be feasible for fixed target experiments using the Fermilab Main 
 Injector \cite{Gee09}. Alternatively, such experiments would be 
 possible in the COMPASS experiment at CERN \cite{Bra08}, provided 
 that one used charged pion rather than muon beams. There exist some 
 data for $\pi^+$ Drell-Yan scattering nuclear targets dating from about 
thirty years ago, Fermilab experiment E444 \cite{E444} and CERN experiment 
 WA39 \cite{WA39}. As a general rule Drell-Yan experiments with $\pi^+$ 
 beams are more difficult than $\pi^-$, since the pions are generally 
 secondary beams arising from proton bombardment and one must be able to 
 separate the $\pi^+$ from protons. In addition, the DY cross sections for 
 $\pi^-$ will generally be larger than the corresponding DY cross section 
 induced by $\pi^+$, as seen from Eq.~(\ref{eq:pipd}).  

 Consider the DY process for a charged pion on a deuteron target. Neglecting 
 for the moment sea quark effects, at sufficiently large $x$ and $x_\pi$ the 
$\pi^{\pm}$-D DY cross sections will be proportional to:
 \bea
 \sigma_{\pi^+D}^{DY}(x,x_\pi) &\sim& {1 \over 9}\left( d^p(x)
 + d^n(x) \right)
 \overline{d}^{\pi^+}(x_\pi)~~, \nonumber \\
  \sigma_{\pi^-D}^{DY}(x,x_\pi) &\sim& {4\over 9} \left( u^p(x) +
 u^n(x) \right)\overline{u}^{\pi^-}(x_\pi). \nonumber \\ 
 \label{eq:pipd}
 \eea

 Consider the ratio $R^{DY}_{\pi D}(x,x_\pi)$, defined by
 \be
 R^{DY}_{\pi D}(x,x_\pi) =  \frac{4 \sigma_{\pi^+D}^{DY}(x,x_\pi) -
 \sigma_{\pi^-D}^{DY}(x,x_\pi)} 
 { \sigma_{\pi^-D}^{DY}(x,x_\pi) -\sigma_{\pi^+D}^{DY}(x,x_\pi) }.
 \label{eq:R}
 \ee
 This ratio will be sensitive to charge symmetry violating (CSV) terms in the
 nucleon valence parton distributions.  Since theoretical CSV 
 effects are no greater than a few percent, sea quark contributions for 
 both nucleon and pion must be included.  To first order in small 
 quantities the DY ratio for 
 pions can be written \cite{Lon94,Lon05a}:
 \bea
 R^{DY}_{\pi D}(x,x_\pi) &\approx & \left( 1 + {2\pis(\xpi)\over 
 \piv(\xpi)}\right) \nonumber \\ &\otimes& \left[ R_{\CS}(x) + 
 R_{\SV}(x,\xpi) \right]; \nonumber \\ 
 R_{\CS}(x) &=& {4 (\deldv(x) - \deluv(x)) \over 3( \uv(x) + \dv(x) ) } .
 \label{eq:Rfinal}
 \eea
 The term $R_{\SV}(x,\xpi)$ in Eq.~(\ref{eq:Rfinal}) contains sea-valence 
 interference terms which are given in Ref.~\cite{Lon94}.

 Eq.~(\ref{eq:Rfinal}) used charge conjugation invariance and  
 assumed charge symmetry for the pion PDFs. In this case one can write 
 \bea
 \piv(x) &=& u_{\V}^{\pi^+}(x) = \bar{d}_{\V}^{\pi^+}(x) = d_{\V}^{\pi^-}(x) = 
  \bar{u}_{\V}^{\pi^-}(x) \ , \nonumber \\ \pis(x) &=& q_{\Ss}^{\pi^+}(x) = 
  \bar{q}_{\Ss}^{\pi^+}(x) = q_{\Ss}^{\pi^-}(x) = \bar{q}_{\Ss}^{\pi^-}(x) 
  \nonumber \\ &\,& [q = u,d] \ .
\label{eq:pidefs}
\eea 
 Eq.~(\ref{eq:Rfinal}) is valid at sufficiently large $x$ and $\xpi$.
 It is expanded to lowest order in both sea quark and CSV terms.  
 In Eq.~(\ref{eq:Rfinal}) a pion CSV term has been neglected; theoretical 
 models predict a very small effect from this term \cite{Lon94}.    

 The relative DY fluxes for charged pions can be obtained by measuring 
 the yield of $J/\psi$'s from $\pi^+-D$ and $\pi^--D$, which should be 
identical to within one or two percent. 
 The nucleon CSV term $R{\CS}(x)$ in Eq.~(\ref{eq:Rfinal}) is
 a function only of $x$.  A number of systematic errors should
 cancel in taking the ratio of cross sections.  In particular, 
 Eq.~(\ref{eq:Rfinal}) is not sensitive to differences between parton
 distributions in the free nucleon and in the deuteron
 \cite{bodek,fs,bicker,mst}, provided that both neutron and proton
 parton distributions are modified in the same way.

\begin{figure}[ht]
\includegraphics[width=3.1in]{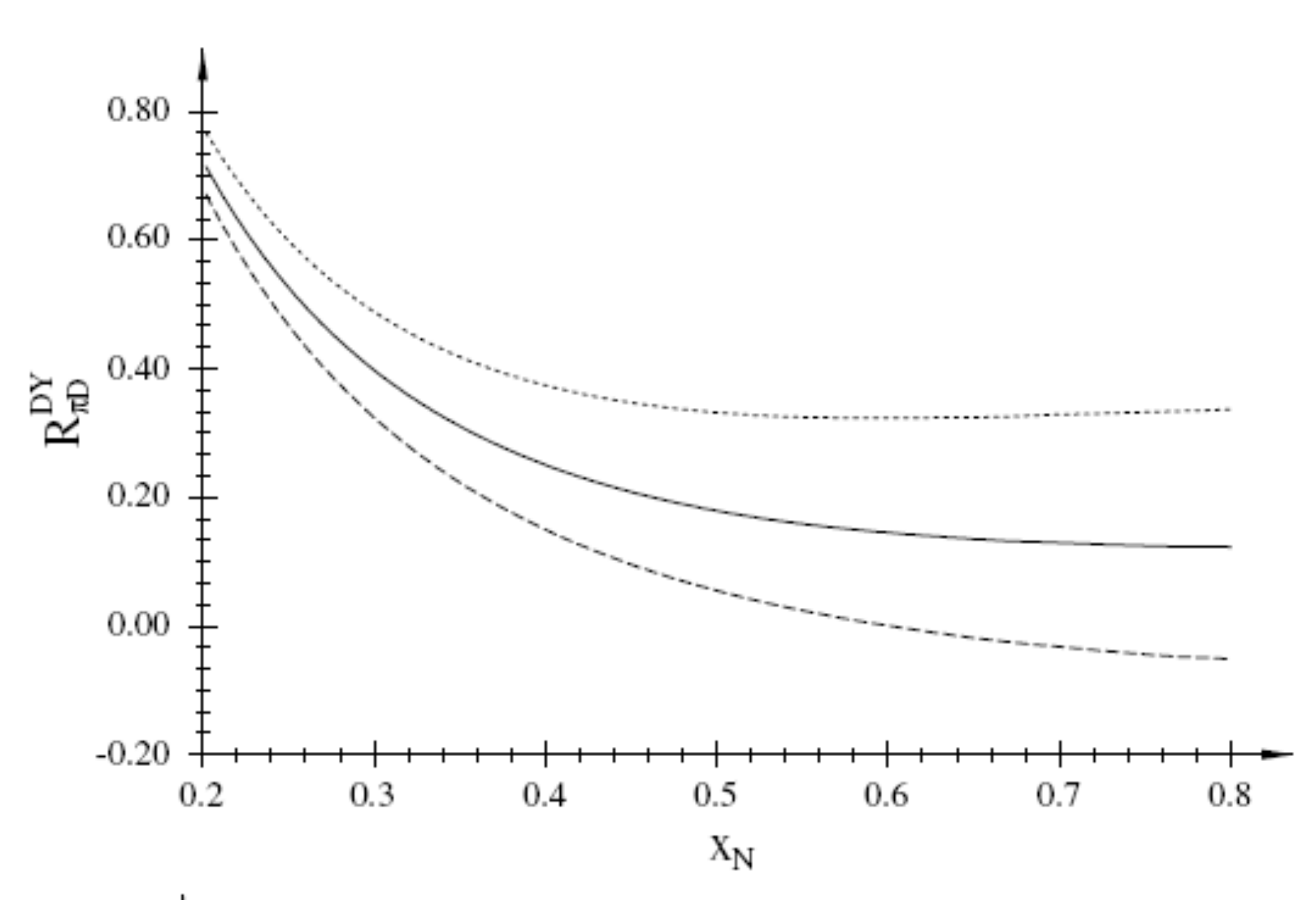}
\caption{Theoretical estimate of nucleon CSV term $R_{\CS}$ of
 Eq.~\protect(\ref{eq:Rfinal}), vs.~$x$, for $\xpi = 0.4$ and $Q^2 = 25$ 
 GeV$^2$ from Ref.~\protect\cite{Lon05a}. Solid curve: no CSV terms, 
$\kappa = 0$; dashed curve: $\kappa = +0.65$; dotted curve: $\kappa = -0.8$.
\label{Fig:rsv25}}
\end{figure}

In Fig.~\ref{Fig:rsv25} we show the ratio $R_{\pi D}^{DY}(x,\xpi)$ vs. $x$ 
 for $\xpi = 0.4$.  For the nucleon PDFs, we used the MRST global fit 
 distributions including a valence CSV contribution of the form of 
 Eq.~(\ref{eq:CSVmrst}).  The pion PDFs were taken from those of Sutton 
 \EA~\cite{Sut92}.  These were fit to older pion DY NA10 and E615 experiments 
 \cite{Bet85,Con89}; these pion distributions can be evolved to higher 
 $Q^2$ using interpolating matrices supplied by the MRST group.  
 Ref.~\cite{Sut92} provided several different pion PDFs; for 
 Fig.~\ref{Fig:rsv25} we used pion PDFs for which 10\% of the pion momentum 
 was carried by the sea.  The pion and nucleon PDFs were 
 evolved to a typical value $Q^2 = 25$ GeV$^2$.  In Fig.~\ref{Fig:rsv25}, the 
 solid curve corresponds to zero CSV contribution, the dashed curve is for 
 $\kappa = +0.65$, and the dotted curve to $\kappa = -0.8$.  These curves 
 represent the 90\% confidence limit on the CSV distributions for the 
 MRST global fit to valence quark CSV \cite{MRST03}, given in 
 Eq.~(\ref{eq:CSVmrst}).  The CSV contribution is surprisingly large.  At 
 $x = 0.5$ the limit of the two CSV terms represents about a 50\% 
 correction to the ratio, while at $x = 0.8$ the contribution is nearly 
 100\%.  

This very large contribution from charge symmetry violation is almost 
certainly an artifact of the fact that the MRST CSV PDFs are independent of 
$Q^2$, while the parton distributions in the denominator depend upon 
$Q^2$. This was discussed in Sect.~\ref{Sec:MRSTcsv}. At large values of 
$Q^2$, DGLAP evolution causes valence parton distributions to move to 
progressively smaller $x$ values. For values $x \ge 0.3$, the numerator 
($Q^2$ independent) will remain large while the denominator will become 
progressively smaller. We expect that the ratios shown in 
Fig.~\ref{Fig:rsv25} would decrease substantially if the CSV parton 
distributions were evolved in $Q^2$. 

 In a Drell-Yan $\pi-D$ experiment, one would first measure DY cross sections 
 over a wide kinematic region, and extract the pion valence and sea 
 distributions. One would then construct the DY ratio of 
 Eq.~(\ref{eq:Rfinal}).  
 The ratio could be predicted from the known nucleon PDFs and the pion PDFs 
 that have been extracted from this experiment (assuming no nucleon CSV).  
 The nucleon CSV distributions can then be extracted by comparing the 
 predicted DY ratio with the observed value.  Since the DY ratio of 
 Eq.~(\ref{eq:Rfinal}) results from  
 subtracting two large and approximately equal terms, it is necessary to 
 determine the relative DY cross sections to a few percent in order for 
 this ratio to be statistically meaningful.  Note that one can also 
 exploit the fact that the CSV contribution to the DY ratio depends only 
 on $x$ while the sea-valence term depends upon both $x$ and $\xpi$. If the 
 CSV term is sufficiently large, the process of extracting the CSV 
 distributions may have to be carried out in an iterative fashion.

\subsubsection{Parity-Violating Asymmetry in Electron Scattering}
 \label{Sec:PVelec}

 The observation of parity violation in the scattering of polarized 
 electrons from the deuteron, carried out in 1978 by Prescott 
 \EA~\cite{Pre78}, played a major role in validating the Standard Model. 
 Recent advances in the technology of parity-violating experiments 
 provide the possibility of repeating such experiments with an increase 
 in precision of better than an order of magnitude \cite{Arr06,You07}.  
 These new experiments would allow a new precision measurement of the Weinberg 
angle, they could probe physics beyond the Standard Model at the multi-TeV 
scale, and could provide tight constraints on nucleon parton distribution 
functions at large Bjorken $x$. In particular, the ratio $d(x)/u(x)$ at 
very large $x$ is not well determined \cite{Bot00}. As we will show, 
parity-violating electron scattering also has the possibility 
of observing parton charge symmetry violation at large Bjorken $x$.   

The parity violating (PV) asymmetry $\Apv$ for electron scattering on a 
nucleon can be written to lowest order in the $\gamma-Z$ interference 
in terms of the structure functions 
 \bea
 \Apv(x,y) &=& \frac{-\Gf Q^2}{4\sqrt{2}\pi\alpha} \left[ 
  \geA r_1(x) + f(y)\geV r_2(x) \right];
   \nonumber \\ 
   f(y) &=& \frac{1- (1-y)^2}{[1+ (1-y)^2]}\ ; \hspace{0.4cm}  
   y \equiv 1- \frac{E'}{E}; \nonumber \\ 
  r_1(x) &\equiv& \frac{F_1^{\gamma Z}(x)}{F_1^{\gamma}(x)} = 
  \frac{2 \sum_q e_q \,\gqV \,q^+(x)}{\sum_q e_q^2 \,q^+(x)} ; \nonumber \\ 
  r_2(x) &\equiv& \frac{F_3^{\gamma Z}(x)}{2F_1^{\gamma}(x)} = 
  \frac{2 \sum_q e_q \,\gqA \, q^-(x)}{\sum_q e_q^2 \, q^+(x)} ;\nonumber \\ 
  \geV &=& -1 + 4\sintW, \hspace {0.6cm} \geA = -1 \ . 
\label{eq:PVdef}
 \eea
In Eq.~(\ref{eq:PVdef}), we have dropped some small corrections to the 
quantity $f(y)$ and we have 
assumed the Bjorken limit where the longitudinal cross section is negligible 
relative to the transverse cross section. The additional terms are included 
in work by Hobbs and Melnitchouk \cite{Hob08}. In the proposed JLab PV 
experiment, the incident electron energy $E$ will be in the range 10-11 GeV 
and the outgoing $E'$ will run from 2 to 4 GeV. The parton model expressions 
for the ratios of structure functions are given by the quantities $r_1$ and 
$r_2$ in Eq.~(\ref{eq:PVdef}).   
  
If we confine our attention to the region of $x$ above 0.3 then the 
contribution to Eq.~(\ref{eq:PVdef}) from sea quarks should be quite small. 
Assuming that electron-deuteron scattering is given by the impulse 
approximation (the sum of scattering from proton plus neutron), and also  
assuming parton charge symmetry the expression for PV $e-D$ scattering 
can be written 
\be
\ApveD(x,y) = \frac{-\Gf Q^2}{4\sqrt{2}\pi\alpha} \left[ \aOd + 
  f(y)\aTd \right] \ .
\label{eq:APVeD}
\ee
In Eq.~(\ref{eq:APVeD}), for couplings at tree level we have 
\bea
 \aOd &=& \frac{6\geA}{5}\left( 2\guv - \gdv \right) \ ; \nonumber \\ 
 \aTd &=& \frac{6\geV}{5}\left( 2\gua - \gda \right) \ ; 
\label{eq:a1d}
\eea
In Eq.~(\ref{eq:a1d}), the quark vector couplings are given in 
Eq.~(\ref{eq:guvdef}), and the quark axial couplings are 
\be 
g^u_{\A} = \frac{1}{2}, \hspace{1.0cm}  g^d_{\A} = -\frac{1}{2} .
\label{eq:qaxial}
\ee
In this region, and with these assumptions, the PV 
asymmetry for $e-D$ scattering 
depends weakly on $y$ (the second term in Eq.~(\ref{eq:APVeD}) is 
significantly smaller than the first term) and is independent of $x$ and  
of quark PDFs. 

We can now include the lowest-order CSV contribution to the parity-violating 
$e-D$ asymmetry. In Eq.~(\ref{eq:APVeD}), the terms $\aOd$ and $\aTd$ 
are modified to  
 \bea
  \aOd &\rightarrow& \aOdz + \delCSV \aOd, \nonumber \\ 
  \aTd &\rightarrow& \aTdz + \delCSV \aTd, \nonumber \\ 
  \frac{\delCSV \aOd}{\aOdz} &=& \left[ -\frac{3}{10} + 
  \frac{2\guv + \gdv}{2(2\guv - \gdv)} \right] 
  \frac{\delta u(x) - \delta d(x)}{u(x) + d(x)}; \nonumber \\ 
  \frac{\delCSV \aTd}{\aTdz} &=& \left[ -\frac{3}{10} + 
  \frac{2\gua + \gda}{2(2\gua - \gda)} \right] 
  \frac{\delta u(x) - \delta d(x)}{u(x) + d(x)}. \nonumber \\ 
\label{eq:PVsimeq}
 \eea
We note that in Eq.~(\ref{eq:PVsimeq}), the largest contribution to the 
CSV effect in the 
parity-violating electron scattering asymmetry comes from the CSV contribution 
to the denominator, \IE~from the structure function $F_1^{\gamma\,D}(x)$ 
(this is the origin of the $3/10$ term). The CSV terms will produce a 
correction to the PV asymmetry which has a characteristic dependence on 
Bjorken $x$.    

\begin{figure}[ht]
\center 
\includegraphics[width=3.2in]{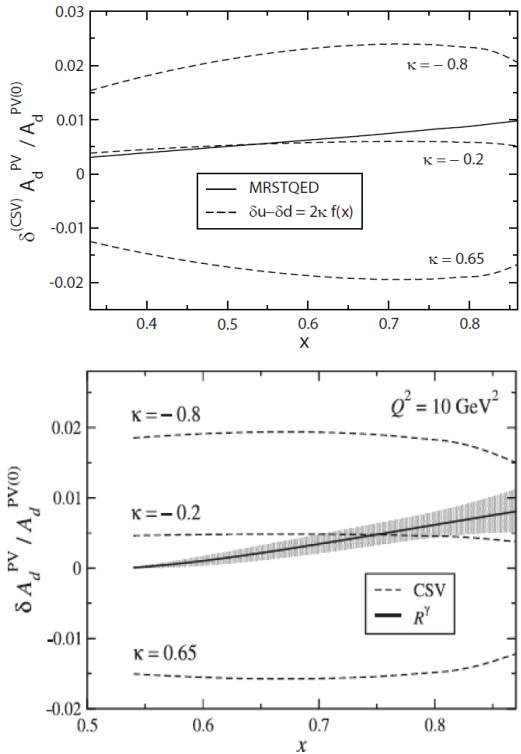}
\caption{The contribution from partonic CSV effects to the PV 
asymmetry for $e-D$ scattering with incident electron energy 10 GeV, 
from Ref.~\protect\cite{Hob08}. 
The CSV contribution is given by Eq.\ \protect({\ref{eq:PVsimeq}}). 
Curves are labeled by the value of $\kappa$ from the phenomenological 
fit of valence CSV distributions determined from the MRST group, 
Ref.\ \protect{\cite{MRST03}} and Eq.\ (\ref{eq:CSVmrst}). The best fit 
$\kappa = -0.2$ and the 90\% confidence limits are represented by 
$\kappa = +0.65$ and $\kappa = -0.8$. Upper figure: $Q^2 = 5$ GeV$^2$. Lower 
figure: $Q^2 = 10$ GeV$^2$.  
\label{Fig:globRcsv}}
\end{figure}

Fig.~\ref{Fig:globRcsv} plots the change in the $e-D$ PV asymmetry 
$\delta \ApveD/\ApveD$ arising from CSV effects, calculated by Hobbs and 
Melnitchouk \cite{Hob08}. This is obtained  
from Eq.~(\ref{eq:PVsimeq}), vs. Bjorken $x$, for electron incident energy 
$10$ GeV. The first graph plots the ratio for $Q^2 = 5$ GeV$^2$ and the 
second graph is for $Q^2 = 10$ GeV$^2$. The CSV PDFs are obtained from the 
phenomenological global fit to high 
energy data from the MRST group \cite{MRST03}.  The valence CSV distributions 
were parameterized using the form of Eq.~(\ref{eq:CSVmrst}).  The three 
dashed curves represent different values for the overall parameter 
$\kappa$. One curve shows the best-fit value $\kappa = -0.2$. The outer curves 
represent the values $\kappa = -0.8$ and $\kappa = +0.65$; these two values 
denote the 90\% confidence 
limit for the valence CSV allowed in the MRST global fit.  

Within the 
90\% confidence limit, the predicted CSV contribution to the PV asymmetry 
tends to increase with increasing Bjorken $x$. The magnitude of the CSV 
contribution increases with increasing $Q^2$; for $Q^2 = 10$ GeV$^2$, the CSV 
contributions allowed within the MRST 90\% confidence level range between 
roughly -0.025 and +0.03 at a value $x = 0.7$. Thus 
if experiments could achieve a precision of about 1\% in the asymmetry, 
it should be possible either to observe effects of partonic CSV in this 
experiment, or alternatively to put strong constraints on upper limits for 
partonic charge symmetry violating effects. Note that our 
results are model-dependent, as the MRST group chose the particular 
functional form given in Eq.~(\ref{eq:CSVmrst}) for their partonic CSV PDFs. 
In addition, for simplicity MRST neglected the 
$Q^2$ dependence of the CSV parton distribution functions in their global 
fits to high energy data. 
      
There is an additional uncertainty  
in the parity-violating asymmetries. Since our predicted PV asymmetry 
becomes significant only at large $x$, we need to account for the fact 
that the $d/u$ ratio in the proton is rather poorly known in this region. This 
uncertainty was studied by Botje \cite{Bot00} who extracted quark PDFs from 
a QCD analysis of combined HERA and fixed-target data. A precise 
determination of $d/u$ at large $x$ comes from the NMC measurements of 
muon DIS on proton and deuteron targets \cite{Ama91,Ama92,NMC97}. 
However, in both cases the limit on accuracy is not the data but the 
theoretical understanding of the EMC effect in deuterium. For example, 
the covariant treatment of Fermi motion and binding by Melnitchouk and 
Thomas \cite{Mel96} show that, contrary to the conclusions in the original 
paper, the SLAC data was consistent with the perturbative QCD predictions 
for the $u/d$ ratio as $x \rightarrow 1$. For $x > 0.4$ the errors on the 
QCD predictions grow fairly rapidly. This occurs because the electromagnetic 
coupling is weighted by the squared charge of the quark flavor.

An independent measurement of $d/u$ in the proton can be obtained by 
measuring $W$ production in $p-\bar{p}$ collisions. In these reactions 
a $W^+$ tends to be produced by annihilation of a $u$ quark from the proton 
and a $\bar{d}$ from the antiproton, while a $W^-$ is produced from a 
$d$ quark in the proton and a $\bar{u}$ from the antiproton. Because the 
$u$ quark carries a larger momentum fraction than the $d$, $W^+$ production 
will tend to be boosted in the proton direction while $W^-$ will be boosted 
in the antiproton direction. One measures a forward backward asymmetry 
$A(y_l)$, where 
\be 
A(y_l) = \frac{ \frac{d\sigma(\ell^+)}{dy_l} - \frac{d\sigma(\ell^-)}{dy_l}}
   {\frac{d\sigma(\ell^+)}{dy_l} + \frac{d\sigma(\ell^-)}{dy_l}} \ .
\label{eq:AyW}
\ee
In Eq.~(\ref{eq:AyW}), $y_l$ is the rapidity of the lepton arising from decay 
of the $W$, and $d\sigma(\ell^{\pm})/dy_l$ is the differential cross section 
for charged lepton production. This is a convolution of the cross section 
for $W$ production, with the relevant $W \rightarrow \ell \nu$ decay 
distribution \cite{Mar90,Melpeng}.  

The forward-backward asymmetries have been measured by the CDF \cite{Aco05b} 
and D0 \cite{Aba08,Aba08b} groups at the Fermilab Tevatron for 
$p-\bar{p}$ collisions at $\sqrt{s} = 1.96$ TeV. The asymmetries tend to be 
particularly sensitive to the slope of the $d/u$ ratio. The higher the 
rapidity, the larger the $x$ range for which $d/u$ can be studied 
\cite{Aba08b}. 
 
The $d/u$ ratio could also be determined 
from large-$x$ parity-violating electron scattering on hydrogen, since the 
PV amplitude preferentially couples to the down quark. This information could 
in principle fix the $d/u$ ratio in the proton and eliminate some of the 
uncertainty in PV DIS reactions on deuterium.

 \subsubsection{Charged Pion Leptoproduction from Isoscalar Targets}
 \label{Sec:seventhreefour}

 In the preceding section it was pointed out that DY processes 
 for charged pions on nucleons can test CSV because the $\pi^\pm$ 
 contain different valence antiquarks.  For this reason,   
 semi-inclusive pion production, from lepton DIS on nuclear targets,
 could be a sensitive probe of CSV effects in nucleon 
 valence parton distributions \cite{Lon05a,Lon96}.  The cross section for 
 this process is given by \cite{Lev91} 
 \bea
 {1\over \sigma_N(x)}{d\sigma^h_N(x,z)\over dz} &=&
 {N^{Nh}(x,z) \over \sum_i e^2_i q^N_i(x)} \ , \nonumber \\ 
  N^{Nh}&\equiv& \sum_i e^2_i q^N_i(x)D^h_i(z) \ .
 \label{cs}
 \eea
 The quantity $N^{Nh}$ in Eq.~(\ref{cs}) is the yield of hadron
 $h$ per scattering from nucleon $N$, and $D^h_i(z)$ is the fragmentation 
 function for a quark of flavor $i$ into hadron $h$.  $D^h_i(z)$ depends on 
 the quark longitudinal momentum fraction $z=E_h/\nu$, where $E_h$ and $\nu$ 
 are respectively the energy of the hadron and the virtual photon.

 For pion electroproduction on an isoscalar target, charge symmetry 
 relates the ``favored'' production of
 charged pions from valence quarks, by
 \begin{eqnarray}
 N^{D\pi^+}_{fav}(x,z) &=& 4\,N^{D\pi^-}_{fav}(x,z) .
 \label{csvlep}
 \end{eqnarray}
 In Eq.~(\ref{csvlep}), $N^{D\pi^+}_{fav}(x,z)$ represents
 the yield of $\pi^+$ per scattering from the deuteron, via
 the favored mode of production (for $\pi^+$ ($\pi^-$) production, the 
 ``favored'' mode of
 charged pion production is from the target up (down) quarks).
 The HERMES collaboration at HERA \cite{HERA}
 has measured semi-inclusive pion production from hydrogen and 
 deuterium.      

 Londergan \EA \cite{Lon96,Lon05a} showed that the ratio  
 $R^{\Delta}(x,z)$ is sensitive to parton CSV effects, where the 
 ratio is defined by
\bea 
R^{\Delta}(x,z) &\equiv& {8\left( {{\textstyle N^{D\pi^-}(x,z)}\over 
 {\textstyle 1 + 4\Delta(z)}} - 
 {{\textstyle N^{D\pi^+}(x,z)}\over {\textstyle 4 + \Delta(z)}}\right) 
  \over N^{D\pi^+}(x,z) - N^{D\pi^-}(x,z) }; \nonumber \\ 
  R^{\Delta}(x,z) &=& C^{\Delta}(z) \left[ R_{\CS}(x) + R_{\SV}(x,z) \right];
  \nonumber \\ 
  \Delta(z) &\equiv& {D_u^{\pi^-}(z) \over D_u^{\pi^+}(z)};  \nonumber \\  
 C^{\Delta}(z) &=& {8(1 + \Delta(z)) \over (1 + 4\Delta(z))(4 + \Delta(z)) }. 
 \label{sigtil}
 \eea
 Eq.\ (\ref{sigtil}) is evaluated at moderately large $x$, where 
 the sea/valence ratio is small.  It has been expanded  
 to first order in the CSV nucleon terms and the sea quark distributions.  
 The term $R_{\CS}(x)$ is given in Eq.~(\ref{eq:Rfinal}), and $R_{\SV}(x,z)$ 
 is a sea-valence interference term given in Ref.~\cite{Lon96}.  
 A CSV part of the fragmentation function has been dropped as 
 theoretical estimates suggest that this term should be very small 
 \cite{Lon96}.

 The ratio $R^{\Delta}$ of Eq.~(\ref{sigtil}) has an overall factor that 
 depends only on $z$; the normalization of the ratio is chosen to make this 
 term close to one for moderate values of $z$.  The remainder of this ratio 
 contains two terms.  The first term depends only on $x$, and is proportional 
 to the nucleon valence CSV fraction; it is identical to the term 
 $R_{\CS}(x)$ defined in Eq.~(\ref{eq:Rfinal}), which could be measured in 
 pion Drell-Yan reactions.  
 The final term in Eq.~(\ref{sigtil}) depends on both $x$ and $z$; it is 
 proportional to the sea quark contributions, and becomes progressively
 less important at large $x$.  

\begin{figure}[ht]
\includegraphics[width=3.0in]{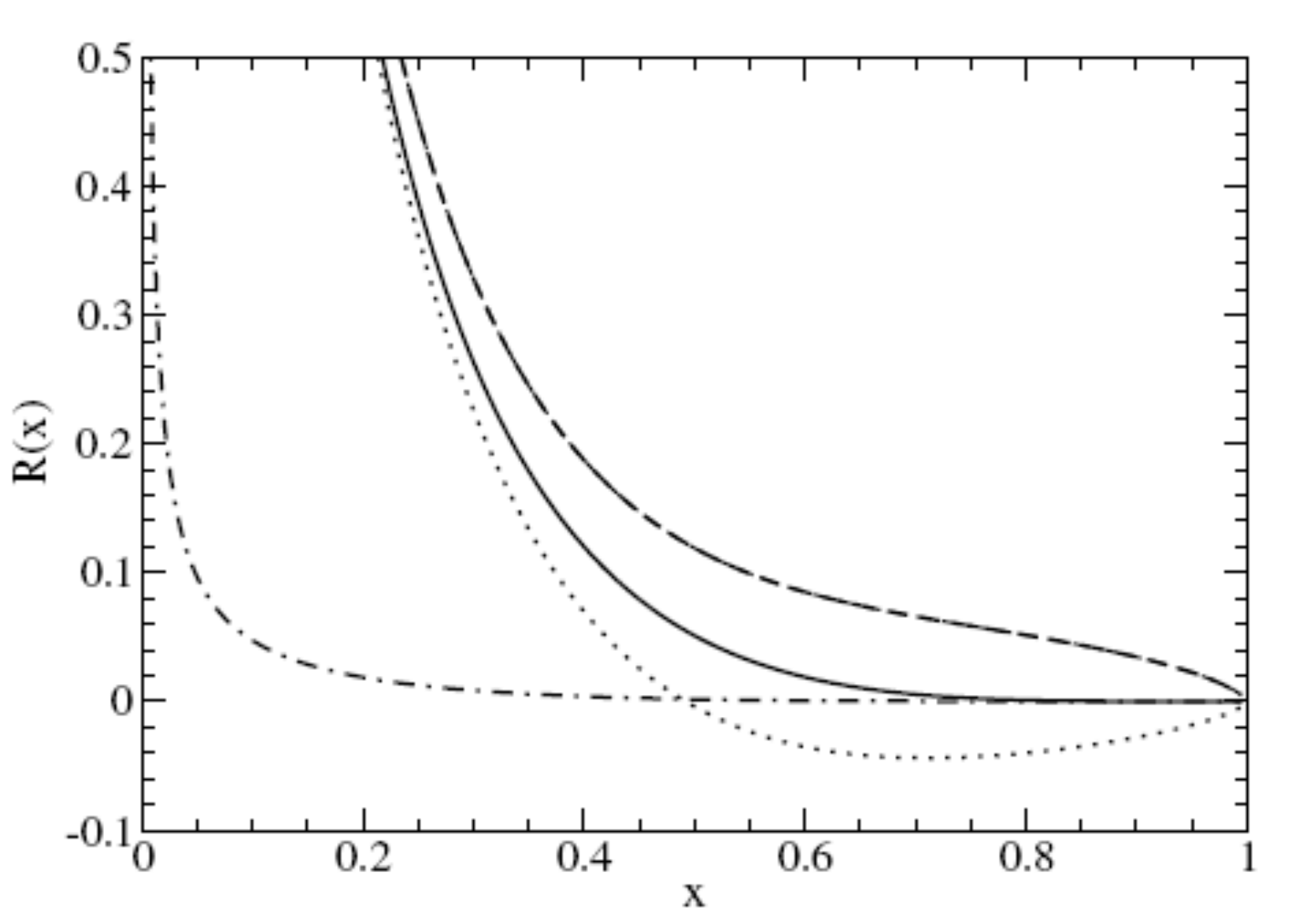}
\caption{Contributions of various terms to the ratio $R^{\Delta}(x,z)$ defined 
in Eq.~(\protect{\ref{sigtil}}) vs.\ $x$ at fixed $z=0.4$. Solid (dot-dashed)
curves: nonstrange and strange sea quark contributions. Long dash-dot (dotted) 
curves: CSV contributions from Eq.~(\protect{\ref{sigtil}}), 
for $\kappa = -0.8$ and $\kappa= +0.65$, respectively.  Curves are calculated 
for $Q^2 = 2.5$ GeV$^2$.
\label{Fig:sidisX}}
\end{figure}

  Fig.~\ref{Fig:sidisX} plots the $x$-dependent contributions to the ratio 
 ${R}^{\Delta}(x,z)$ of Eq.~(\ref{sigtil}) vs.~$x$ at fixed $z=0.4$ 
 \cite{Lon05a}.  The solid (dot-dashed) curves show the 
 non-strange (strange) sea contributions to the ratio. The strange quark 
 contribution is negligible except at extremely low $x$.  The long dash-dot 
 and dotted curves show the contributions from quark CSV contributions from 
 the MRST global fit with $\kappa = -0.8$ and $+0.65$, respectively; these 
 represent the 90\% confidence limits for the MRST CSV PDFs.  At this value 
 of $z$, the coefficient $C^{\Delta}(z)$ from Eq.~(\ref{sigtil}) has 
 a value very close to 1. The CSV terms are substantial only for large $x \ge 
 0.4$.  The ratio requires precise experimental measurements of the
 $x$-dependence of $R^{\Delta}(x,z)$ for fixed $z$.  Note that this depends 
 critically on the validity of the factorization hypothesis for the 
 semi-inclusive yields, as given by Eq.~(\ref{cs}).   

For $x \ge 0.4$, the contributions from charge symmetry violating PDFs 
are substantial, and they rapidly become the dominant contribution at 
larger $x$.  Thus, at the levels determined by the MRST global fit, 
it would appear that precise measurements of charged pion production 
in semi-inclusive DIS electroproduction reactions on deuterium 
have the possibility of observing these isospin-violating effects, 
or they would be able to lower the current allowed limits on partonic 
CSV effects.  Theoretically it would be possible to observe such effects in 
measurements of $e + D \rightarrow \pi^{\pm} + X$ at Jefferson 
Laboratory. However, the validity of Eq.~(\ref{sigtil}) requires 
that factorization (the fragmentation function for quarks into pions) be 
valid to within a few percent. It would be necessary to demonstrate that 
factorization is obeyed to a very high degree, at energies available at 
Jefferson Lab. 

\begin{figure}[ht]
\includegraphics[width=3.0in]{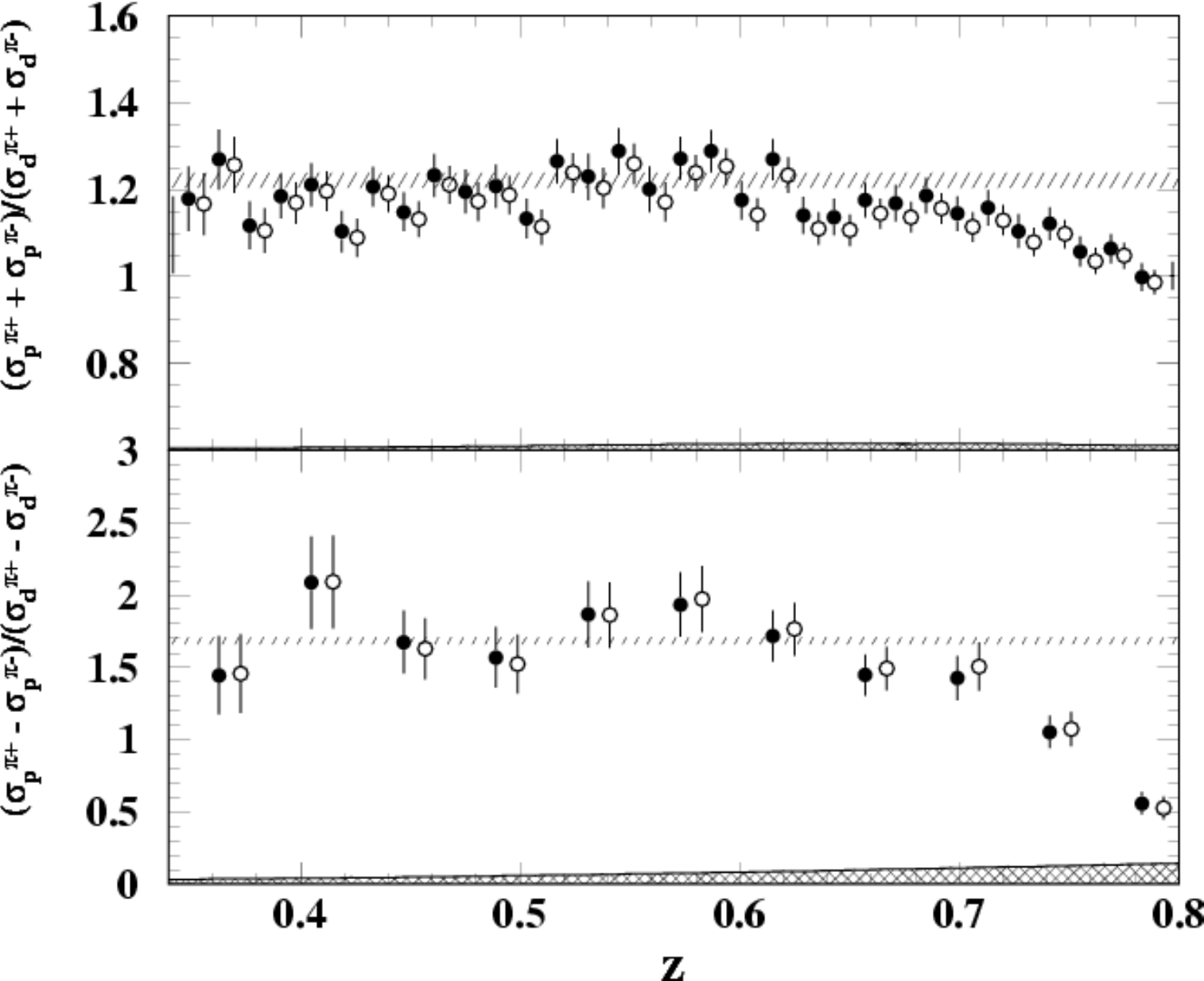}
\caption{Ratios of proton to deuteron semi-inclusive charged pion 
electroproduction cross sections on proton and deuteron at fixed $x= 0.32$ as 
a function of $z$, from Ref.~\protect\cite{Nav07}. Solid (open) symbols 
reflect data after (before) subtraction of coherent $\rho$ production 
events. Symbols are offset slightly in $z$ for clarity. 
Top curve: sum of $\pi^+$ and $\pi^-$ cross sections (first equation in 
Eq.~(\protect\ref{eq:Xsecrat})); bottom curve: difference of cross 
sections (second equation in Eq.~(\protect\ref{eq:Xsecrat})). The hatched 
area in the bottom curve indicates systematic uncertainties. The shaded 
bands represent a variety of calculations in both leading and next-to-leading 
order QCD from CTEQ \protect\cite{Lai00} and GRV \protect\cite{GRV98}. 
\label{Fig:Bosted}}
\end{figure}

The validity of factorization has been checked for semi-inclusive 
deep inelastic reactions at Jefferson Lab energies. Navasardyan 
\EA~\cite{Nav07} measured charged pion electroproduction from $p$ and $D$ 
 with 5.5 GeV electrons. They were compared with a factorization hypothesis 
\bea 
 d\sigma &\sim& \sum_q e_q^2 \, q(x,Q^2)\, D_{q\rightarrow \pi}(z,Q^2) 
  e^{-b p_T^2} \nonumber \\ &\times& \left[ \frac{1 + A\cos \phi + 
  b\cos(2\phi)}{2\pi}\right] 
\label{eq:factor}
\eea 
With factorization, the semi-inclusive cross section appears as the product 
of a parton distribution function depending on $x$ but not $z$, times a 
fragmentation function for a quark to a pion that depends on $z$ but not $x$. 
Assuming factorization, one can derive expressions for ratios of the 
pion electroproduction cross sections on protons and deuterium, 
\bea
\frac{\sigma_p(\pi^+) + \sigma_p(\pi^-)}{\sigma_{\D}(\pi^+) + 
  \sigma_{\D}(\pi^-)} &=& \frac{4u^+(x) + d^+(x)}
  {5(u^+(x) + d^+(x))} \nonumber \\ 
\frac{\sigma_p(\pi^+) - \sigma_p(\pi^-)}{\sigma_{\D}(\pi^+) - 
  \sigma_{\D}(\pi^-)} &=& \frac{4\uv (x) - \dv (x)}
  {3(\uv (x) + \dv (x))} \ . \nonumber \\ 
\label{eq:Xsecrat}
\eea

Fig.~\ref{Fig:Bosted} shows ratios of pion electroproduction cross sections 
on $p$ and $D$ at fixed $x= 0.32$ vs.~$z$. From Eq.~(\ref{eq:Xsecrat}), 
these linear combinations should be independent of $z$. Now, a number of 
assumptions have gone into Eq.~(\ref{eq:Xsecrat}); in addition to 
factorization, this relation assumes parton charge symmetry, neglects 
contributions from heavy quarks and also contributions from any $p_T$ 
dependence of parton distributions. Nevertheless, to within about ten percent 
the ratios show very little dependence on $z$ for $z < 0.7$. The deviation 
from these curves for $z > 0.7$ results from the $N \rightarrow \Delta$ 
transition region. Furthermore, the shaded bands show the ratio that is 
expected from phenomenological parton distributions from the CTEQ 
collaboration \cite{Lai00} and from GRV \cite{GRV98}. The experimental 
results show that factorization is a rather good approximation at Jefferson 
Laboratory energies. 

Despite the perhaps surprisingly good agreement with the factorization 
hypothesis in this energy region, nevertheless factorization is not 
sufficiently accurate to carry out tests of charge symmetry violation in pion 
electroproduction reactions, at current Jefferson Lab 
energies. The relations in Eq.~(\ref{sigtil}) would be more reliable at 
a possible future electron-ion collider, where factorization 
 should be assured to a high degree. As Eq.~(\ref{sigtil}) was derived in 
 lowest order QCD, it is necessary to check whether the results remain 
 essentially unchanged in NLO.    

 \subsubsection{Test of Weak Current Relation $F_2^{W^+ N_0}(x) =
 F_2^{W^- N_0}(x)$}
 \label{Sec:seventhreethree}

 From Eq.~(\ref{eq:Fccdef}), at high energies the $F_2$ 
structure functions for charge-changing neutrino and antineutrino interactions 
 on an isoscalar target are equal except for contributions from valence quark 
 CSV, plus strange and charm quark terms, \IE 
 \bea
&\,& F_2^{W^+ N_0}(x,Q^2) - F_2^{W^- N_0}(x,Q^2) \nonumber \\ &=& x\left[ 
 \deldv(x) - \deluv(x) +  2\left( s^-(x) - c^-(x)\right) \right] .\nonumber \\ 
 \label{f1diff}
 \eea
 These cross sections might be measured at various experimental facilities.  
 At a high energy electron collider, weak interaction processes such as 
$e^- p \rightarrow \nu_e X$ are no longer completely negligible with respect 
to the electromagnetic process $e^- p \rightarrow e^- X$. 
Charged-current cross sections in $e^{\pm}-p$ reactions were made at 
HERA \cite{Adl03}, where precise structure functions and parton distributions 
were measured for momentum transfers $Q^2 > 100$ GeV$^2$. Tests of 
parton charge symmetry would require collisions of electrons with an 
isospin-zero nucleus such as the deuteron. Then by 
comparing charge-changing weak interactions induced by electrons and positrons
 the quantity in Eq.~(\ref{f1diff}) could be measured. This might be 
feasible at a future electron-ion collider. 

Alternatively with very high energy neutrinos, the ratio of 
Eq.~(\ref{f1diff}) could be 
 measured by comparing $W$ boson production on an isoscalar target induced 
 by neutrinos and antineutrinos.  Theoretical estimates of this process were 
 made by Londergan, Braendler and Thomas \cite{herathy}.  One constructs 
the ratio
 \bea
 R_W(x) &\equiv& { 2\left( F_2^{W^+D}(x) - F_2^{W^-D}(x)\right) \over
  F_2^{W^+D}(x) + F_2^{W^-D}(x)} \cr &\approx& \frac{ \deldv(x)- \deluv(x) 
 + 2(s^-(x) - c^-(x))}{\sum_j q_j^+(x)} \nonumber \\ 
 &\equiv& R_{CSV}(x) + R_S(x) \ . 
 \label{Rcsv}
 \eea
 At sufficiently high energies, the only quantities contributing to the 
 ratio $R_W$ are ``valence'' strange and charm distributions, or valence 
 quark CSV terms.  The ratio could also be checked for any isoscalar nuclear
 target, replacing the nucleon parton distributions by their
 nuclear counterparts.

 \begin{figure}[ht]
\center 
\includegraphics[width=2.7in]{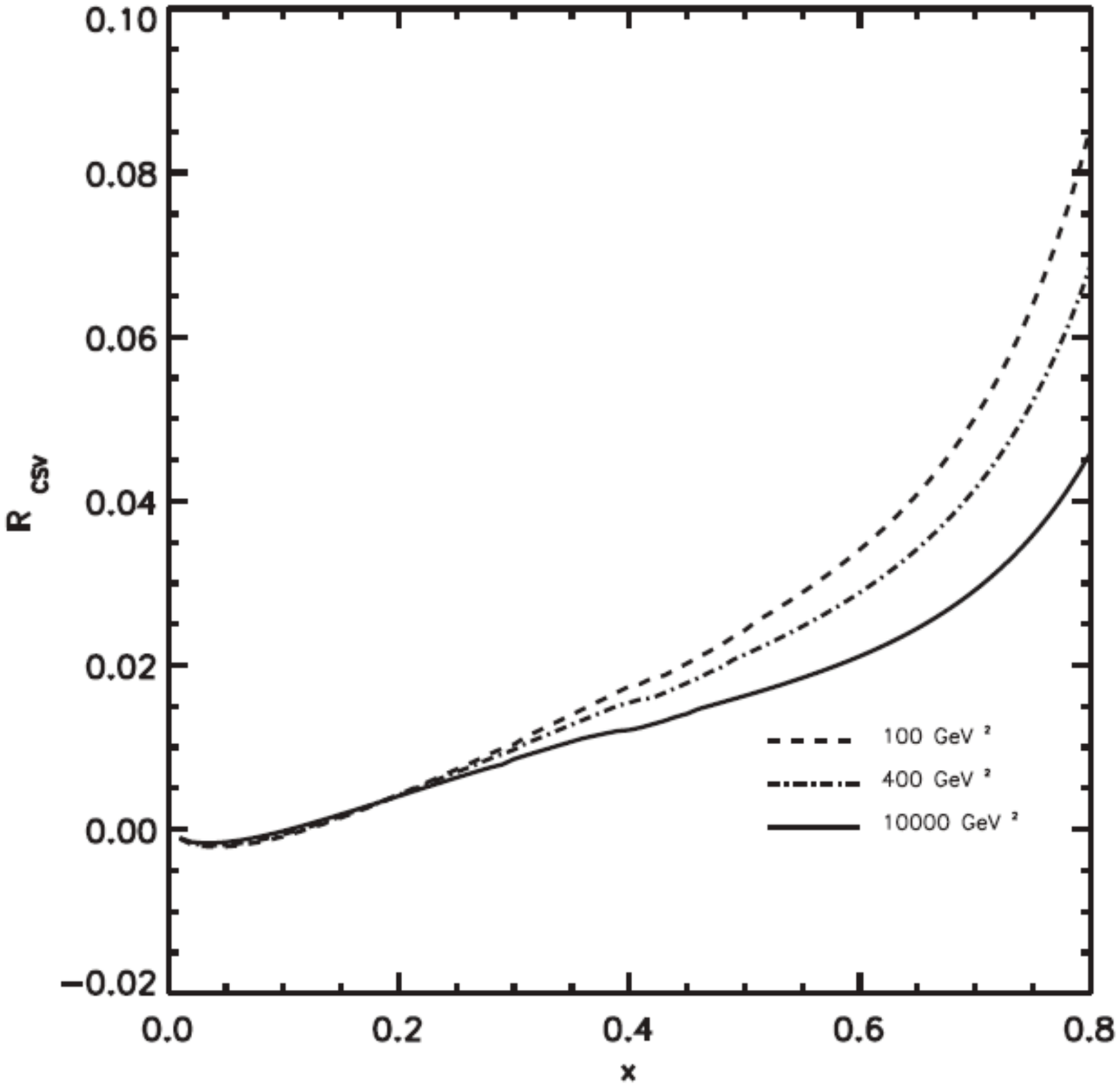}
 \includegraphics[width=3.0in]{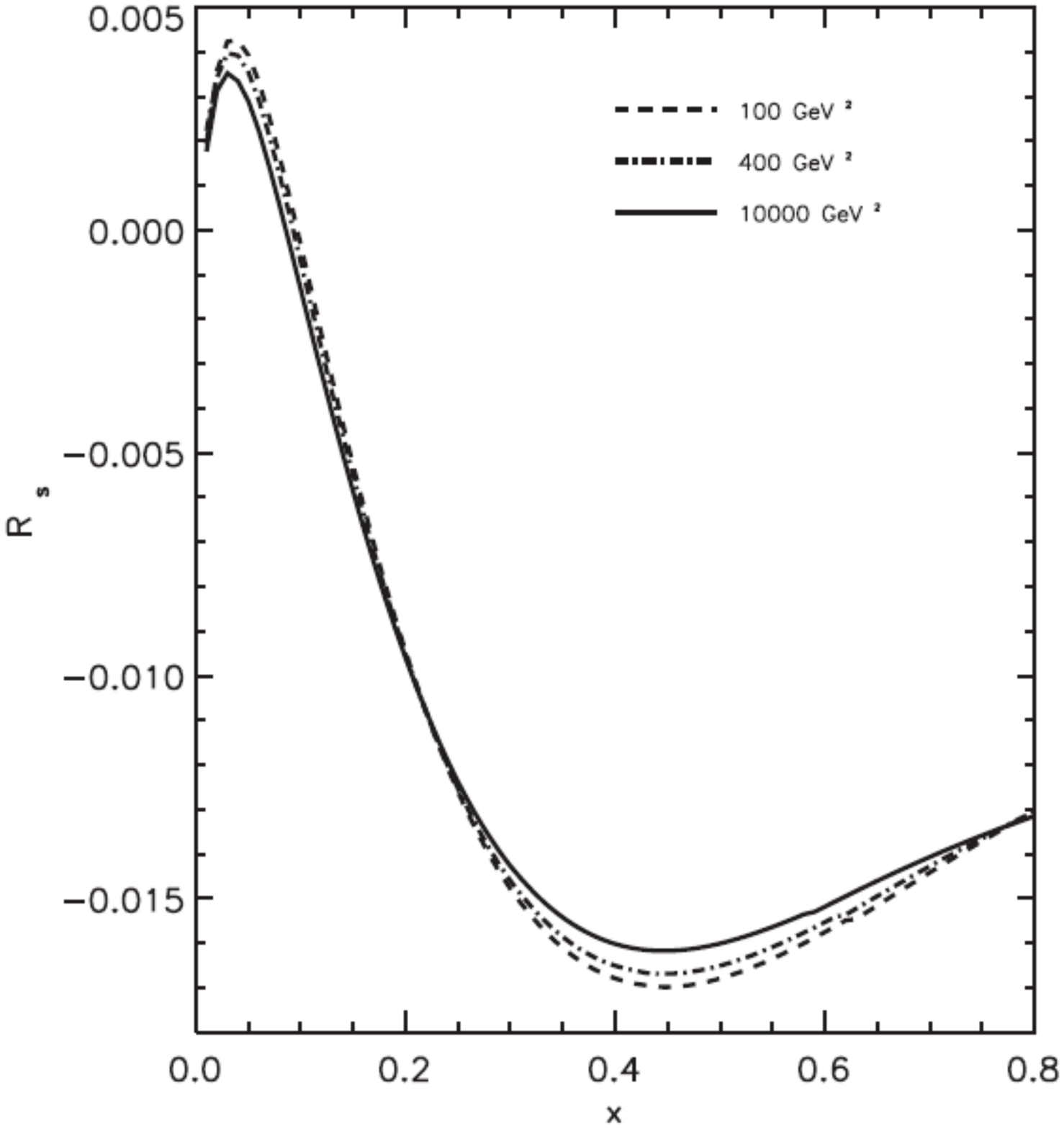}
 \caption{Top: theoretical estimates of CSV contribution $R_{CSV}$ to
 the ratio $R_W(x)$ of Eq.~\protect(\ref{Rcsv}) vs.\ $x$, for various
 values of $Q^2$.  b) Bottom: theoretical estimates of $s-\bar{s}$ difference,
 $R_s(x)$ of Eq.~\protect(\ref{Rcsv}). From Ref.\ \protect\cite{herathy}.}
 \vspace{0.1truein}
 \label{fig42cc}
 \end{figure}

 The top figure in Fig.\ \ref{fig42cc} shows the theoretical CSV contribution,
 $R_{CSV}(x)$ from Eq.~(\ref{Rcsv}).  The dashed curve is calculated for
 $Q^2 =100$ GeV$^2$, the dot-dashed curve for $Q^2 =400$ GeV$^2$, and the
 dash-triple dot curve for $Q^2 =10,000$ GeV$^2$.  The quantity
 $R_{CSV}(x)$ is predicted to be greater than 0.02 provided $x > 0.4$, 
 using CSV estimates of Rodionov \EA~Ref.\cite{Rod94}.
 All theoretical calculations predict that in the valence region, 
 $\delta d_{\rm v}(x)$ is positive and $\delta u_{\rm v}(x)$ negative,
 so their effects should add, producing several percent effects at the 
 largest values of $x$. The term $R_s$ of Eq.~(\ref{Rcsv}), proportional
 to the difference between strange quark and antiquark distributions (we 
 neglect possible contributions from charm quarks), 
 is shown in the bottom figure in Fig.\ \ref{fig42cc}. 

\begin{figure}[ht]
\center
\includegraphics[width=2.5in]{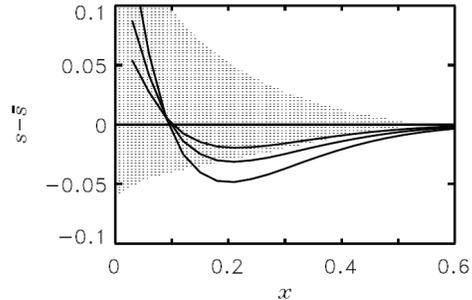}
\caption{The strange quark asymmetry $s(x) - \bar{s}(x)$ calculated in 
 the meson-cloud model of Melnitchouk and Malheiro, 
 Ref.\protect\cite{Mel97}. The curves correspond to different values for 
 the $NKY$ form factor. The shaded region is an eetimate of the uncertainty 
 in the calculations. 
\label{Fig:MelMal}}
\end{figure}

 As we mentioned in Sec.~\ref{Sec:sixtwo} and expanded upon in 
 Sec.~\ref{Sec:sixthreetwo}, there are now experimental 
 measurements from which one can extract the strange quark asymmetry 
 $s^-(x) = s(x)-\bar{s}(x)$. These come from 
 production of opposite-sign dimuon pairs in reactions initiated by 
 $\nu$ or $\bar{\nu}$ beams, from the CCFR and NuTeV groups 
 \cite{Baz95,Gon01}. The first moment $\langle s^-(x) \rangle$ must 
 be zero since there is no net strangeness in the nucleon. This means that 
 if there is a nonzero strange quark asymmetry it must have at least one 
 node in 
 $x$. The latest analysis of these results by Mason \cite{Mas07} gives a 
 positive value for the second moment $S^- = \langle xs^-(x) \rangle$; 
 this is in agreement 
 with analyses of these experiments by the CTEQ group \cite{Kre04}. 

 Note that these results have the opposite sign for the strange quark 
 asymmetry from the calculations of Braendler \EA~shown in Fig.~\ref{fig42cc}. 
 Those results were calculated in the framework of `meson-cloud' models 
 \cite{Sig87,Ji95,Hol96,Bro96,Mel97,Spe97} 
 extended to include strange quarks. In these models the nucleon fluctuates 
 to a configuration $N \rightarrow K + Y$, where $Y$ represents a $\Lambda$ or 
 $\Sigma$ baryon. The $\bar{s}$ is associated with the virtual kaon 
 production, while the $s$ quark resides with the residual strange baryon 
 \cite{Sig87}. Fig.\ref{Fig:MelMal} shows the quantity $s(x) - \bar{s}(x)$ 
 calculated 
 using the model of Melnitchouk and Malheiro\cite{Mel97}. The curves 
 show values of $s - \bar{s}$ calculated using various values for the 
 $N \rightarrow KY$ form factor, and the shading represents the uncertainty 
 in the calculations. In the Melnitchouk-Malheiro calculation the 
 $s(x)-\bar{s}(x)$ difference also has the opposite sign from the 
 experimental determination, although it should be noted that to within one 
 standard deviation the experimental result is consistent with zero. For both 
 the meson-cloud and experimental values for $s(x)-\bar{s}(x)$, the magnitude 
 of the strange quark contribution to the 
 quantity $R_W$ in Eq.~(\ref{Rcsv}) is comparable to the contribution 
 arising from CSV, although the $x$-dependence of the two contributions is 
 quite different.

 \section{Charge Symmetry Violation for Sea Quarks}
 \label{Sec:eight}

 For valence quark charge symmetry, several theoretical models 
 give quantitatively similar predictions for charge 
 symmetry violating PDFs. Estimates of valence quark CSV by Sather 
 \cite{Sat92} and Rodionov \EA~\cite{Rod94} are in rather good agreement, 
 and both the magnitude and shape of those 
 valence CSV PDFs agree quite well with the best phenomenological 
 global fit from the MRST group \cite{MRST03}. The situation is much different 
 for charge symmetry in the sea quark sector.  It is considerably more 
 difficult to construct reliable theoretical models to estimate sea quark CSV 
 effects, and until recently the phenomenological 
 situation was less certain.  As we have seen, one problem is 
 that the tests of charge symmetry generally combine effects from heavy  
 quarks in addition to CSV terms.  At sufficiently large $x$ the 
 contributions from heavy quarks should become extremely small relative to 
 valence quark CSV effects.  However, at small $x$ strange quark contributions 
 are significant.  Unless these contributions are known quite precisely, it 
 is difficult to determine the upper limits on parton CSV in the sea.  
 
 \subsection{Estimates of Sea Quark CSV}
 \label{Sec:phenCSV}

The MRST group also searched for the presence of charge 
symmetry violation in the sea quark sector \cite{MRST03}.  They chose a 
specific functional form for sea quark CSV, dependent on a single parameter
$\WTdelt$, 
\bea 
\bar{u}^n(x) &=& \bar{d}^p(x)\left[ 1 + \WTdelt \right] ; \nonumber \\  
\bar{d}^n(x) &=& \bar{u}^p(x)\left[ 1 - \WTdelt \right] . 
\label{eq:seaCSV}
\eea  
 With the form chosen by MRST, the net momentum carried by antiquarks in the 
 neutron and 
proton are approximately equal; although this quantity is not conserved 
in QCD evolution, the change in momentum carried by antiquarks in the 
neutron was found to be 
very small in the kinematic region of interest. Using Eq.~(\ref{eq:seaCSV}) to 
represent sea quark CSV effects, the MRST group performed a global fit to a 
wide array of high-energy data, where the coefficient $\WTdelt$ was varied to 
obtain the best fit. [The MRST group refer to the sea quark CSV parameter 
as $\delta$; however we use the notation $\WTdelt$ to avoid confusion with 
our definitions of partonic charge symmetry.] 

\begin{figure}[ht]
\center
\hspace{-0.5cm} 
\includegraphics[width=2.8in]{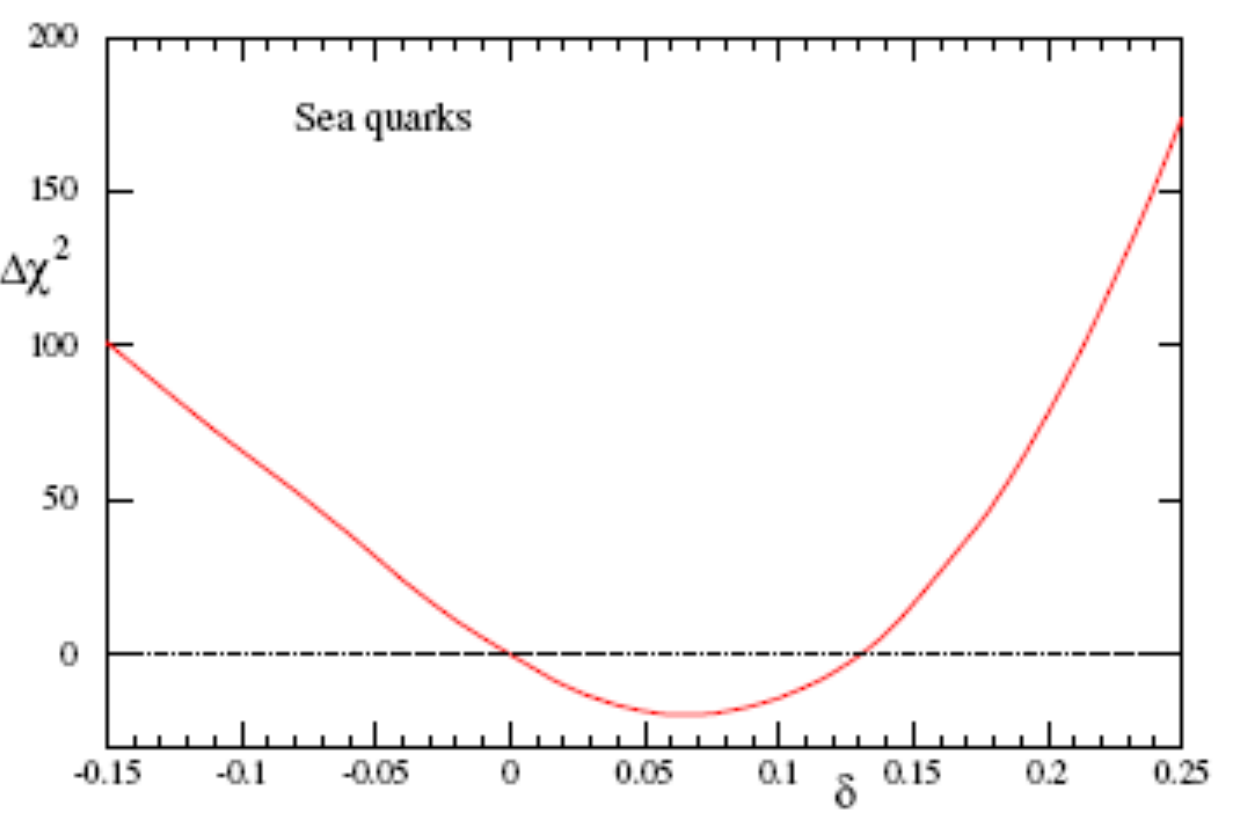}
\caption{[color online] The $\chi^2$ obtained by MRST, 
 Ref.~\protect{\cite{MRST03}} for a 
global fit to high energy data of parton distribution functions including 
sea quark CSV with the functional form defined in 
Eq.~(\protect\ref{eq:seaCSV}). $\chi^2$ is plotted vs.~the free parameter 
$\WTdelt$ (which MRST label $\delta$).  
\label{Fig:MRSTsea}}
\end{figure}

Somewhat surprisingly, evidence for sea quark CSV in the MRST global fit 
was substantially stronger than for valence quark CSV. The $\chi^2$ they 
obtain is plotted vs.~the parameter $\WTdelt$ in Fig.~\ref{Fig:MRSTsea}. 
The best fit was obtained for $\WTdelt = 0.08$, \IE, an 8\% violation of 
charge symmetry in the nucleon 
sea. The $\chi^2$ corresponding to this value is substantially 
better than with no charge symmetry violation, primarily because of 
the improvement in the fit to the NMC $\mu-D$ DIS data 
\cite{Ama91,Ama92,NMC97} when $\bar{u}^n$ is increased.  The fit to the E605 
Drell-Yan data \cite{E605} was also substantially improved by the sea quark 
CSV term.  

Note that the MRST parameterization does not satisfy the ``weak form'' 
of charge symmetry as described in Eq.~(\ref{eq:weakform}).  The weak form
of charge symmetry would require that the first moments of sea quark CSV 
should vanish, \IE 
\be
\langle\dubar \rangle = \langle\ddbar \rangle = 0 \ .
\label{eq:seaweak}
\ee
 In fact, using the MRST parameterization of Eq.~(\ref{eq:seaCSV}) for sea 
quark CSV and either the MRST or CTEQ phenomenological sea quark parton 
distribution functions, then both of the first moments in 
Eq.~(\ref{eq:seaweak}) would be infinite. 

 Benesh and Londergan used quark models to estimate the magnitude of 
 sea quark CSV\cite{Ben98}. They included sea quarks in quark model 
 wavefunctions and attempted to calculate the sign and magnitude of 
 sea quark CSV in such models. Reasonably
 model-independent estimates have been made for valence quark CSV, but 
 calculations of sea quark CSV require additional
 assumptions, and such calculations are likely to have substantial 
 model dependence. Benesh and Londergan predicted very small 
 CSV effects for antiquarks. They estimated that the fractional amount of 
 CSV in the sea, $\delta \bar{q}/\bar{q}$, should be at least an order of 
 magnitude smaller than the corresponding fractional CSV effects for valence 
 quarks. In quark model calculations, 
 qualitative arguments would suggest that sea quark CSV effects
 should be small. The relative magnitude of CSV effects will be given
 approximately by
 \be
 \frac{\delta \bar{q}}{\bar{q}} \sim \frac{\delta M}{\langle M \rangle} ~~,
 \ee
 where $\langle M \rangle$ is the energy of the lowest contributing 
 intermediate states, and $\delta M$ is the mass difference for
 intermediate states related by charge symmetry.  For antiquarks,
 the lowest energy states are four-quark states,
 whose energy is roughly twice the energy of the lowest diquark
 states that contribute for valence quarks. The mass difference between 
 charge symmetric four quark states is approximately $\delta M \approx 
 M_n - M_p = 1.3$ 
 MeV, or three times smaller than the mass difference for minority valence 
 quarks. This naive estimate suggests that sea quark CSV effects should be
 roughly an order of magnitude smaller than for valence quarks. Cao and Signal 
 carried out meson-cloud calculations of sea quark CSV \cite{Cao00}; they 
 suggest that the bag model calculations of Benesh and Londergan neglect 
 higher-order contributions that might be substantial.  

 In their quark model estimates of sea quark CSV, Benesh and Londergan 
 found that the sea quark CSV contributions $\delta\bar{u}(x)$ and 
 $\delta\bar{d}(x)$ tended to be roughly the same magnitude and to 
 have opposite sign; this is similar to the situation with the valence 
 quark CSV. However, from Eq.~(\ref{eq:seaCSV}) the phenomenological sea 
 quark CSV form assumed by the MRST group obeys the relation 
\be
  \delta\bar{u}(x) + \delta\bar{d}(x) = -\WTdelt [ \bar{d}(x) - 
 \bar{u}(x)] \ .
 \label{eq:deldbar}
\ee 
The quantity $\bar{d}(x) - \bar{u}(x)$ has been measured by the E866 
group \cite{Haw98,Tow02}, who compared Drell-Yan $pD$ and $pp$ experiments. 
Their measurements are in agreement with measurements of the same quantity 
at HERMES \cite{Ack98}. At HERMES this quantity was extracted from 
semi-inclusive DIS experiments of charged pion production from $e-p$ and $e-D$ 
reactions. We will discuss the measurements of this quantity in 
Sec.~\ref{Sec:eightfourone}.   

 \subsection{Limits on Sea Quark CSV}
 \label{Sec:eighttwo}

There are relatively few experimental limits on charge symmetry violation 
for sea quark parton distributions. As we mentioned in 
Sec.~\ref{Sec:seventwoone}, one can obtain rather strong experimental 
constraints on sea quark PDFs from experimental measurements of the $W$ 
charge asymmetry in $p\bar{p}$ reactions \cite{Abe98}. Such measurements 
put some limits on the magnitude of charge symmetry violating sea quark 
PDFs \cite{Bod99}, but they primarily rule out very large CSV for sea 
quarks. As we have mentioned previously, most tests of parton charge 
symmetry have contributions from heavy quark PDFs. Strong tests of 
parton CSV will require rather precise knowledge particularly of strange 
quark parton distributions. Since sea quark parton distributions increase 
quite rapidly at small $x$, it is difficult to separate contributions 
from partonic CSV in this region from effects due to strange quarks. 
The results obtained from the global fits of the MRST group to high 
energy data \cite{MRST03}, discussed in the preceding section, appeared 
to show a several percent CSV effect. 

Perhaps the most promising area to search for sea quark CSV is to 
look for contributions to DIS sum rules \cite{Ste95,Hin96}. The lowest-order 
sum rules often involve integrals of parton distributions over all $x$. As we 
discussed in Sec.~\ref{Sec:sixtwo}, valence quark normalization requires 
that the first moment of the valence quark CSV terms vanish. 
Consequently, when one integrates parton distributions over all $x$, 
the only remaining charge symmetry violating contribution will arise 
from sea quarks. As we will discuss in the following section, sea quark 
CSV terms will contribute to various sum rules, particularly the 
Gottfried sum rule \cite{Got67}, and possibly also the Adler sum 
rule \cite{Adler}.    

\subsubsection{W Production Asymmetry at a Hadron Collider}
 \label{Sec:seventhreeone}

 Production of $W$ bosons resulting from the scattering of protons on an 
 isospin-zero target 
 represents an area where, in principle, one could test parton charge 
 symmetry. On an isospin-zero target, \textit{e.g.}, the deuteron, we are 
 interested in semi-inclusive reactions of the type $p + D \rightarrow W^+ 
 + X$ and $p + D \rightarrow W^- + X$. We can then define the sum of $W^+$ 
 and $W^-$ cross sections and the forward-backward asymmetry
\bea
  \sigS(\xF) &\equiv& \left(\frac{d\sigma(\xF)}{d\xF}\right)^{W^+} 
  + \left(\frac{d\sigma(\xF)}{d\xF}\right)^{W^-} \ ; \nonumber \\ 
  A(\xF) &=& \frac{\sigS(\xF) - \sigS(-\xF)}{\sigS(\xF) + \sigS(-\xF)} .
\label{eq:Wasymm}
\eea
 In Eq.~(\ref{eq:Wasymm}) the Cabibbo-favored terms in $\sigS$ are invariant 
 under the transformation $\xF \rightarrow -\xF$ for Feynman $\xF = x_1 
 - x_2$. In the forward-backward asymmetry $A(\xF)$, the only terms that 
 survive are charge symmetry violating terms plus heavy quark terms in the 
 Cabibbo-unfavored sector.  

 As was discussed in Sec.~\ref{Sec:seventwoone}, when the CCFR group 
 performed its initial analysis \cite{Sel97,Sel97a} of the ``charge ratio'' 
 $R_c$ defined in 
 Eq.~(\ref{Rc}), at small $x$ the charge ratio appeared to deviate from one,
 with the deviation growing with decreasing $x$. It was pointed out by 
 Boros \EA~\cite{Bor97,Bor99} that, if the analysis of this data was 
 accurate, a likely explanation of this discrepancy was a surprisingly large 
 violation of charge symmetry in the nucleon sea quark PDFs.  This  
 CSV term in the sea quark distributions was sufficiently large that 
 one would expect the forward-backward asymmetry $A(\xF)$ of 
 Eq.~(\ref{eq:Wasymm}) to be as large as several percent \cite{Bor99}.  

 However, as was discussed in Sec.~\ref{Sec:seventwoone}, the low-$x$ 
 discrepancy disappeared upon re-analysis of the CCFR experiment 
 \cite{Yan01}. Londergan \EA~then analyzed the prospects for $W$ 
 forward-backward asymmetry \cite{LMT06} using the (much smaller) sea quark 
 CSV obtained in the MRST global fit to high energy data \cite{MRST05}. 
 The asymmetries obtained were extremely small, generally less than 
 one percent. The sea quark CSV terms were substantially 
 smaller than suggested by the original CCFR analysis.  In addition, the 
 strange quark contributions, which were originally much smaller than the 
 CSV terms, were now no longer negligible, and they tended to 
 cancel the CSV contribution. As a result of these newer results, it 
 was concluded that these $W$ production asymmetries no longer represent a 
 promising prospect to determine charge symmetry violation in quark PDFs.   

 \subsubsection{Limits on Charge Symmetry Violation for Gluon Distributions}
 \label{Sec:glueCSV}

It is possible that gluon distributions are different for proton and 
neutron. In fact, if sea quark distributions are charge asymmetric, then 
this would lead to a small charge symmetry violation in gluon distributions 
since sea quark and gluon distributions are coupled through the DGLAP 
evolution equations \cite{Dok77,Gri72,Alt77}. It is 
also conceivable that some other mechanism might give rise to charge 
symmetry violation in gluon distributions. 

Piller and Thomas \cite{Pil96} pointed out that CSV in gluon distributions 
might be probed through measurements of heavy quarkonium production. For 
example, if one considers $J/\psi$ production arising from nucleon-nucleon 
collisions, then the differential cross section can be written as the 
sum of two terms 
\bea
\frac{d^2 \sigma (c\bar{c})}{d\xF dM^2} &=& \sum_{i,j} 
  \frac{1}{s\sqrt{\xF^2 + \frac{4M^2}{s}}} \Biggl[ \widehat{\sigma}_{gg} 
  g_i(x_b)g_j(x_t) \nonumber \\ &+& \widehat{\sigma}_{qq}[ 
  q_i(x_b)\bar{q}_j(x_t) + \bar{q}_i(x_b)q_j(x_t) ] \Biggr] . 
  \nonumber \\ 
\label{eq:Jpsi}
\eea
In Eq.~(\ref{eq:Jpsi}), the first term represents the contribution from 
gluon-gluon fusion and the second term is from quark-antiquark annihilation 
leading to heavy quarkonium. The corresponding quantities 
$\widehat{\sigma}_{gg}$ and $\widehat{\sigma}_{qq}$ represent the 
subprocess cross sections for gluon-gluon fusion and quark-antiquark 
annihilation, respectively. 

If one focuses on $J/\psi$ production in proton-neutron collisions then 
charge symmetry violation in either the quark or gluon distributions will 
produce a forward-backward asymmetry in the resulting cross sections. Hence 
such a forward-backward asymmetry in $J/\psi$ production would be sensitive 
to partonic CSV. If we define the forward-backward asymmetry in terms of 
Feynman $\xF = x_b - x_t$, we obtain 
\be 
\Delta \sigma_{pn}(\xF) \equiv \frac{d\sigma (J/\psi)}{d\xF}|_{\xF} 
 - \frac{d\sigma (J/\psi)}{d\xF}|_{-\xF} \ , 
\label{eq:FBasymm}
\ee
then it is straightforward to show that 
\bea
\Delta \sigma_{pn}(\xF) &\sim& \widehat{\sigma}_{gg}g(x_t)\delta g(x_b) 
  \nonumber \\ 
 &+& \widehat{\sigma}_{qq} \bigl[ u(x_t)\ddbar(x_b) + \ubar(x_t)\deld(x_b) 
 \nonumber \\ &+& d(x_t)\dubar(x_b) + \dbar(x_t)\delu(x_b) \bigr] \nonumber \\ 
  &-& \left[ x_b \leftrightarrow x_t \right] \ .
\label{eq:FBPDF}
\eea
In Eq.~(\ref{eq:FBPDF}), $\delta g(x) = g^p(x) - g^n(x)$. 
Piller and Thomas showed that CSV contributions in the gluon distribution 
and in sea quark PDFs would be most important at small $\xF$ while 
contributions from valence quark CSV should dominate at large values of $\xF$. 

\begin{figure}
\center
\includegraphics[width=2.85in]{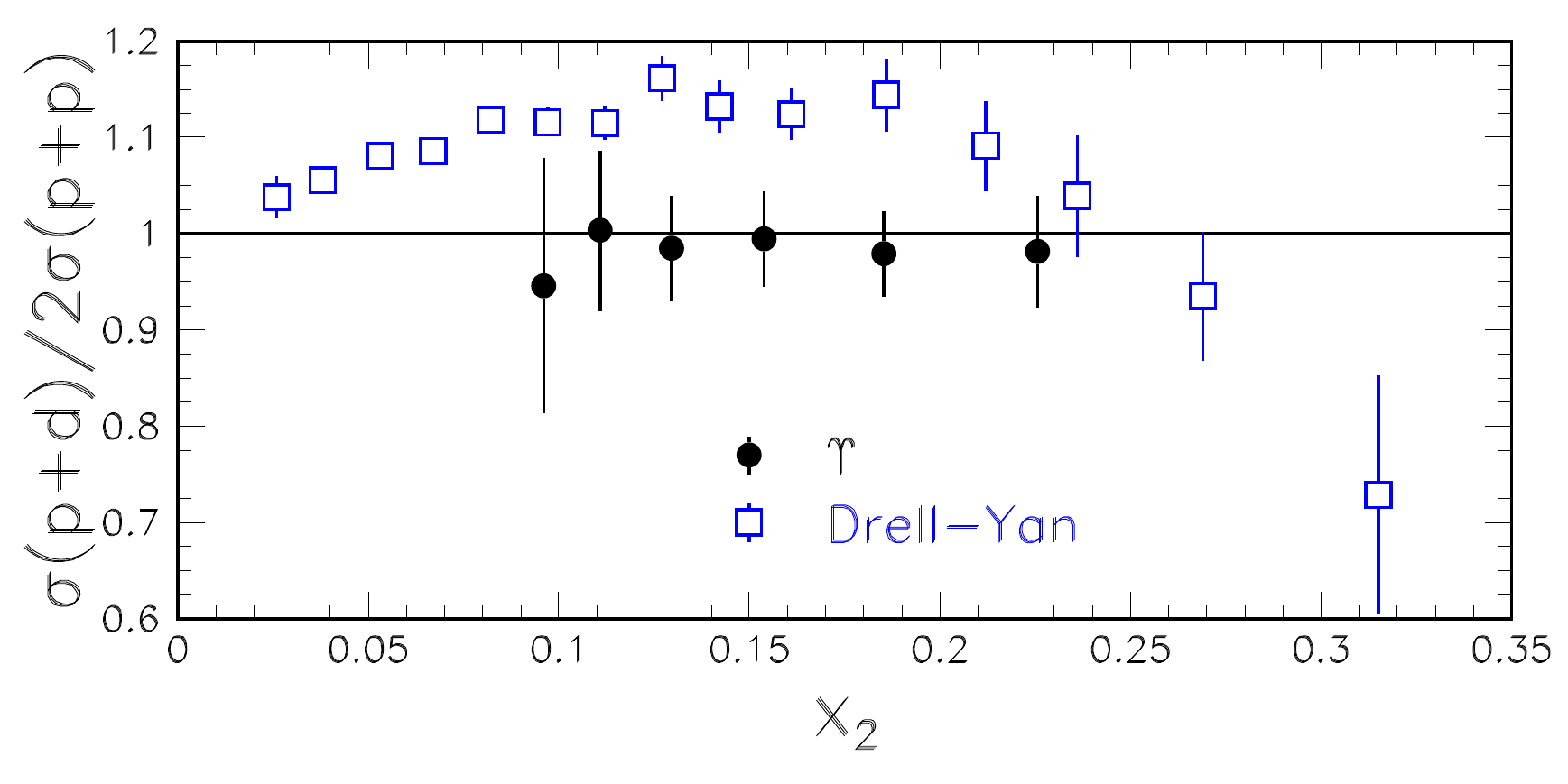}
 \caption{[color online] The ratio $\sigma(p+D)/2\sigma(p+p)$ as a function 
 of target 
 $x$, for $\Upsilon$ resonance production cross sections for $p+D$ and $p+p$ 
 reactions from the E866/NuSea Collaboration, Ref.~\protect\cite{Zhu08}. Solid 
 circles are the $\Upsilon$ production cross sections. For comparison the 
 open squares are the corresponding ratios for the E866 Drell-Yan cross 
 sections from Ref.~\protect\cite{Tow02}.}
 \vspace{0.1truein}
 \label{Fig:Upsilon}
 \end{figure}

The E866/NuSea group has recently measured $\Upsilon$ production in 
reactions arising from 800 GeV protons on hydrogen and deuterium 
targets \cite{Zhu08}. 
At these energies, the dominant contribution to $\Upsilon$ production 
comes from gluon-gluon fusion. In this case one expects the resonance 
cross-section ratio 
\be
\frac{\sigma(p+D \rightarrow \Upsilon)} {2\sigma(p+p \rightarrow \Upsilon)} 
  \rightarrow  \left[1 - \frac{\delta g(x_t)}{2g(x_t)}\right] 
\label{eq:Upsrat}
\ee
Thus CSV in gluon distributions would be manifested by a deviation of 
the $\Upsilon$ production ratio in Eq.~(\ref{eq:Upsrat}) from one. 
In Fig.~\ref{Fig:Upsilon} the E866/NuSea collaboration plot the 
cross section ratio for $\Upsilon$ production in $pp$ and $pD$ reactions vs. 
target $x$. These are plotted as the solid circles in 
Fig.~\ref{Fig:Upsilon}. Within statistics they find no measurable deviation 
from one. This is in contrast to significant deviations which were seen for 
the ratio of $pD$ and $pp$ Drell-Yan processes (these are shown as the open 
squares in Fig.~\ref{Fig:Upsilon}). The asymmetry in DY processes arose from 
differences between $\bar{u}$ and $\bar{d}$ distributions in the 
proton, as will be discussed in Sect.~\ref{Sec:eightfourone}. 
From the $\Upsilon$ measurement we can put upper limits of roughly ten 
percent on possible CSV effects in gluon distributions. 

 \subsection{Charge Symmetry Contributions to DIS Sum Rules
 \label{Sec:eightfour}}

 Sum rules can provide extremely useful information on
 a single moment of parton distributions.  If we choose the first moment 
 of quark PDFs over $x$, we can invoke the quark normalization conditions 
 on both the valence quark distributions and on valence quark CSV, 
 as expressed in Eqs~(\ref{qnorm}) and (\ref{csvnorm}), respectively.  
 Consequently, terms which contribute to lowest-order sum rules 
 will depend on integers that represent valence quark normalizations, plus 
 the first moments of antiquark distributions and sea quark CSV. Sum rules 
 that measure the first moments of various structure functions are 
 extremely useful in that if one chooses appropriate linear combinations 
 of structure functions, contributions from heavy quark CSVs will cancel 
 out. This removes a major source of uncertainty, since for sea quarks  
 the contributions from sea quark CSV and from heavy 
 quark distributions are generally difficult to separate.    

 Two sum rules, the Adler sum rule \cite{Adler} and
 Gross-Llewellyn Smith sum rule \cite{GLS}, can be directly
 related to linear combinations of quark normalization integrals.  
 We will discuss the Gottfried sum rule (GSR) \cite{Got67} in the 
 following section.  Unlike the Adler or Gross-Llewellyn Smith
 sum rules, the ``naive'' Gottfried sum rule expectation
 $S_G = 1/3$ is obtained only if one assumes both charge symmetry
 for parton distributions, and equality of light sea quark distributions 
 in the proton sea, $\bar{u}^p(x) = \bar{d}^p(x)$, or more precisely the 
 equality of the first moment of these light sea quark PDFs. The 
 expectation that the light quark sea distributions should be equal is 
 often referred to as $SU(2)$ flavor symmetry. This is an unfortunate 
 connotation since the $\bar{u}^p(x)$ and $\bar{d}^p(x)$ distributions are 
 not related by any underlying dynamical symmetry. However, if all of the 
 light sea quarks were generated through gluon radiation, then one would 
 expect the sea quark distributions to be identical except for effects due 
 to light quark mass differences and small electromagnetic effects.  

 As we have discussed in \ref{Sec:seven}, theoretical expectations for 
 CSV effects in parton distributions are expected 
 to be no larger than a few percent. In Sect.~\ref{Sec:seventwo} we showed 
 that current experimental upper
 limits on CSV are of the order of several percent for $x \le 0.4$, and 
 larger than 10\% for $x > 0.4$.  It would therefore be useful to construct 
 sum rules which could in principle distinguish between CSV effects and 
 effects arising from sea quark flavor asymmetry.
 In this Section, we will review the current status of the Adler, 
 Gross-Llewellyn Smith and Gottfried sum rules, with particular attention 
 to the contributions from partonic CSV effects. Good reviews of DIS 
 sum rules up to about 1996 can be found in the Handbook of Perturbative 
 QCD \cite{Ste95} and in the review by Hinchliffe and Kwiatkowski 
 \cite{Hin96}. We reviewed sum rules and parton CSV in 
 detail in our previous review article \cite{Lon98a}. There has been little 
 change in the status of the experimental Adler and Gross-Llewellyn Smith sum 
 rules since the publication of that review article.

 \subsubsection{Gottfried Sum Rule: Sea Quark Flavor and Charge Symmetry 
\label{Sec:eightfourone}}

 In the past fifteen years  we have obtained a great deal of quantitative 
 information 
 on sea quark flavor asymmetry ($\bar{d}^p(x) \ne \bar{u}^p(x)$) in the 
 nucleon. Information on the first moment of these distributions can be 
 extracted from measurements of the Gottfried Sum Rule (GSR) \cite{Got67}, 
 obtained from the difference of $F_2$ structure functions from charged lepton 
 DIS on neutrons and protons. The Gottfried sum rule is also known as the 
\textit{valence isospin sum rule} \cite{Hin96}. Using Eq.~(\ref{Gttfrd}), 
we obtain  
 \bea
 S_{\G} &\equiv& \int_0^1 \,dx\, \frac{\left[ F_2^{\ell p}(x) -
   F_2^{\ell n}(x) \right]}{x} \nonumber \\ 
   &=& \frac{1}{3} - \frac{2}{3}\int_0^1 \,dx 
  \left[ \bar{d}^p(x) - \bar{u}^p(x) \right] \nonumber \\ &+& 
  \frac{2}{9} \int_0^1 \,dx \left[ 4\delta\bar{d}(x) +
 \delta \bar{u}(x)\right] \ .
 \label{GSRi}
 \eea
 If the nucleon sea is charge symmetric, and the first moment of the proton 
 antiquark distributions are equal, then we obtain the ``naive'' 
 expectation $S_{\G} = 1/3$.  Earlier measurements of the GSR 
 \cite{whitlow,SLAC,emc,emcb,BCDMS} obtained results that appeared to be less 
 than the naive expectation of $1/3$, but with significant error bars. The 
 first really precise GSR value was obtained by the New Muon Collaboration 
(NMC) \cite{Ama91,Ama92}. The NMC result \cite{Ama91,Ama92,Arn94}, 
$S_{\G} = 0.235 \pm 0.026$, was more than four standard 
 deviations below $1/3$. Assuming charge symmetry, this
 implies a substantial excess $\bar{d}^p > \bar{u}^p$.  This effect is 
 much larger than can be
 accommodated by perturbative QCD. Next-to-leading order (NLO) and
 next to next to leading order (NNLO) QCD calculations predict
 very small effects \cite{sachrajda}. However, the NMC result could be
 due to a combination of charge symmetry and flavor symmetry violating effects.

 Note that all three of the sum rules which we discuss here -- the 
 Gottfried, Adler and Gross-Llewellyn Smith sum rules -- involve 
 dividing the $F_2$ or $xF_3$ structure functions by $x$. This emphasizes 
 the contributions from small $x$. In all of these sum rules we invoke 
 quark normalization conditions as given in Eqs.~(\ref{qnorm}) and 
 (\ref{csvnorm}). These normalization conditions hold only after 
 integration over all $x$. In reality one measures the sum rule down to 
 some smallest value $x_{min}$. In that case one must estimate the 
 contributions from the unmeasured region $ 0 < x \leq x_{min}$.  

 The GSR provides information only on the first moment of proton sea quark 
 differences. It was proposed to make a `direct' measurement
 of sea quark flavor asymmetry by comparing Drell-Yan processes initiated by
 protons, on proton and deuteron targets. This was suggested first by Ericson 
 and Thomas \cite{Ericson:1984vt} in the context of the pionic explanation 
 of the EMC effect, and later by Ellis and Stirling \cite{Ell91}. In the 
 Drell-Yan [DY] process 
 \cite{DY70} hadronic collisions produce opposite sign lepton pairs 
 with large invariant mass.  The charged leptons are formed 
 from the decay of a virtual photon arising from annihilation of a 
 quark (antiquark) in the projectile with an antiquark (quark) of the 
 same flavor from the target.

 The experiments measure the ratios of Drell-Yan (DY) cross sections 
 for incident protons on deuteron and proton targets.  Assuming
 the validity of the impulse approximation, the ratio is given by
 \begin{eqnarray}
 &\,& R^{DY}(x_1,x_2) \equiv \frac{\sigma_{DY}^{pD}}{2\sigma_{DY}^{pp}} 
  \nonumber \\ &\rightarrow& \frac{1}{2} \left( 1+ \frac{\bar{d}(x_2) - 
  \delta\bar{d}(x_2)}{\bar{u}(x_2)} \right) \ .
 \label{eq:sigDYpd}
 \end{eqnarray}
 The last line of  Eq.~(\ref{eq:sigDYpd}) follows in the limit of large 
 Feynman $x_F = x_1 - x_2$ assuming that  
 \be 
 \frac{d(x)}{u(x)} \rightarrow 0, \hspace{0.4cm} x \rightarrow 1 \, ,
 \label{eq:doveru}
 \ee
 where $x_1$ and $x_2$ are respectively the longitudinal momentum fractions 
 carried by the projectile (target) quarks or antiquarks. If charge symmetry 
 holds then from Eq.~(\ref{eq:sigDYpd}), in the limit of large $x_F$, the 
 ratio of $pD$ to $pp$ Drell-Yan cross sections would 
 directly measure the ratio of the down antiquark to up antiquark
 distributions in the proton, at a given value of $x_2$.

\begin{figure}
\center
\includegraphics[width=2.85in]{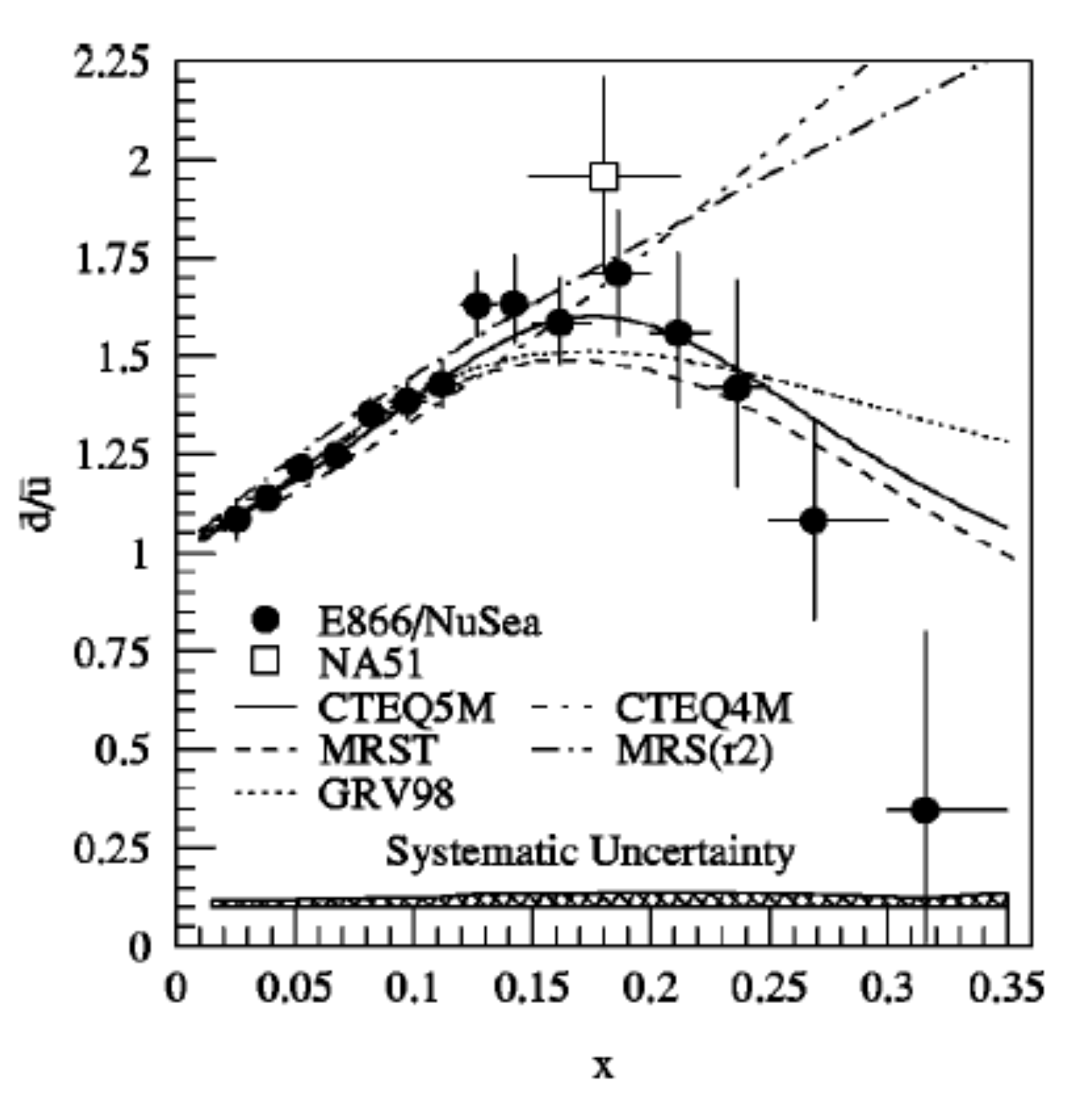}
 \caption{The ratio $\bar{d}^p(x)/\bar{u}^p(x)$ as a function of target 
 $x$, obtained from Drell-Yan cross sections for $pp$ and $pD$ reactions. 
 Solid circles are data from the E866 Collaboration, 
 Ref.~\protect\cite{Haw98,Tow02}; the open square is the NA51 point, 
 \protect\cite{Bal94}. Curves are results of calculations using various 
 phenomenological PDFs.}
 \label{Fig:Towellfig9}
 \end{figure}

 Experiment NA51 at CERN \cite{Bal94} measured Drell-Yan
 processes for 450 GeV protons on proton and deuteron
 targets, obtaining a ratio $\bar{u}^p/\bar{d}^p = 0.51 \pm 0.04 (stat)
 \pm 0.05 (syst)$ for a single averaged point $\langle x\rangle = 0.18$.
 The E866 group at Fermilab \cite{Haw98,Pen98,Tow02} compared Drell-Yan
 processes for 800 GeV protons on liquid hydrogen and deuterium
 targets.  Fig.~\ref{Fig:Towellfig9} shows results from E866. For values
 $x_2 < 0.2$ the ratio is greater than one, and appears
 to decrease at higher values of $x_2$, perhaps becoming less than one 
 at $x_2 \sim 0.3$.  Both the NA51 and E866 experiments show a substantial 
 excess $\bar{d}^p > \bar{u}^p$ at small $x$. This measurement was confirmed 
 by subsequent semi-inclusive DIS (SIDIS) measurements, comparing yields of 
 positive and negative pions from scattering of energetic electrons on proton 
 and deuteron targets at HERMES \cite{Ack98}.  The E866 group obtained 
 the first moment of the light sea quark difference 
\be
  \langle \bar{d}-\bar{u} \rangle = 0.118 \pm 0.012 \ .
\label{eq:Delmom}
\ee 
Comprehensive review articles on light sea quark asymmetries have been 
produced by Kumano \cite{Kum98} and by Garvey and Peng \cite{Gar01}. 

 However, in order to obtain the result of Eq.~(\ref{eq:Delmom}) the E866 
 group assumed parton charge symmetry. Eq.~(\ref{eq:sigDYpd}) shows that the 
 Drell-Yan ratios could also have 
 contributions from sea quark CSV effects. One cannot extract the magnitude 
 of flavor symmetry violating effects without assuming sea quark charge 
 symmetry, as was emphasized by Ma \cite{Ma92,Ma93}. Ma claimed that in 
 principle one could explain the entire effect on the Gottfried sum rule 
 from parton charge symmetry violation, even if parton flavor symmetry 
 was exact. However, this would require sea quark CSV effects an order of 
 magnitude larger than those obtained in the MRST phenomenological fit to 
 high energy data \cite{MRST03}. A more natural source of the 
 $\bar{d}-\bar{u}$ difference 
 was predicted by Thomas \cite{Thomas:1983fh}. This incorporates effects 
of the pion cloud of the nucleon; the proton predominantly emits a 
 $\pi^+$, which contains a valence $\bar{d}$ quark, leading to an excess 
 $\bar{d} > \bar{u}$. 

 \begin{figure}
\center
\includegraphics[width=2.85in]{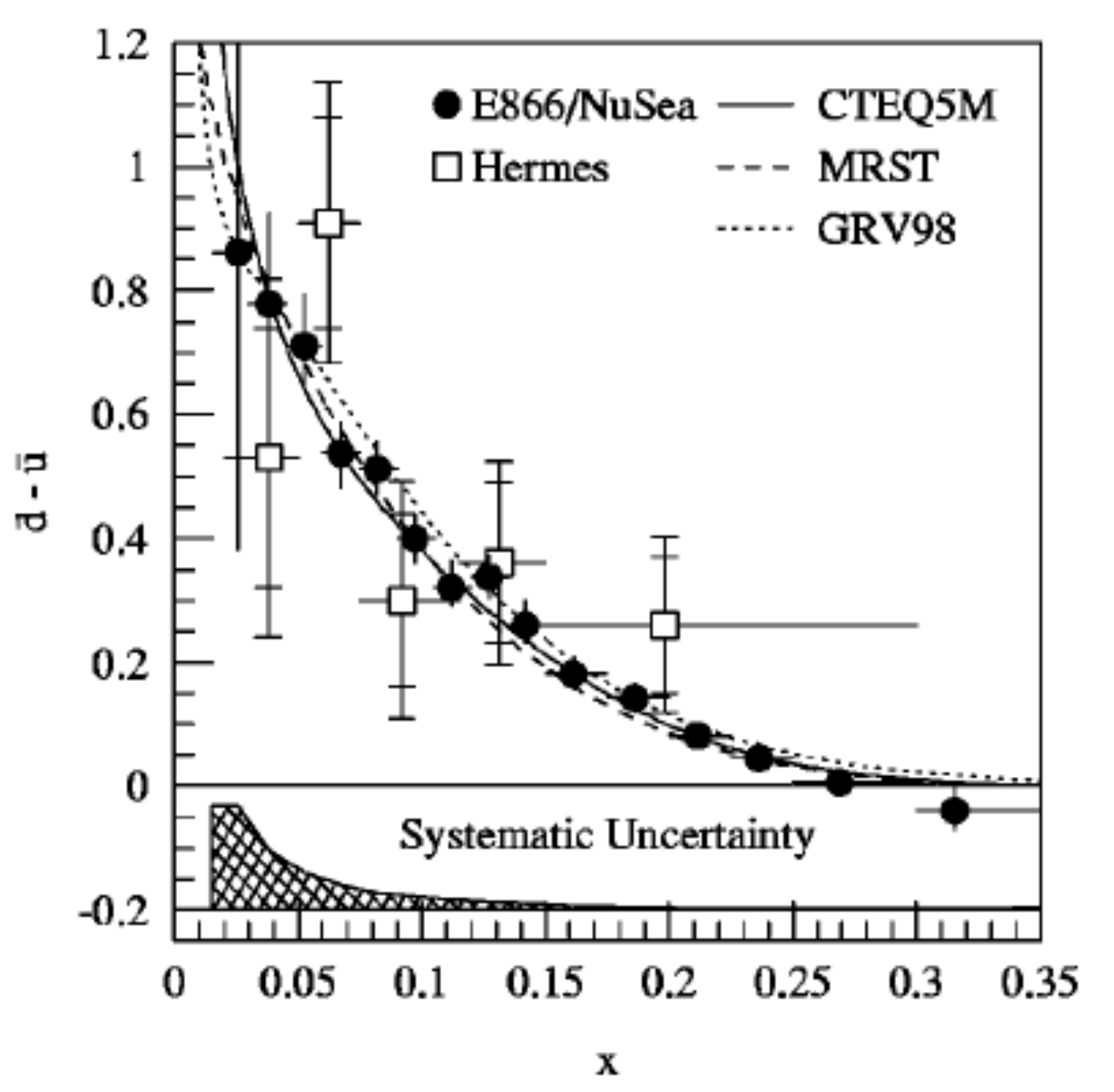}
 \caption{The quantity $\bar{d}^p(x) - \bar{u}^p(x)$ as a function of target 
 $x$. Solid circles: data from the E866 DY cross sections scaled to fixed 
 $Q^2 = 54$ GeV$^2$, Ref.~\protect\cite{Haw98,Tow02}. Open squares: 
 results from SIDIS measurements at HERMES, \protect\cite{Ack98}, 
 corresponding to $\langle Q^2 \rangle = 2.3$ GeV$^2$. Curves 
 are calculations using various phenomenological PDFs.}
 \label{Fig:Towellfig10}
 \end{figure}

In Fig.~\ref{Fig:Towellfig10} we plot the quantity 
$\bar{d}^p(x)-\bar{u}^p(x)$. The solid circles are the E866 DY points from 
Ref.~\cite{Haw98,Tow02} scaled to fixed $Q^2 = 54$ GeV$^2$. 
The open squares are the results from the SIDIS measurements at HERMES 
\cite{Ack98}, which correspond to an averaged value 
$\langle Q^2 \rangle = 2.3$ GeV$^2$. The curves are calculations using various 
 phenomenological PDFs from GRV98 \cite{GRV98} (dotted curve), 
 MRST \cite{MRST98} (dashed curve) and CTEQ5M \cite{Lai00} (solid curve).   

We first review results obtained for the Gottfried sum rule with the 
MRST2001 \cite{MRST01} global fit PDFs with no isospin violation.  In 
this parameterization, the sea quark distributions are obtained at a 
starting scale $Q_0^2 = 1$ GeV$^2$ and values at higher $Q^2$ can be 
obtained through DGLAP evolution.  The MRST2001 fit obtains 
\be 
S_{\G} = 1/3 - 2/3\langle \bar{d}-\bar{u} \rangle = 0.266 \ , 
\label{eq:SgMRST} 
\ee   
The MRST result for the Gottfried sum rule is just over $1\sigma$ 
above the NMC value. The E866 $pp$ and $pD$ DY data essentially determine 
the MRST value for the $\bar{d}-\bar{u}$ asymmetry in the proton. These 
results are basically consistent with the NA51 DY point \cite{Bal94} 
and with the HERMES semi-inclusive DIS measurements of positive and 
negative pion production in $ep$ and $eD$ scattering \cite{Ack98}.    

 If one includes the phenomenological sea quark CSV effects obtained by 
 MRST \cite{MRST03}, then the contribution to the Drell-Yan ratio in the 
 limit of large $x_F$ will be 
 \be
  R^{DY}(x_1,x_2) \rightarrow \frac{1}{2} \left( 1+ 
 \frac{1.08 \bar{d}(x_2)}{\bar{u}(x_2)} \right) \ .
 \label{eq:DYcsv}
 \ee
 Thus, including the MRST sea quark CSV term would decrease the extracted 
 sea quark flavor asymmetry by roughly 8\% (one needs to take care since 
 Eq.~(\ref{eq:DYcsv}) is true only in the limit of large Feynman $x$). 
In principle, one could insert the sea quark distributions 
from the 2003 MRST global fit including CSV, and calculate the 
contribution from sea quark CSV to the Gottfried sum rule.  However, using 
the MRST functional form from Eq.~(\ref{eq:seaCSV})  
gives an infinite result for $S_{\G}$.  This is due to the 
fact that MRST choose the sea quark CSV proportional to 
the sea quark PDFs, which have infinite first moment.  The 
MRST group plans to carry out future global fits  
assuming a modified functional form for sea quark CSV \cite{ThornePC};    
this would produce finite CSV contributions to the Gottfried sum rule.    

From Eq.~(\ref{eq:sigDYpd}) it is apparent that in comparing $pp$ and $pD$ 
 Drell-Yan cross sections, one has contributions from both sea quark flavor 
asymmetry, \IE, $\bar{d}(x) \ne \bar{u}(x)$, but also from parton charge 
symmetry violation. Peng and Jansen \cite{Pen95} pointed out that one 
can obtain information on $\bar{d}/\bar{u}$ from measurements of $W$ or 
$Z$ production in $pp$ collisions, which have no CSV contributions. For 
example, if one measures the ratio of $W^+$ and $W^-$ production then 
one obtains 
\bea
 R(\xF) &\equiv& \frac{d\sigma/d \xF (p+p \rightarrow W^+)}
  {d\sigma/d \xF (p+p \rightarrow W^-)} \ , \nonumber \\ 
  R(\xF)_{(\xF = 0)} &\approx& \frac{u(x)}{d(x)}\,
  \frac{\bar{d}(x)}{\bar{u}(x)} \nonumber \\ 
  R(\xF)_{(\xF >> 0)} &\approx& \frac{u(x_1)}{d(x_1)}\,
  \frac{\bar{d}(x_2)}{\bar{u}(x_2)}
\label{eq:RWprod}
\eea
In Eq.~(\ref{eq:RWprod}), the final two equations are true in the limit 
where one neglects strange quark contributions. Peng and Jansen showed 
that these $W$-production ratios were quite sensitive to different 
phenomenological predictions for light sea quark distributions.

 \subsubsection{Adler Sum Rule}
 \label{Sec:eightfourtwo}

 The Adler Sum Rule \cite{Adler} is given by the integral of the 
 $F_2$ structure functions for charged current $\nu$ and $\ovnu$ 
 DIS on the proton.  The Adler Sum Rule, $S_A$, is defined (in the limit 
 $Q^2 \rightarrow \infty$) as
 \bea
&\,& \SA \rightarrow \int_0^1\,dx\, 
  \left[ \frac{F_2^{W^- p}(x,Q^2) - F_2^{W^+ p}(x,Q^2)}{2x} \right]
 \nonumber \\ &=& \int_0^1\,dx\, \left[  u^p_{\V}(x) - d^p_{\V}(x)\left(1- 
  |V_{td}|^2 \right) - s^-(x) \right] \nonumber \\ &\approx& 1 \ . 
 \label{eq:Adlerp}
 \eea
 We obtain the result $\SA = 1$ if we neglect the term $|V_{td}|^2
 \approx 1\times 10^{-4}$.  The Adler sum rule thus requires
 subtracting the $F_2$ structure function for antineutrinos and
 neutrinos on protons, and dividing by $x$ (this emphasizes the
 contribution from very small $x$).  The Adler sum rule then
 follows from the normalization of the valence quark distributions.
 As a consequence of the algebra of SU(2) charges,
 the Adler sum rule has no QCD corrections. Since the Adler sum rule 
involves measurements only on the proton, it has no CSV corrections. 

Another name for the Adler sum rule is the \textit{isospin sum rule} 
\cite{Hin96}. Note: different overall normalizations for the Adler sum 
rule appear in the literature. An alternative normalization is a factor 
of two larger than ours \cite{Lea96,Hin96}. Our normalization agrees with 
that used by the WA25 experimental group \cite{WA25,WA25b}. 

The best experimental data to date are from the WA25 experiment 
\cite{WA25,WA25b}, who used the CERN-SPS wide band neutrino beams in the
BEBC H and D bubble chambers. Note that the WA25 measurements involve 
neutrino CC measurements on neutrons (\EG deuterons) and protons,
 \textit{and not} $\bar{\nu}$ and $\nu$ on protons, as given
 in the definition of the Adler sum rule, Eq.~(\ref{eq:Adlerp}).  This 
 was done because $\nu$ beams generally have much higher fluxes 
 than $\ovnu$.  The WA25 experiment substituted neutrinos on neutron 
 targets using the relation 
 \bea
&\,&  F_2^{W^+ n}(x,Q^2) = F_2^{W^- p}(x,Q^2) \nonumber \\ &+& 2x 
  \,[ s^-(x) - \delta u(x) - \delta \bar{d}(x) ] ~~,
 \eea 
 as was discussed in Sect~\ref{Sec:six}. Thus, the WA25 group does not 
measure the integral $\SA$ of Eq.~(\ref{eq:Adlerp}), but instead measures 
a different quantity
 \bea
 \widetilde{S}_{\A} &\equiv& \int_0^1\,dx\, \left[ \frac{F_2^{W^+ n}(x,Q^2) -
 F_2^{W^+ p}(x,Q^2)}{2x} \right]
 \nonumber \\ &=& \int_0^1\,dx\,
 \left[  u^p_{\rm v}(x)- d^p_{\rm v}(x) - \delta u(x)
  - \delta\bar{d}(x) \right] \nonumber \\
 &=&  \SA -\int_0^1\,dx\, \left[ \delta\bar{u}(x)+
 \delta\bar{d}(x) \right] .
 \label{Adlercsv}
 \eea
From Eq.~(\ref{Adlercsv}) we see that the difference between the Adler 
sum rule and the measurements from WA25 involves the first moment of 
contributions from sea quark CSV. If charge symmetry is exact, or if the 
``weak form'' of charge symmetry holds (see Eq.~(\ref{eq:weakform})), then 
the Adler sum rule would be identical to what was measured by WA25, 
\IE, $\SA = \widetilde{S}_{\A}$.  
 
 \begin{figure}
\center
\includegraphics[width=2.75in]{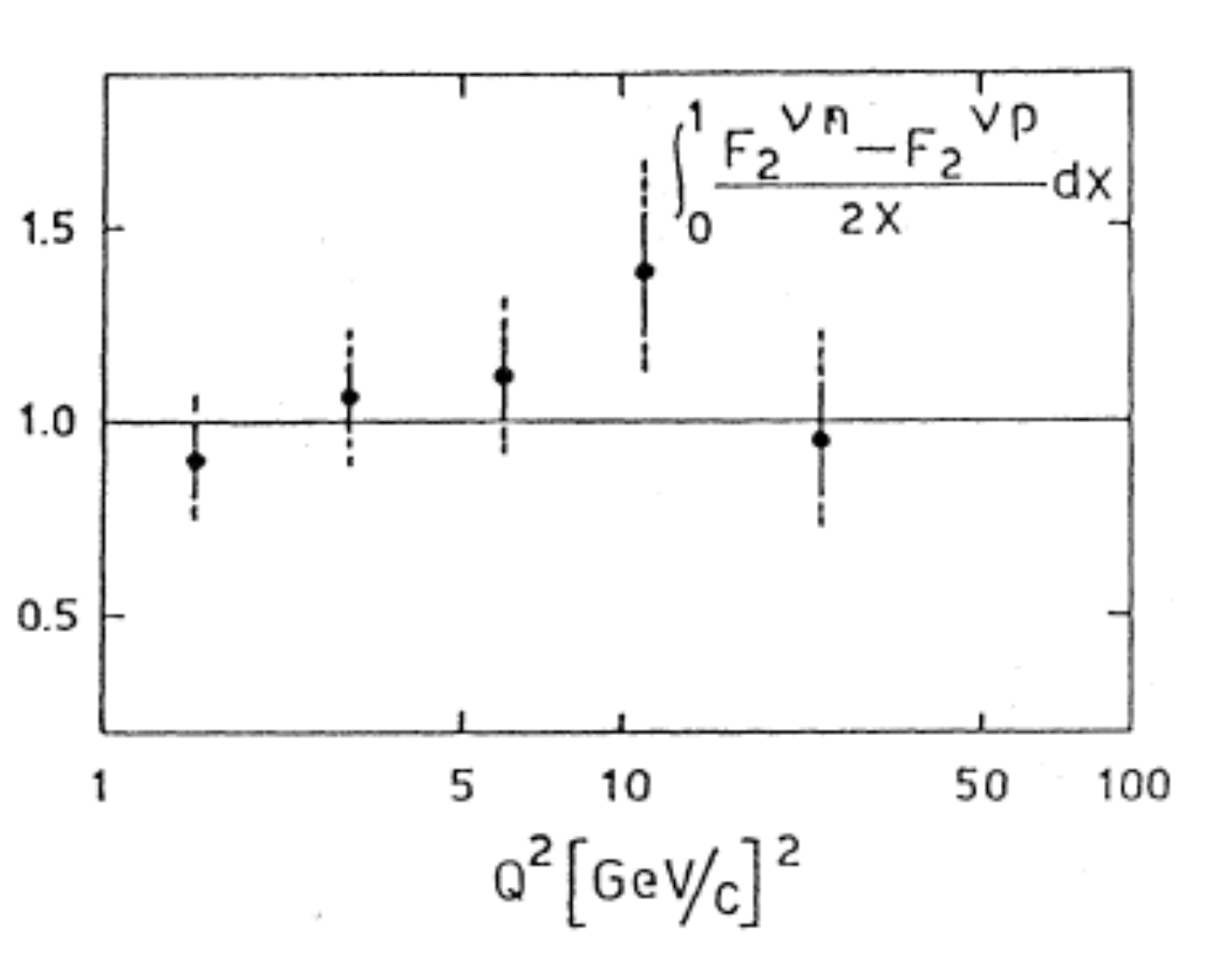}
 \caption{Experimental results for the ``Adler sum rule'' from the WA25 
 group, Ref.\ \protect\cite{WA25,WA25b}. Note that what is measured is given 
 by Eq.\ (\protect\ref{Adlercsv}) and not the Adler sum rule of 
Eq.\ (\protect\ref{eq:Adlerp}).}
 \label{fig52}
 \end{figure}

 Fig.~\ref{fig52} shows the experimental situation regarding the Adler sum 
 rule. The experimental points are from the WA25 experiment \cite{WA25,WA25b}. 
 The experimental data are shown for several values of $Q^2$.  The average 
 value is $\widetilde{S}_{\A} = 1.01 \pm 0.08 \ (stat) \pm 0.18\ (syst)$.  
 However, the total $\nu N$ cross section used by the WA25 group is smaller 
 than the presently accepted value \cite{blair,berge}.  If the WA25 value is 
 readjusted to fit this total cross section their result becomes 
 $\widetilde{S}_{\A} = 1.08 \pm 0.08 \ (stat) \pm 0.18\ (syst)$.

 The results show no significant $Q^2$ dependence.  The large errors arise
 from the factor $1/x$ in the integral, Eq.~(\ref{eq:Adlerp}), which 
 gives a heavy weighting to the data at small $x$.
 The paucity of data in this region and the
 relatively large error bars there give a large uncertainty
 in the sum rule value.  There are currently efforts underway to 
 develop new generation neutrino experiments, with substantially  
 higher fluxes than were available in the past.  If these 
 efforts come to fruition, it might be possible to test 
 the Adler sum rule with $\nu$ and $\ovnu$ beams on a proton target.

We can use the 
phenomenological sea quark CSV amplitudes determined by MRST 
\cite{MRST03} to estimate the CSV contribution to the WA25 measurement.  
Assuming that the quark normalization integral is indeed one, 
then inserting the MRST sea quark function of Eq.~(\ref{eq:deldbar}) 
into the Adler sum rule, one obtains  
 \bea
 \widetilde{S}_{\A} &=&  \SA -\int_0^1\,dx\, \left[ \delta\bar{u}(x)+
 \delta\bar{d}(x) \right] \nonumber \\ 
 &=& 1 + \WTdelt\, \langle \bar{d}^p - \bar{u}^p \rangle 
 = 1.008 \ .  
 \label{eq:AdlMRST}
 \eea
The MRST result for sea quark CSV implies a difference of 
less than 1\% between the Adler sum rule and the quantity measured 
in the WA25 experiment.  Note that this result could be strongly model 
dependent because of the functional form assumed by MRST. 
   
One could also in principle measure a sum rule using antineutrino beams 
on protons and deuterium, and obtain 
 \bea
 \ovSA &\equiv& \int_0^1\,dx\, \left[ \frac{F_2^{W^- p}(x,Q^2) -
 F_2^{W^- n}(x,Q^2)}{2x} \right] \nonumber \\  &=& 
 \SA +\int_0^1\,dx\, \left[ \delta\bar{u}(x)+
 \delta\bar{d}(x) \right] \nonumber \\ 
 &=& 1 - \WTdelt\, \langle \bar{d}^p - \bar{u}^p \rangle 
 = 0.992 \ .  
 \label{eq:Adlanti}
 \eea
The last line of Eq.~(\ref{eq:Adlanti}) holds if we assume the MRST 
result for sea quark CSV. Note that if the ``weak form'' of charge symmetry 
holds (see Eq.~(\ref{eq:weakform}), then 
 $\SA = \widetilde{S}_A$ and $\SA = \ovSA$. We will discuss
 this more in Sect.~\ref{Sec:eightfourfour} in connection with a  
 ``charge symmetry sum rule.''

One could also imagine measuring a similar quantity on a nucleus, rather 
than the proton. If one compares the integral of the $F_2$ structure 
function for a nucleus with $Z$ protons and $N = A-Z$ neutrons, then one 
would expect 
 \bea
&\,& S_{\A}^{\A} = \int_0^{M_{\A}/M}\,dx\, 
  \biggl[ \frac{F_2^{W^- A}(x,Q^2)}{2x} \nonumber \\ 
  &-&\frac{ F_2^{W^+ A}(x,Q^2)}{2x} \biggr] = \frac{Z-N}{A} \ . 
 \label{eq:AdlerA}
 \eea
In Eq.~\ref{eq:AdlerA} the quantity $M_{\A}$ is the nuclear mass and $M$ the 
nucleon mass, the structure functions $F_2$ are normalized per 
nucleon, and we have assumed the impulse approximation. Nuclear effects 
in neutrino DIS have been evaluated at length by Kulagin and Petti 
\cite{Kul06,Kul07,Kul07b}. They calculate nuclear modifications to 
structure functions arising from three general sources. The first category 
of effects, incoherent scattering from bound nucleons, tends to affect 
structure functions mainly at large Bjorken $x$. These are evaluated by 
Kulagin and collaborators through Fermi motion and nuclear binding effects. 
It can be shown that Fermi motion and binding effects give zero contribution 
to the nuclear 
Adler sum rule. The second types of corrections tend to affect the 
structure functions in the region $x \sim 0.1$. These can arise from 
off-shell effects or from nuclear modifications of the meson cloud. 
The third type of corrections arise from coherent nuclear shadowing or 
anti-shadowing effects. These predominantly affect the structure functions 
in the region of low $x < 0.1$. 

Because of the isovector nature of the Adler sum rule, nuclear pion 
corrections give zero contribution. Kulagin and Petti \cite{Kul07b} 
require cancellation of the isovector off-shell and nuclear shadowing 
corrections to the Adler 
sum rule. This provides constraints on these corrections. Kumano and 
collaborators \cite{Kum02,Hir03,Hir04} have also made systematic 
considerations of nuclear corrections to structure functions and nuclear 
parton distributions.

 \subsubsection{Gross-Llewellyn Smith sum rule}
 \label{Sec:eightfourthree}

 The Gross-Llewellyn Smith (GLS) Sum Rule \cite{GLS} is derived from the
 $F_3$ structure functions for neutrinos and antineutrinos. This is also 
 called the \textit{baryon sum rule}~\cite{Hin96}. If we 
 sum the $F_3$ structure functions for neutrinos and antineutrinos 
 on an isoscalar target, then from Eq.~(\ref{F3plus}) we obtain 
 \bea
&\,& S_{\GLS} \equiv \int_0^1 \frac{dx}{2x} \,\left[ xF_3^{W^+ N_0}(x) + 
  xF_3^{W^- N_0}(x) \right] \nonumber \\ &=& \int_0^1 \biggl[ u_{\V}(x) 
  + d_{\V}(x) + s^-(x)  \nonumber \\ &-&  \frac{(\delta u_{\V}(x) + 
  \delta d_{\V}(x))}{2} \biggr] \, dx \nonumber \\ &=& 3 \, 
  \left[ 1 - \frac{\alpha_s(Q^2)}{\pi} - 
  a(n_f)\left(\frac{\alpha_s(Q^2)}{\pi}\right)^2 \right. \nonumber \\ &-& 
  \left. b(n_f)\left(\frac{\alpha_s(Q^2)}{\pi}\right)^3 \right] 
  + \Delta HT \ .
 \label{SGLSdef}
 \eea
 The result $S_{\GLS} =3$ follows from the normalization of the quark 
 valence distributions.  An identical prediction would be obtained using 
 either a proton or neutron target in the sum rule.  The Gross-Llewellyn 
 Smith sum rule holds only in leading twist approximation, and only to 
 lowest order in the strong coupling constant $\alpha_s$. Our expression for 
 the GLS sum rule thus includes a QCD correction (the term in square
 brackets in Eq.~(\ref{SGLSdef})), which was derived
 by Larin and Vermaseren \cite{larin}, and the quantity $\Delta HT$
 represents a higher twist contribution \cite{braun}.

 As is the case for the Adler and Gottfried sum rules, the
 Gross-Llewellyn Smith sum rule requires that the structure
 function be divided by $x$ in performing the integral.  This
 gives a strong weighting to the small-$x$ region, such that
 as much as 90\% of the sum rule comes from the region $x \leq 0.1$.
 Of the three sum rules we discuss in this review, the GLS sum rule is 
 experimentally the best determined.  The most precise value has
 been obtained by the CCFR collaboration \cite{CCFR}, which
 measured neutrino and antineutrino cross sections on iron
 targets, using the quadrupole triplet beam (QTB) at Fermilab.
 A summary of experimental details for precision measurements using  
 high-energy neutrino beams is given in the review article by Conrad, 
 Shaevitz and Bolton \cite{Con98}. 

\begin{figure}
\center
\includegraphics[width=3.2in]{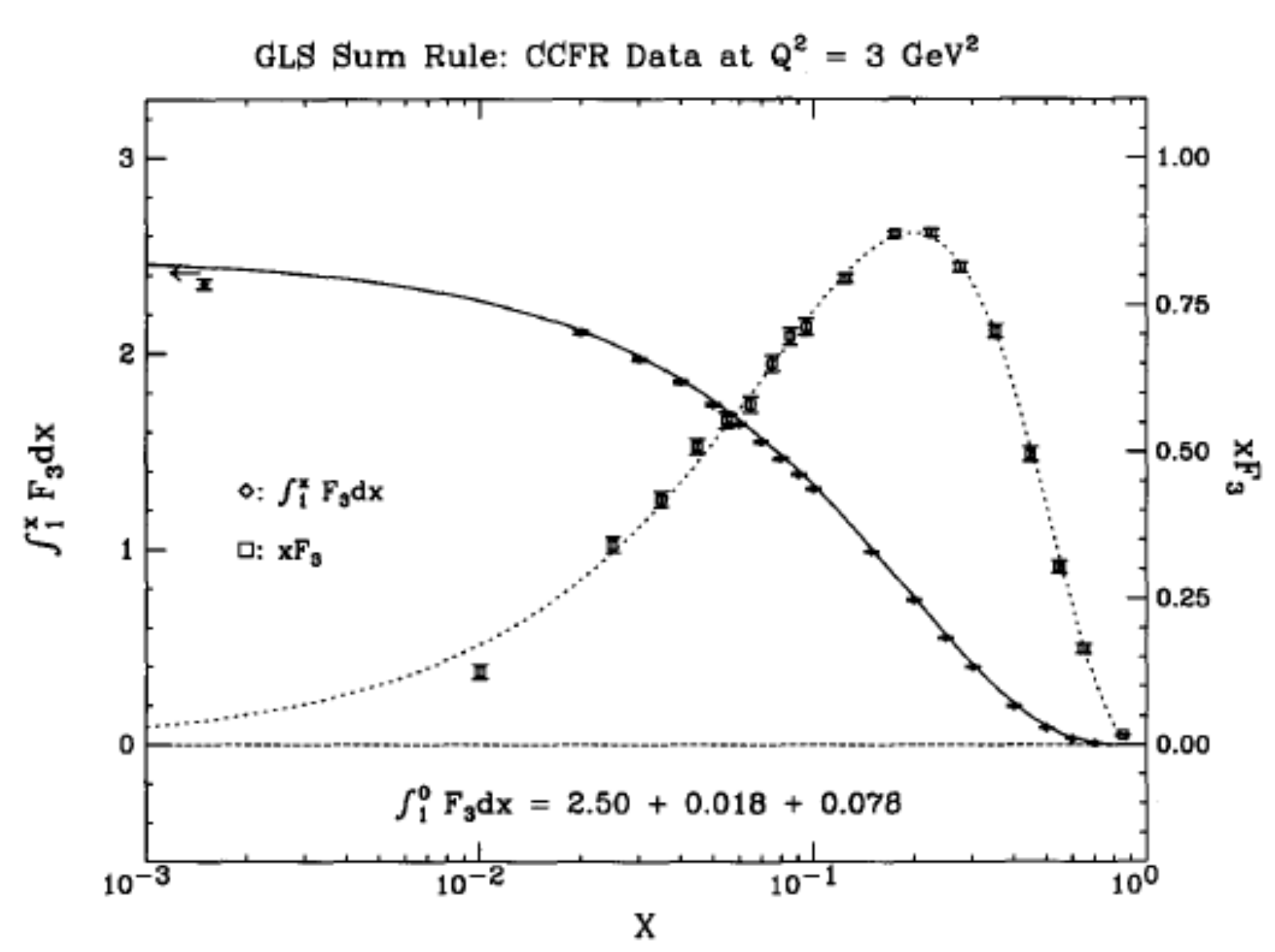}
 \caption{Experimental results for Gross-Llewellyn Smith sum rule,
 Eq.\ \protect\ref{SGLSdef}, from CCFR group, Ref.\ \protect\cite{CCFR}. 
 Squares: $x\overline{F}_3(x)$, sum of neutrinos plus antineutrinos, at 
 $Q^2 = 3$ GeV$^2$. Dashed curve: analytic fit to $x\overline{F}_3$. Diamonds: 
 approximation to the integral $S_{\GLS}(x)$ of Eq.~\protect\ref{eq:SGLSx}. 
 Solid line: fit to the integral $S_{\GLS}(x)$.}
 \label{fig53}
 \end{figure}

 In Fig.~\ref{fig53} we show the CCFR measurements and the experimental values 
 of $x\overline{F}_3(x)$ (the sum of $xF_3$ for neutrinos plus that for 
 antineutrinos) vs.~$x$.  They obtain cross sections at several values of 
 $x$ and $Q^2$. The squares give the value of $x\overline{F}_3(x)$ 
 interpolated to an average momentum 
 transfer $Q^2 = 3$ GeV$^2$ (this is the mean $Q^2$ for the lowest $x$-bin 
 in the CCFR experiment, since the lowest $x$ values contribute the 
 greatest amount to the GLS sum rule). The dashed curve is the best fit to 
 $x\overline{F}_3$ of the form $Ax^b(1-x)^c$. The CCFR reported value for 
 the sum rule at 
 this $Q^2$ value is $S_{GLS} = 2.50 \pm 0.018 \ (stat) \pm 0.078 \ (syst)$.  
 The GLS sum rule
 is therefore known to 3\%. Because of the large contribution to the GLS 
 sum rule from small $x$, one measures $xF_3$ at various values of $x$, 
 and evaluates the integral 
\be
S_{\GLS}(x) = \int_x^1 \frac{dy}{2y} \,\left[ yF_3^{W^+ N_0}(y) + 
  yF_3^{W^- N_0}(y) \right].
\label{eq:SGLSx} 
\ee
The Gross-Llewellyn Smith sum rule is then obtained by estimating the 
limit  
\be
S_{\GLS} = {\rm lim}_{x \rightarrow 0} \ S_{\GLS}(x) \ .
\label{eq:GLSlim} 
\ee
The solid curve in Fig.~\ref{fig53} is $S_{\GLS}(x)$.  

\begin{figure}
\center
\includegraphics[width=2.3in]{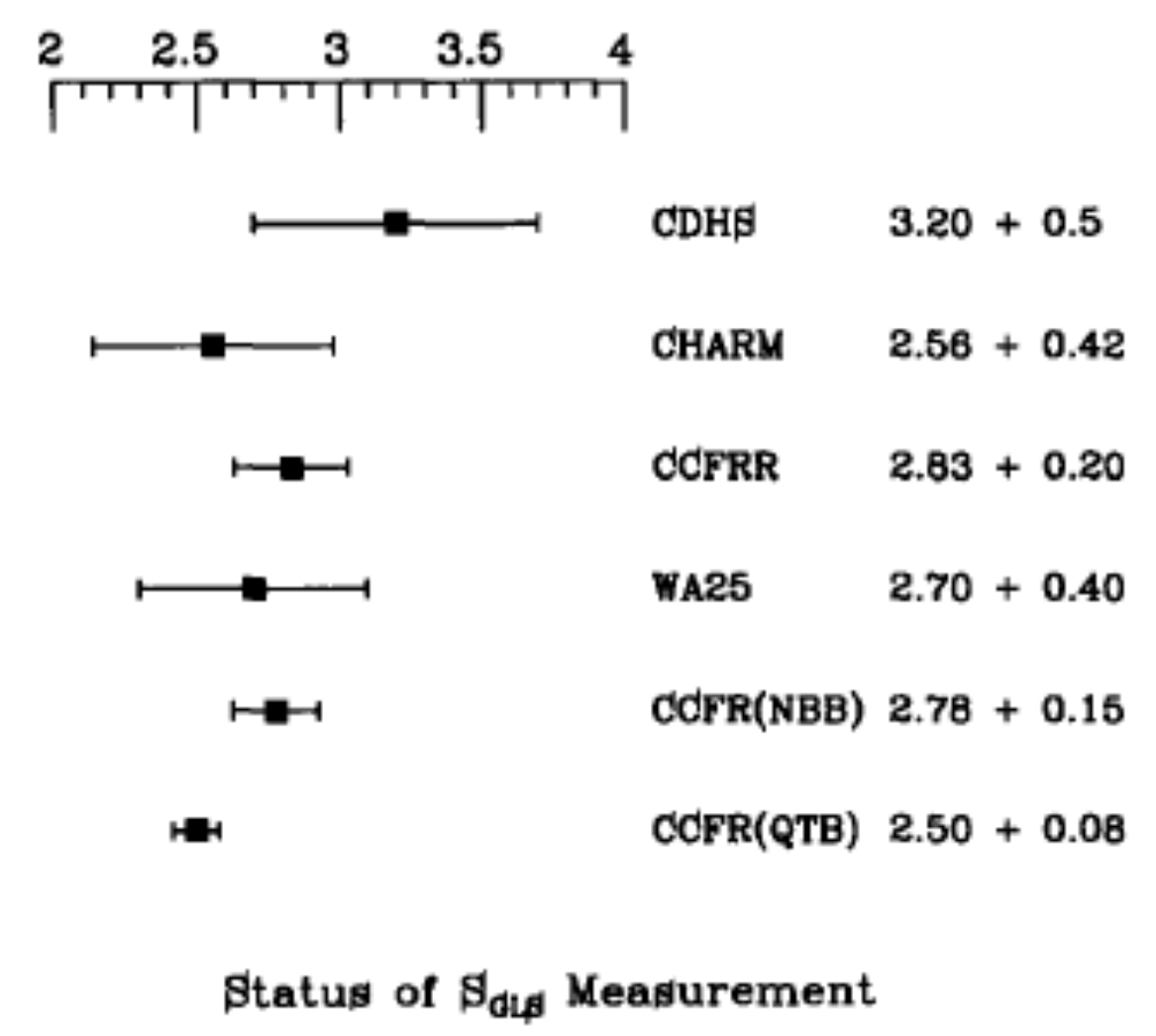}
 \caption{Experimental results for Gross-Llewellyn Smith sum rule,
 and their errors, for a series of experiments, in chronological order
 from top to bottom.}
 \label{fig54}
 \end{figure}

 A theoretical value for the Gross-Llewellyn Smith sum rule requires 
 evaluating the QCD corrections.  Calculations of the GLS sum rule include 
 next-to-leading order QCD corrections, using a QCD scale parameter
 $\Lambda_{QCD} = 213 \pm 50$ MeV.  With this scale parameter and NLO QCD 
 corrections, one obtains a theoretical prediction $S_{GLS} = 2.63 \pm 0.04$ 
 \cite{Mis90}. The theoretical prediction is two standard 
 deviations above the experimental value.  In Fig.~\ref{fig54} we show the 
 evolution over time of the GLS sum rule value. The measurements shown 
 are from the CDHS \cite{CDHS2}, CHARM \cite{CHARM}, CCFRR \cite{Macfar}, 
 and WA25 \cite{WA25} collaborations. There are also two points from 
 the CCFR measurements, the first using the Narrow Band Beam (NBB) 
 neutrino data \cite{NBB,Mishra} and the second using the QTB 
 data \cite{CCFR} from the Fermilab Tevatron. 

 The errors on the GLS sum rule are now at a level where the value
 of the strong coupling constant $\alpha_s$ is a major source of
 error.  The CCFR group now have data on $xF_3$ over a wide enough
 range of $Q^2$ that, together with renormalized data from several other
 experiments, they may be able to evaluate the GLS sum rule
 without extrapolation for a large range of $Q^2$ values.
 This raises the hope that one can calculate the Gross-Llewellyn Smith
 sum rule as a function of $Q^2$, and use the resulting $Q^2$
 dependence of the sum rule to determine $\alpha_s(Q^2)$.
 The CCFR group has recently re-calculated both
 the GLS sum rule, and the strong coupling constant $\alpha_s$
 \cite{harris}.  With data of this quality over a large $Q^2$ range,
 it may be possible to use the $Q^2$ dependence to put constraints on
 the strong coupling constant.  Additional information regarding this
 procedure can be found in the thesis by Seligman \cite{Sel97a}.

 The structure functions $xF_3^{W^+ N_0}(x) + xF_3^{W^- N_0}(x)$, which
 form the integrand for the GLS sum rule, are obtained by
 taking the difference between cross sections for neutrinos
 and antineutrino charged-current processes on isoscalar targets 
 \cite{Con98}.  In the limit of exact charge 
 symmetry, the $F_2$ structure functions exactly cancel in this 
 subtraction, and only the $F_3$ structure functions survive.  However,
 CSV effects make additional contributions to this integrand,
 \IE~using Eq.~(\ref{eq:Fccdef}),
 \bea
&\,& \frac{3\pi}{2 G^2M_NE} \left( d\sigma^{\nu N_0}/dx - 
  d\sigma^{\bar{\nu} N_0}/dx \right) \nonumber \\ &=&
  \frac{1}{2}\left( xF_3^{W^+ N_0}(x,Q^2) + xF_3^{W^- N_0}(x,Q^2)
  \right) \nonumber \\
  &+& F_2^{W^+ N_0}(x,Q^2) - F_2^{W^- N_0}(x,Q^2)
  \nonumber \\ &=& x\,\biggl[ u^p_{\V}(x) + d^p_{\V}(x) +
  3s^-(x) - c^-(x) \nonumber \\ &-&  \frac{3}{2}\delta u_{\V}(x)  + 
  \frac{1}{2}\delta d_{\V}(x) \biggr] \ . 
 \label{F3csv}
 \eea

 In addition to the light valence quark distributions, Eq.~(\ref{F3csv}) 
 contains additional contributions from strange and CSV `valence' 
 quark distributions.  However, the CSV amplitudes have no effect
 on the GLS sum rule value.  Since quark valence distributions
 obey the normalization conditions of Eqs.~(\ref{qnorm}) and 
 (\ref{csvnorm}), the contributions from valence strange and CSV terms 
 must integrate to zero. Note, however, that the valence quark CSV 
 effects contribute to the integral $S_{\GLS}(x)$ at any finite value 
 of $x$, and that the CSV effects vanish only upon integration over all $x$.  

\begin{figure}
\center
\includegraphics[width=2.8in]{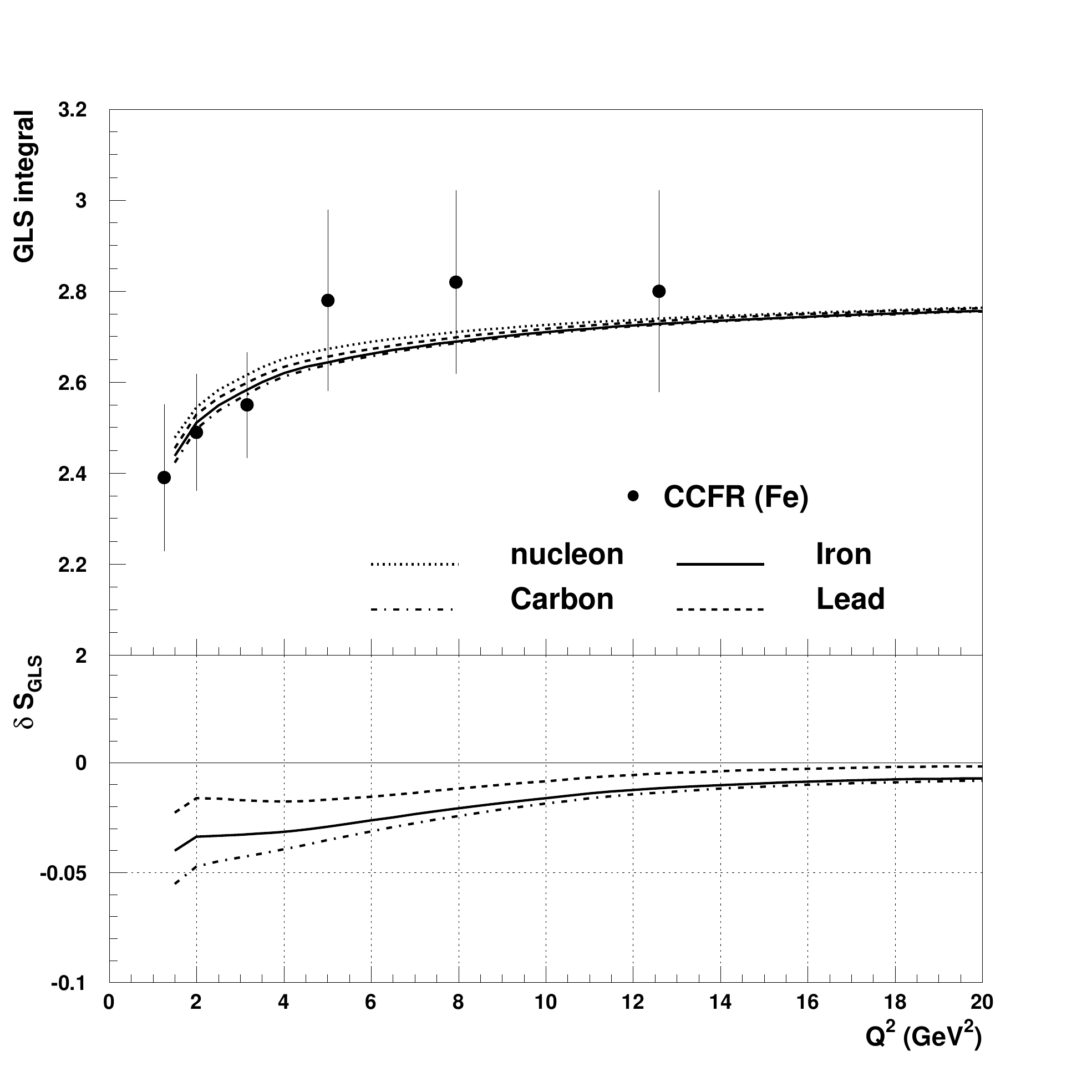}
 \caption{Nuclear corrections to the Gross-Llewellyn Smith sum rule, as 
 a function of $Q^2$, from Ref.~\protect\cite{Kul07b}. Data points are from 
 the CCFR measurements on iron, Ref.~\protect\cite{CCFR}. Dotted curve: 
 nucleon results; dot-dashed curve: results for carbon; solid curve: results 
 for iron; dashed curve: results for lead.}
 \label{Fig:GLSQ2}
 \end{figure}

 Since the GLS sum rule is evaluated on nuclear targets (for the CCFR 
 measurement, on iron), we need to consider nuclear modifications of 
 structure functions and their potential effect on the GLS sum rule. 
 Such an investigation has been carried out by Kulagin and Petti 
 \cite{Kul07b}, using methods that were summarized in our discussion 
 of the Adler sum rule (see Sect.~\ref{Sec:eightfourtwo}). As for 
 the Adler sum rule, the Fermi motion and nuclear binding corrections 
 have zero effect on the GLS sum rule. In the limit $Q^2 \rightarrow 
 \infty$, the off-shell and nuclear shadowing corrections also tend to 
 cancel. In Fig.~\ref{Fig:GLSQ2} we show the GLS sum rule as a function 
 of $Q^2$. The experimental points are the CCFR measurements on iron, 
 vs.~$Q^2$. The curves are nuclear calculations of Kulagin and Petti 
 \cite{Kul07b} for various nuclei. The dotted curve is for the nucleon 
 and the solid curve for iron. This gives an idea of the magnitude of 
 nuclear corrections to the GLS sum rule, and their $Q^2$ dependence. 
 In addition there may be additional corrections from the pseudo-CSV effects 
 suggested by Clo{\"e}t \EA~\cite{Cloet:2009qs} and discussed in 
 Sect.~\ref{Sec:pseudoCSV}.  

 \subsubsection{A ``Charge Symmetry Sum Rule''}
 \label{Sec:eightfourfour}
 \mb{.5cm}

 In Sect.~\ref{Sec:eightfourone} we showed that the Gottfried sum rule 
 contains contributions from both charge symmetry violation and from 
 sea quark flavor asymmetry.  If sufficiently accurate experimental data 
 can be obtained, one could define sum rules which could differentiate between
effects due to parton charge symmetry violation and those arising from 
differences in light sea PDFs, \IE, $\bar{d}^p(x) \ne \bar{u}^p(x)$.  
Ma \cite{Ma92} defined a ``charge symmetry'' sum rule in terms of
 the $F_2$ structure functions for charged current neutrino
 and antineutrino interactions on the neutron and proton,
 \bea
 S_{\CS} &\equiv& \,\int_0^1 \,\frac{dx}{x}\, \left[
 F_2^{W^+ p}(x) + F_2^{W^- p}(x) \right. \nonumber \\ &\,& \left. 
  - F_2^{W^+ D}(x)-  F_2^{W^- D}(x) \right] \nonumber \\ 
  &=&  2\int_0^1\,dx\, \left[ \delta\bar{u}(x)+
 \delta\bar{d}(x) \right] \quad .
 \label{eq:MaCSrule}
 \eea
 In Eq.~(\ref{eq:MaCSrule}), $F_2^{W^+ D}(x)$ is the $F_2$ structure 
 function per nucleon for neutrino charged-current DIS on the 
 deuteron.  From Eq.~(\ref{eq:MaCSrule}) we see that if either the strong
 form or weak form of charge symmetry holds for the nucleon sea
 quark distributions, then $S_{\CS}$ will be zero. A deviation of 
 this sum rule from zero would signal either a violation of parton charge 
 symmetry, or a contribution from higher-twist terms. The higher twist terms 
 would be expected to become progressively smaller with increasing $Q^2$.  
 Just as for the Adler sum rule, there are no QCD corrections to the
 charge symmetry sum rule.

 The charge symmetry sum rule of Eq.~(\ref{eq:MaCSrule}) is closely related 
 to what has been measured as the Adler sum rule (see the discussion in 
 Sect.~\ref{Sec:eightfourtwo}).  We can easily see that
 \bea
 \widetilde{S}_{\A} &=& \int_0^1\,dx\, \left[ \frac{F_2^{W^+ n}(x,Q^2) -
 F_2^{W^+ p}(x,Q^2)}{2x} \right] 
 \nonumber \\ &=& \SA - \frac{S_{\CS}}{2} \ .
 \label{ScSa}
 \eea 
  The WA25 group \cite{WA25} measured cross sections from 
 neutrinos on protons and deuterium, so their integral should 
 give the Adler sum rule minus one-half the charge symmetry
 sum rule.  However, as is shown in Fig.~\ref{fig52}, errors in the WA25 
 measurement are of the order of 20\%, so the charge symmetry sum rule at 
 present is consistent with zero at the 40\% level, if we assume that the 
 Adler sum rule is one. We can obtain an 
 estimate of the charge symmetry sum rule from the MRST global fit 
 including sea quark CSV discussed in Sect.~\ref{Sec:eighttwo}. Using the 
 MRST form for sea quark CSV, Eq.~(\ref{eq:AdlMRST}) with best value 
 $\widetilde{\delta} = 0.08$ predicts $S_{\CS} \sim -0.016$.  The predicted 
 quantity is extremely small; however, we stress that this result is 
 based upon the (strongly model-dependent) functional form for sea quark CSV 
 chosen by MRST (see Eq.~(\ref{eq:seaCSV})).

 \section{Summary and Outlook}
 \label{Sec:Summary}
 \mb{.5cm}

 We have reviewed the features of charge symmetry, an approximate symmetry 
 in particle and nuclear systems, as it relates to parton distributions. 
 First, we reviewed the relation between high energy cross sections and 
 parton distributions, in terms of structure functions. Then we wrote 
 the structure functions in terms of parton distribution functions. Until 
 recently, phenomenological parton distribution functions assumed the  
 validity of charge symmetry. This meant that PDFs for the neutron could 
 be given in terms of those for the proton. If we relax this assumption, 
 we must differentiate between neutron and proton PDFs. This requires the 
 introduction of charge symmtry violating PDFs. 

 In Sect.~\ref{Sec:PDFs} 
 we expanded the structure functions in terms of parton distributions 
 without making the assumption of charge symmetry. We then derived relations 
 between these structure functions. Some relations that exist in the limit 
 of exact parton charge symmetry must be modified if we relax that assumption. 

 In Sects.~~\ref{Sec:MRSTcsv} and \ref{Sec:sevenone}, we reviewed the 
 phenomenological and theoretical  
 situation regarding parton charge symmetry. One theoretical method for 
 predicting partonic CSV contributions is to examine the dependence of quark 
 models for parton distributions on variations in quark and nucleon mass. 
 For valence quarks, we showed that this leads to predictions for the 
 magnitude and sign of the CSV terms. The quantities $\deluv (x) = \uv^p(x) 
 - \dv^n(x)$ and $\deldv (x) = \dv^p(x) - \uv^n(x)$ are found to be opposite 
 in sign and roughly equal in magnitude. Since at large Bjorken $x$ one 
 has $\dv (x) << \uv (x)$, the fact that the valence CSV distributions are 
 predicted to be roughly equal implies that the percent charge symmetry 
 violation for the ``minority'' valence quark distribution should be 
 substantially greater than for the ``majority'' valence quark distribution. 

 There now exist phenomenological valence CSV PDFs from the MRST group 
\cite{MRST03}. They 
 assumed a particular functional form with one overall free parameter. That 
 parameter was varied in a global fit of high energy experimental data. MRST 
 obtained a very shallow minimum in fitting the high energy data. At the 
 90\% confidence level the free parameter $\kappa$ multiplying this 
 phenomenological form could range between $-0.8$ and $+0.65$, with a best 
 fit value $\kappa = -0.2$. MRST chose a functional form such that $\deluv$ 
 and $\deldv$ were required to be equal and opposite. This assured that the 
 total momentum carried by valence quarks in the neutron and proton were 
 equal (this quantity is reasonably well fixed by experiment). This choice 
 by MRST agrees reasonably well with theoretical valence quark CSV PDFs 
 obtained from quark models. In fact the best fit of MRST is in surprisingly 
 good agreement with theoretical valence CSV parton distributions from 
 Sather \cite{Sat92} and Rodionov \EA~\cite{Rod94}. 

 In Sect.~\ref{Sec:QEDsplt} we discussed an additional mechanism for charge 
 symmetry 
 violation. This occurs when a quark radiates a photon, in analogy with the 
 well-known case where a quark radiates a gluon. If one incorporates these 
 photon radiation terms into the QCD evolution equations, these give rise to 
 a new type of charge-symmetry violation. These ``QED splitting'' terms have 
 been analyzed by two groups \cite{MRST05,Glu05} with qualitatively 
 similar results.  

 In Sect.~\ref{Sec:seventwo} we reviewed experimental limits on parton 
 charge symmetry. Parton charge symmetry violation has never been directly 
 observed. The strongest upper 
 limits on parton CSV come from comparison of the $F_2$ structure functions 
 obtained from charged lepton DIS with those extracted from 
 charged-current DIS arising from neutrinos and antineutrinos, with both 
 taken on isoscalar targets. We reviewed the upper limits that can be 
 extracted by comparison of the NMC $\mu-D$ reactions \cite{Ama91,Ama92,NMC97} 
 with charged-current DIS for $\nu$ and $\bar{\nu}$ on Fe, from the CCFR 
 \cite{Sel97} and NuTeV \cite{Yan01} experiments. Comparison of these 
 experiments requires  
 a number of corrections, from absolute cross sections normalizations to 
 nuclear effects to shadowing corrections. However, the most thorough studies 
 to date place upper limits of parton CSV at about the $6-10\%$ level in the 
 region $0.03 \le x \le 0.4$. 

 The NuTeV group has obtained an independent measurement of the Weinberg 
 angle by measuring charged-current and neutral-current DIS for $\nu$ and 
 $\bar{\nu}$ on an iron target \cite{Zel02a,Zel02b}. They obtain a value 
 for $\sintW$ that differs by three standard deviations from the value 
 obtained at the $Z$ pole. In Sect.~\ref{Sec:sixthreetwo} we review this 
 situation in considerable detail. We reviewed possible contributions from 
 a number of ``QCD effects'' on the NuTeV measurement. In particular, we 
 showed that valence parton CSV effects had the possibility to make 
 substantial contributions to the NuTeV Weinberg angle  
 measurement. At the 90\% confidence level obtained in the MRST 
 phenomenological fit, partonic CSV could completely remove the NuTeV anomaly, 
 or alternatively could make it twice as large. We showed that theoretical 
 CSV effects predicted by quark models and obtained from QED splitting 
 are likely to remove approximately two-thirds of the NuTeV anomaly in 
 the Weinberg angle.  

In Sect.~\ref{Sec:pseudoCSV} we discussed a new nuclear reaction mechanism. 
This concerns the differential effect of $\rho$ exchange on protons and 
neutrons. Cl{\"o}et \EA~\cite{Cloet:2009qs} have pointed out that this will 
produce effects that mimic those of CSV in a nucleus with $N > Z$. As a 
result we refer to this as \textit{pseudo CSV}. It was pointed out that 
this should produce a characteristic $A$ dependence of the EMC effect; 
experiments have been proposed to look for evidence of this effect. It is 
predicted that this effect would account for up to two-thirds of the 
NuTeV discrepancy in the Weinberg angle. 

The current upper limits on partonic CSV for valence quarks are still 
 reasonably large. Both the 90\% confidence level of the MRST phenomenological 
 fit, and the best direct measurement from comparison of the $F_2$ structure 
 functions for charged-lepton and charged-current DIS, give upper limits on 
 valence quark CSV of the order of a few percent. Medium and high-energy 
 facilities have now reached a precision where one 
 could envision dedicated experiments that would either measure parton 
 charge symmetry violation, or significantly reduce the upper limits on 
 partonic CSV. 

 In Sect.~\ref{Sec:seventhree} we discussed four such experiments. 
 Since tests of parton charge symmetry require comparison of PDFs in the 
 neutron and proton, all of these 
 experiments require isoscalar targets. The first of these is a comparison 
 of Drell-Yan cross sections for $\pi^+$ and $\pi^-$ projectiles on a 
 target such as deuterium. Studies in the valence region for both pion 
 and nucleon have the possibility of revealing valence quark CSV effects. 
 A second experiment would be parity-violating DIS in $e-D$ reactions. 
 Such an experiment is currently being planned for Jefferson Laboratory 
 following the 12 GeV upgrade; it would have the potential to measure 
 CSV effects if the PV asymmetry could be measured to roughly one percent. 

 A third experiment that might test parton charge symmetry is a comparison 
 of $\pi^+$ and $\pi^-$ electroproduction from deuterium. Sufficiently 
 precise experiments also have the possibility of revealing CSV effects. 
 Identification of CSV effects requires that factorization be valid to a few 
 percent, so such experiments might be most reliably carried out at a future 
 electron-ion collider. One final experiment would be a comparison of 
 $W^+$ and $W^-$ production from neutrino and antineutrino charged-current 
 DIS  on an isoscalar target. In all of these cases we have made estimates
 of the magnitude of CSV effects expected in these reactions.
 
 In Sect.~\ref{Sec:eight} we reviewed the situation regarding sea quark CSV 
 effects. Theoretically, the situation for sea quark charge symmetry is 
 not nearly as well founded as for valence quarks. In the valence region, sea 
 quark contributions are quite small, whereas they are substantial  
 at small $x$. It is very difficult to disentangle heavy quark and CSV 
 effects at small $x$. On rather general grounds one can argue  
 that the magnitude of partonic CSV effects should be given by  
 \be
 \frac{\delta q}{q} \sim \frac{\delta m}{\langle M \rangle}
\label{eq:seaCSV2}
\ee
 In Eq.~(\ref{eq:seaCSV2}), the quantity $\delta m$ would represent quark 
 mass differences, in the range 1 and 5 MeV, and $\langle M \rangle$ 
 denotes an effective mass of the system after removal of a quark. For 
 valence quark CSV one would estimate   
 $\langle M \rangle \sim 500$ MeV, or a typical diquark mass, while for 
 sea quark CSV one would expect something like $\langle M \rangle \sim 1.3$ 
 GeV, a typical mass for a three quark-one antiquark state. From this rather 
 general argument one would expect that sea quark CSV effects should be 
 significantly smaller than those for valence quark CSV. 

  In Sect.~\ref{Sec:phenCSV} we discussed the phenomenological sea quark 
 CSV studies by the MRST group \cite{MRST03}. They assumed a functional 
 form for sea 
 quark CSV with an overall free parameter which was varied in a global fit to 
 high energy data. Their best fit was obtained with a surprisingly large 
 value, about 8\%, for sea quark CSV. As we mentioned, at small values of 
 $x$ sea quarks, gluons and heavy quarks all contribute, making it 
 difficult to isolate sea quark CSV effects. 

 Since sea quark and gluon distributions are connected through QCD evolution 
 equations, sea quark CSV should in principle lead to charge symmetry 
 violation in gluon distributions. In Sect.~\ref{Sec:glueCSV} we reviewed 
 possibilities for testing CSV in gluon distributions. A recent experiment 
 measuring $\Upsilon$ production in $pp$ and $pD$ scattering could place 
 upper limits on gluonic charge symmetry violation \cite{Zhu08}.  

  In Sect.~\ref{Sec:eightfour} we reviewed one possibility to search for sea 
 quark CSV, which is to look for contributions to DIS sum rules. Several of 
 these sum rules involve the first moments of parton distributions. The 
 valence CSV distributions, and heavy quark ``valence'' 
 distributions, must give zero first moment in order to respect 
 valence quark normalization. In some of these sum rules the only terms 
  whose first moment survives are sea quark CSV contributions.  
 
 We showed that CSV contributions to the
 nucleon sea have no effect on the Gross-Llewellyn Smith or Adler 
 sum rules, but in principle they affect the Gottfried sum rule. 
 The Gottfried sum rule has contributions both from asymmetries in 
 the light quark sea (the fact that $\bar{d}(x) \ne \bar{u}(x)$), and also 
 from sea quark CSV. We also pointed out that the best existing experimental
 ``test'' of the Adler sum rule actually measures a somewhat different 
 quantity, one that contains a non-zero contribution from sea quark CSV. The 
 ``Adler sum rule'' measurement can thus be used to place an upper limit on 
 parton sea quark CSV. Finally, we introduced a new sum rule, a ``charge 
 symmetry'' sum rule which would be zero if either the ``strong form'' or 
 the ``weak form'' of charge symmetry holds.  A test of this charge symmetry 
 sum rule would require measuring the structure functions for neutrino
 and antineutrino charged-current reactions on protons and deuterium, with 
particular attention
 to the small-$x$ region. If one assumes the validity of the Adler sum rule 
 then similar information could be obtained from measurements of either 
 neutrinos or antineutrinos on an isoscalar target. 

In conclusion, in recent years much progress has been made in precision 
measurements of structure functions. From these one can extract parton 
distribution functions which are now known to considerable accuracy. 
Recent experiments are now able to focus on specific questions such as 
the gluon distributions (both spin-independent and spin-dependent) and 
the flavor content of spin structure functions. We know that parton 
distributions should have small charge symmetry-violating components. 
Recently, global fits of parton distributions have been carried out where 
one drops the assumption of charge symmetry. This gives indirect indications 
of the magnitude and shape of parton CSV. We have also discussed a series of 
experiments that could in principle reveal charge symmetry violation in 
parton distributions. Such experiments require great precision, coupled with 
an accurate knowledge of heavy quark distributions. However, we are 
optimistic that such experiments can either lower the upper limits on 
parton CSV, or can find direct experimental evidence for parton charge 
symmetry violation.  

 \vspace{0.6cm}
 {\bf Acknowledgments}
 \mb{.5cm}

 Research by one of the authors [JTL] was supported in part by the US National
 Science Foundation under research contract NSF-PHY0555232. Research by one 
 author [AWT] was supported by the US Department of Energy under Contract 
 No. DE-AC05-06OR23177, under which Jefferson Science Associates, LLC operates 
 Jefferson Laboratory. Research by one author [JCP] was supported in part by 
 the US National
 Science Foundation under research contract NSF-PHY0601067. The authors
 would like to acknowledge discussions with and contributions by 
 C. Boros, W. Melnitchouk, D.J. Murdock and G.A. Miller. One of 
 the authors [JTL] acknowledges several discussions with S.E. Vigdor regarding 
 this review, and also discussions with C. Benesh, S. Gottlieb, S. Kulagin, 
 K. Kumar, E.J. Stephenson and R.S. Thorne. One of the authors [AWT] wishes 
 to acknowledge discussions with I. Clo{\"e}t and W. Bentz. One author [JCP] 
 acknowledges discussions with G.T. Garvey and J.W. Moss. This paper is 
 Jefferson Laboratory preprint JLAB-THY-09-954.

 \end{document}